# WASHINGTON UNIVERSITY

# SCHOOL OF LAW

# A NEW SOLUTION TO MARKET DEFINITION: AN APPROACH BASED ON MULTI-DIMENSIONAL SUBSTITUTABILITY STATISTICS

BY

Yan Yang

A dissertation submitted
in partial fulfillment of the requirements
for the degree of Juris Scientiae Doctoris (JSD)
at Washington University School of Law

**For my beloved parents and grandma**

**December 2018**

**Saint Louis, Missouri**

A DISSERTATION PRESENTED TO

THE FACULTY OF WASHINGTON UNIVERSITY SCHOOL OF LAW
IN SAINT LOUIS, MISSOURI

AND TO
THE DISSERTATION COMMITTEE

JOHN N. DROBAK, GEORGE ALEXANDER MADILL PROFESSOR OF
REAL PROPERTY & EQUITY JURISPRUDENCE, PROFESSOR OF
ECONOMICS

GERRIT DE GEEST, CHARLES F. NAGEL PROFESSOR OF
INTERNATIONAL AND COMPARATIVE LAW

PETER K. CRAMER, ASSISTANT DEAN GRADUATE PROGRAMS,
LECTURER IN LAW

_______________________
JOHN N. DROBAK
SUPERVISING PROFESSOR

_______________________
GERRIT DE GEEST
EXAMINING PROFESSOR

_______________________
PETER K. CRAMER
EXAMINING PROFESSOR

IN PARTIAL FULFILLMENT OF THE REQUIREMENTS
FOR THE DEGREE OF JURIS SCIENTIAE DOCTRIS (JSD)

# ABSTRACT OF THE DISSERTATION

A New Solution to Market Definition:

An Approach Based on Multi-dimensional Substitutability Statistics

By

Yan Yang

Juris Scientiae Doctoris

Washington University School of Law

Saint Louis, Missouri

December 2018

John Drobak, George Alexander Madill Professor of Real Property & Equity Jurisprudence,

Professor of Economics, Committee Chair

Market definition is an important component in the premerger investigation, but the models used in the market definition have not developed much in the past three decades since the Critical Loss Analysis (CLA) was proposed in 1989. The CLA helps the Hypothetical Monopolist Test to determine whether the hypothetical monopolist is going to profit from the small but significant and non-transitory increase in price (SSNIP). However, the CLA has long been criticized by academic scholars for its tendency to conclude a narrow market. Although the CLA was adopted by the 2010 Horizontal Merger Guidelines (the 2010 Guidelines), the criticisms are likely still valid.

In this dissertation, we discussed the mathematical deduction of CLA, the data used, and the SSNIP defined by the Agencies. Based on our research, we concluded that the narrow market conclusion was due to the incorrect implementation of the CLA; not the model itself. On the other hand, there are other unresolvable problems in the CLA and the Hypothetical Monopolist Test. The SSNIP test and the CLA are bright resolutions for market definition problem during their time, but we have more advanced tools to solve the task nowadays. In this dissertation, we propose a model which is based directly on the multi-dimensional substitutability between the products and is capable of maximizing the substitutability of product features within each group. Since the 2010 Guidelines does not exclude the use of models other than the ones mentioned by the Guidelines, our method can hopefully supplement the current models to show a better picture of the substitutive relations and provide a more stable definition of the market.

# TABLE OF CONTENTS

# CHAPTER 1: INTRODUCTION

Market definition is an important component in the premerger investigation. It is the basis of coordinated effect analysis and calculation of the Herfindahl-Hirschman Index. Although it does not come up with a direct prediction of post-merger price, it is useful in assessing the market power on material price, product price, the possibility of collusion, limitation of entry, etc. However, new methods in market definition have not been developed for long. The Antitrust Division of the United States Department of Justice, and the Federal Trade Commission (in this dissertation, we are going to refer them as "the Agencies") adopted the Critical Loss Analysis (CLA) in the 2010 Horizontal Merger Guidelines, but the CLA itself was proposed in 1989 to help the Hypothetical Monopolist Test to determine whether the hypothetical monopolist is going to profit from the small but significant and non-transitory increase in price (SSNIP). The Hypothetical Monopolist Test and the CLA constitute the current approach in defining the relevant product market.

However, the CLA has been criticized by academic scholars for tending to conclude a narrow market. It was also distrusted by the courts in some cases, because of concluding unreasonably narrow markets. Even though the CLA was eventually adopted in the Guidelines, the criticisms still seem to be valid. However, with research in the deduction of CLA, the data used, and the SSNIP defined by the Agencies, we conclude that the narrow market conclusion is not due to the CLA model but the problem in its implementation.

On the other hand, we also find out that there are other unresolvable problems in the CLA and the Hypothetical Monopolist Test. The Hypothetical Monopolist Test and the CLA was a compromise given the limitation of the data and calculation capacity at the time it was proposed. They are bright resolutions in their background, but we have more advanced tools to solve the problem nowadays. The quantification of substitutability was believed to be impossible decades ago, but it is no longer hard now. In this dissertation, we are going to propose a model which is based directly on the substitutability between the products and is capable of minimizing the difference of product features within each group.

In Chapter 2, we discussed about the background of the research and other aspects of the premerger investigation, like the unilateral effect and the models used to assess it. We also included the coordinated effects, the background of the Hypothetical Monopolist Test, and the HHI in Chapter 2 as well. We discussed the background and the mathematic deduction of the CLA, and the different data required by different definitions of the SSNIP in Chapter 3. We also analyzed some of the incorrect criticisms of the CLA and discussed the limitation of the CLA in Chapter 3 as well. In Chapter 4, we proposed a method to define the relevant market based on the substitutability. We first introduced the background and the rationality of using quantified substitutability as the standard of market definition. We then discussed the data collection and the practibility of the models. We conclude Chapter 4 with two example of applying the model we proposed to define the relevant product market. The Chapter 5 is the conclusion.

# CHAPTER 2: BACKGROUND OF THE STUDY

## 2. 1 THE LAWS

Antitrust law seeks to maintain market competition by regulating anti-competitive conduct by market participants. Since Sherman Act was enacted in 1890, the United States antitrust law has been the trend-leader of the world's competition law for more than a hundred years. In the United States, the federal antitrust laws consist of statutory provisions including Sections 1 and 2 of the Sherman Act, Section 7 of the Clayton Act, and Section 5 of the Federal Trade Commission Act.

The Sherman Antitrust Act was passed by the United States Congress in 1890 is the world's first competition law statute. Congress passed the law as a "comprehensive charter of economic liberty aimed at preserving free and unfettered competition as the rule of trade"[1]. "The antitrust laws proscribe unlawful mergers and business practices in general terms, leaving courts to decide which ones are illegal based on the facts of each case."[2] Since 1890, the goal of U.S. Antitrust law has been "to protect the process of competition for the benefit of consumers, making sure

---

[1] Northern Pac. Ry. Co. v. United States, 356 U.S. 1, 78 S. Ct. 514, 2 L. Ed. 2d 545 (1958).

[2] Federal Trade Commission, Guide to Antitrust Laws – The Antitrust Laws, (Dec. 20, 2016, 7:23 PM), https://www.ftc.gov/tips-advice/competition-guidance/guide-antitrust-laws/antitrust-laws, url.

there are strong incentives for businesses to operate efficiently, keep prices down, and keep quality up"[3]

In 1914, Congress created the Federal Trade Commission by the Federal Trade Commission Act and the Clayton Act. The Section 7 of Clayton Antitrust Act sought to prevent anticompetitive practices in their incipiency. "To block a merger, the agency must convince a court that the merger's effect [']may be substantially to lessen competition, or to tend to create a monopoly. [']"[4] This act prohibits a merger if it "in any line of commerce or in any activity affecting commerce in any section of the country, the effect of such acquisition may be substantially to lessen competition, or to tend to create a monopoly."[5] The Hart–Scott–Rodino Antitrust Improvements Act supplemented the Clayton Antitrust Act and stipulated the parties of acquisitions which can possibly lessen the competition in its relevant market shall file notifications to Federal Trade Commission or the Assistant Attorney General in charge of the Antitrust Division of the Department of Justice pursuant to the requirement of the act thereof[6]. It also sets up threshold for mergers to be reviewed. The threshold is revised annually based on the gross national product.

---

[3] id.

[4] Federal Antitrust Act (Clayton Act), 63 P.L. 212, 38 Stat. 730, 63 Cong. Ch. 323 (1914).

[5] The 2010 Horizontal Merger Guidelines, Section 1 Overview, Washington, D.C.: U.S. Dept. of Justice (2010).

[6] Hart-Scott-Rodino Antitrust Improvements Act of 2000, 146 Cong Rec S 10848 (2000).

Section 5 of the Federal Trade Commission Act declares unfair methods of competition in or affecting commerce, and unfair or deceptive acts or practices in or affecting commerce are unlawful. Meanwhile, it sets the power to prohibit unfair practice and the scope of business under that power.

This dissertation discusses antitrust under the realm of anticompetitive mergers prohibited by Section 7 of the Clayton Antitrust Act. The prohibition results from potential anticompetitive effect of a merger, which leads to an inherent need for prediction in the execution of the law. As the congress intended, the merger enforcement should "interdict competitive problems in their incipiency and that certainty about anticompetitive effect is seldom possible and not required for a merger to be illegal." This makes antitrust premerger investigation a very special field of law, contradicting the ex post relief nature of law.

One of the most important amendment of Clayton Act Section 7 is the Clayton Act Amendments 1950 (Celler-Kefauver Act). In this amendment the Congress broadened the scope of mergers by adding stock acquisition and assets acquisition consideration. Section 8 of the Clayton Act added a situation when dealing with people who are directors of both companies that are merging. The statute also set up a monetary threshold for review. The threshold value is usually in accordance with the threshold set in the Hart–Scott–Rodino Antitrust Improvements Act. The value is set by the Federal Trade Commission.

"[I]t is highly disruptive to ['] unscramble the eggs['] by separating two firms after they have joined, merger review is usually prospective."[7] Thus, premerger investigation borrows predictive tools from economics and statistics to analyze what will likely happen if a merger proceeds as compared to what will likely happen if it does not.

---

[7] Joseph Farrell & Carl Shapiro, Antitrust Evaluation of Horizontal Mergers: An Economic Alternative to Market Definition (unpublished)

## 2. 2 THE HORIZONTAL MERGER GUIDELINES

Since economic analysis has rationalized the antitrust policies a century ago, quantitative analysis has consistently been in symbiosis with antitrust practice. A lot of questions under antitrust were analyzed by econometric models under interdisciplinary research. At the same time, the court also adopted some of the approaches in academia. To make merger analytical techniques more transparent to the public, the Department of Justice and the Federal Trade Commission (the "Agencies") promulgated the first Horizontal Merger Guidelines in 1968. The Merger Guidelines were revised for several times, with the most recent version released in August 19, 2010.

The current Merger Guidelines "reflect the ongoing accumulation of experience at the Agencies" [8]They are used to clarify existing policy and to reflect new learning in premerger investigation. While the Horizontal Merger Guidelines issued in 1992 was replaced by the current one, and thus are no longer effective, "[t]he Commentary on the Horizontal Merger Guidelines issued by the Agencies in 2006 remains a valuable supplement" [9] to the current guidelines. In contrast with Section 7 of Clayton Act, and the precedents in antitrust, the Guidelines are about the practical aspects of the merger review which are done by the Agencies before bringing a merger review suit to the court. Two aspects of impact are considered by the Agencies at this stage. They are **"unilateral effects"** and **"coordinated effects"**.

---

[8] The 2010 Horizontal Merger Guidelines, Footnote 1, Washington, D.C.: U.S. Dept. of Justice (2010).

[9] id

"A picture worths a thousand words." Before we go further about the unilateral effects and the coordinated effects, we provided a figure (Figure 1) to describe the relations between all the quantitative models in antitrust premerger investigations.

## 2.2.1 Unilateral Effects and Merger Simulation Models

Though the Guidelines did not specify the definition of unilateral effect, there are a lot definitions we can find in academic scholarship. In "Vertical restraints and the effects of upstream horizontal mergers" written by Froeb, Tschantz and Werden in 2007, defines it as "effects of a change in the merging firms' incentives, holding fixed other firms' reaction functions."[10] The unilateral effect has been defined more narrowly in many other papers. For example, Joseph Farrell and Carl Shapiro defined unilateral effect as the effect which will arise "if the merger would give the merged entity a unilateral incentive to raise prices (or otherwise harm consumers)."[11] Jonathan Baker and David Reitman defined it as "effects within a static oligopoly model."[12]

The first definition is the most consistent one to the Guidelines. We summarized the Guidelines' content and defined the unilateral effects as the possible effect of a merger caused by the merging companies alone (without the reaction of the other market participants). In other words, unilateral effects are the predicted effects caused by the possible further actions of the merging companies, under the condition that all the other market participant keep acting like the status quo. Besides unilateral effects on price, the Guidelines also discuss the unilateral effects on diminishing innovation and reducing product variety, which are neither limited to oligopoly situation or effects on price. Therefore a boarder definition is what the Guidelines are intended.

---

[10] Luke Froeb, Steven Tschantz, and Gregory J. Werden, *Vertical Restraints and the Effects of Upstream Horizontal Mergers*, Contribution to Economic Analysis 282, 369-381 (2007).

[11] Joseph Farrell & Carl Shapiro, *Antitrust Evaluation of Horizontal Mergers: An Economic Alternative to Market Definition*, BE J. Theor. Econ. 10, 1-41 (2010).

[12] Jonathan B. Baker & David Reitman, *Research Topics in Unilateral Effects Analysis*, Research Handbook on the Economics of Antitrust Law (2009).

The Agencies consider more aspects of the unilateral effect than merely the increase in price. However, the unilateral effect on price is the primary outcome of merger simulations. Merger simulation used to assess the price increase after the merger. Merger Simulation is based on several game theoretic models. In most cases, the simulation is about the manufacturers of differentiated products changing the product price according to the cost, price-demand relations and the other participants' reaction to their pricing strategy. As a result, a Bertrand oligopoly model[13] is used (market form of oligopoly is assumed). Antitrust Logit Model[14] (ALM) is one of the variations among the Bertrand Oligopoly Model. It uses a Logit regression model to "predict price and welfare effects of horizontal merger in differentiated product industries"[15]. We will discuss more about the Logit demand model in the next chapter[16] since it is also used in assessing the actual loss in Critical Loss Analysis. The Bertrand model is the most widely used merger simulation model. Analyzing a merger with a Bertrand oligopoly model requires the following steps:

---

[13] The Bertrand oligopoly model is a model derived from the Bertrand duopoly model. The classic Bertrand duopoly model uses price as the variable and calculates the manufacturers' equilibrium price when the demand changes along with the price as the demand function suggests. Different demand function applied to the Bertrand model will result in different equilibriums. The MLA use a Logit demand model within a framework of Bertrand oligopoly model.

[14] Gregory J. Werden & Luke M. Froeb. *The Antitrust Logit Model for Predicting Unilateral Competitive Effects*, Antitrust L.J. 70, 257 (2002).

[15] Gregory J. Werden, Luke M. Froeb, & Timothy J. Tardiff, *The Use of the Logit Model in Applied Industrial Organization*, Int. J. Econ. Bus. 3.1, 83-105 (1996).

[16] See, Section 2.4.2 Logit Demand Model

1.  Collect High Frequency Purchase data collected from cashers of retailers (the retail profit will be deducted from the selling price), manufacturer's cost data, consumer behavior and preference survey data and other data related to the analysis.

2.  Decide the type of demand estimation model being used. It depends on the availability of data and the state of the market. In fact, the Agencies will not only use one demand model. Although in each simulation only one model can be used, the Agencies will conduct the merger simulation with different demand curves, and assess the unilateral effect considering all the results. Figure 2[17] is an example which compares the trace of equilibrium price calculated by different demand curves. The demand curve also involves the calculation of actual loss in Critical Loss Analysis; therefore we will discuss further about it in Chapter Two.

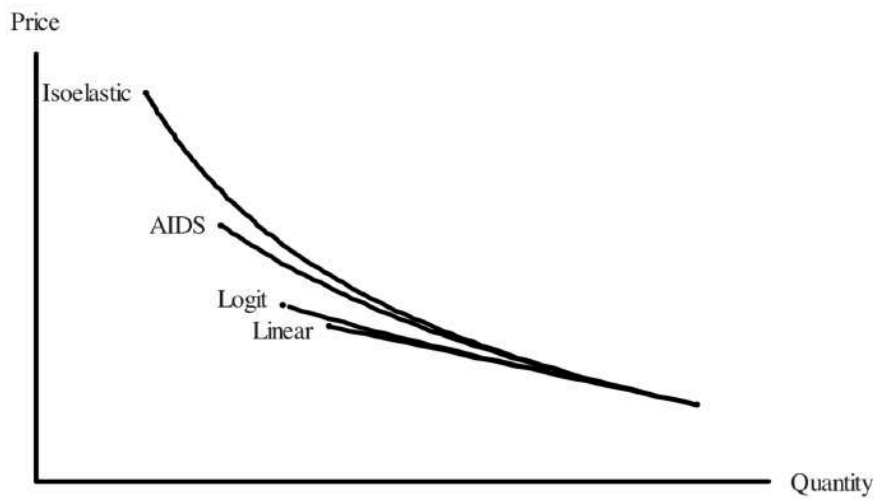

Figure 2. Four Demand Curves Plotted
between the Competitive and Monopoly Prices

---

[17] Gregory J. Werden, *Unilateral Competitive Effects of Horizontal Mergers I: Basic Concepts and Models*, Issues in Competition Law and Policy (2010).

3.  Apply the manufacturer's cost data into the viable cost function and fixed cost function. Combine the two functions to deduct a cost function using demand as the independent variable.

4.  Use the Bertrand Model to model the price-profit relation. If we set product i's price as $P_i$ ; the vector of prices for its competitors is $P_{-i}$ , the demand of product i's product is $D_i(P_i, P_{-i})$. Therefore the cost of brand i's production is $C_i(D_i(P_i, P_{-i}))$. Profit of product i $\prod^I$ will be as shown in Equation 1.

$$\prod{}^{I}(P_i, P_{-i}) = P_i \cdot D_i(P_i, P_{-i}) - C_i(\, D_i(P_i, P_{-i})\, )$$

*Equation 1*

5.  Use Equation 1 to calculate the equilibrium price $P_i{}^*$. The equilibrium will be the place where the $\prod^I$ gets its maximum value. Therefore the equilibrium condition will be as shown in Equation 2.

$$\frac{\partial \Pi^I \left( P_i, P_{-i} \right)}{\partial P_i} = D_i(P_i, P_{-i}) + [P_i - C_i'(D_i(P_i, P_{-i}))] \cdot \frac{\partial D_i \left( P_i, P_{-i} \right)}{\partial P_i} = 0$$

*Equation 2*

In the situation where the products are homogeneous or can be regarded as homogeneous (utility supply, for example), the products has full elasticity. In this case, the manufacturers can only change supply amount to influence the market price and use an upward sloping price-quantity pairs to decide the best strategic quantity[18][19]; hence a Cournot and supply function model[20] is used. In this model, the merger simulation will calculate the equilibrium supply and use the price at the equilibrium supply as the effect of the merger.

The first three steps of performing a Cournot and supply function model will be similar with the first three steps to perform a Bertrand model during Merger Simulation, except that in the second step we use supply and price function instead of using demand forms. In step 4, we will use a different profit function. Here, $\prod^I$ is still the profit of product i. The quantity produced will be $q_i$. Therefore the cost function will be $C_i(q_i)$ The aggregated production quantity of all the homogeneous products like product i is Q. Here, an inverse demand-price function is used. The price of the homogeneous products as well as product i is P = D(Q). Thus, we get the profit function of this industry as shown in Equation 3.

$$\prod{}^I(q_i, Q) = q_i \cdot P - C_i(q_i) = q_i \cdot D(Q) - C_i(q_i)$$

*Equation 3*

Therefore, the Nash, non-cooperative equilibrium quantity $q_i$ can be calculated using Equation 4, which maximizes the value of $\prod^I$.

$$\frac{\partial \Pi^I(q_i, Q)}{\partial q_i} = p + q_i \cdot D'(Q) - C_i'(q_i) = 0$$

---

[18] De Maa, Jan, and Gijsbert Zwart, Modeling the electricity market: Nuon-Reliant (2005).

[19] Peter A. G. van Bergeijk and Erik Kloosterhuis (eds.), Modeling European Mergers, Theory, Competition Policy and Case Studies, Cheltenham: Edward Elgar, 150-171 (2005).

[20] The Cournot and supply function model is a model derives from the Cournot competition model. The classic Cournot competition model uses quantity produced as the variable and calculates the manufacturers' equilibrium quantity when the price changes along with the demand as the demand function suggests.

*Equation 4*

The Bertrand model and Cournot model mentioned above are in their basic forms. The Agencies calibrate them to achieve a better simulation result of the merger according to other data collected from the companies and from customer survey. There are also a few papers in academia further discussing their application in merger[21][22][23].

There are other industries where the price is decided neither by the amount of supply nor by having homogeneous products. Industries fitting the assumptions of the auction model[24], like the ones involving one-on-one negotiation to decide a personalized service/project or the ones involving pure auctions like government procurement, are examples of these. "In many industries, especially those involving intermediate goods and services, buyers and sellers negotiate to determine prices and other terms of trade. In that process, buyers commonly negotiate with more than one seller, and may play sellers off against one another… A merger between two competing sellers prevents buyers from playing those sellers off against each other in negotiations."[25]

---

[21] Daniel F. Spulber, *Bertrand Competition When Rivals' Costs are Unknown*, J. Ind. Econ. 43, 1-11 (1995).

[22] Gregory J. Werden & Luke M. Froeb, *The Effects of Mergers in Differentiated Products Industries: Logit Demand and Merger Policy*, J. Law Econ. Organ. 407-426 (1994).

[23] Gregory J. Werden & Luke M. Froeb, *The Entry-Inducing Effects of Horizontal Mergers: An Exploratory Analysis*, *J. Ind. Econ.* 46, 525-543 (1998).

[24] An auction model is a game where the players are the buyers and the sellers. The bidders place their bids based on certain functions. The utility is the payoff and it follows certain functions as well. The players aim at maximizing the utility and place bit according to their best interest. The seller sets up the rules to maximize the buyers' bid amount.

[25] The 2010 Horizontal Merger Guidelines, Section 6.2 Bargain and Auctions, Washington, D.C.: U.S. Dept. of Justice (2010).

Buyers' playing off the sellers is a major way to keep price low in those industries, because the negotiation and personalized plan will incur significant cost. Only the "winner" of the auction can avoid the loss. Therefore, there is also a strong incentive for the sellers to keep the service/ project price as low as possible and the service/project as attractive as possible. However, this also creates a strong incentive to coordinate. The paper "*Collusion in second price auctions with heterogenous bidders*"[26] has shown that even when the bidders are heterogeneous, the bidder's net payoff from participating in collusion is surely positive and independent from the bidder's valuation. Therefore, from the perspective of coordinate effect, since the auction market usually will not have a large number of suppliers, a merger can incur a significant increase in market concentration and risk of coordination. From the perspective of unilateral effect, a bidder's net payoff means a decrease in the consumer surplus. The Guidelines comment accordingly that "[T]his alone can significantly enhance the ability and incentive of the merged entity to obtain a result more favorable to it, and less favorable to the buyer, than the merging firms would have offered separately absent the merger."[27]

According to different market rules, the auction model being applied varies from ascending oral auction model, sealed-bid auction model, second-price auction model, etc. The idea of an auction model is different from the earlier two. In the auction model, we assume the probability of winning an auction with different prices follows a certain distribution plus error. Therefore with the company cost and profit from the auction, we can assess the winning probability and price of market participants according to the distribution, and calculate the price change before and after the merger.

---

[26] George J. Mailath & Peter Zemsky, *Collusion in Second Price Auctions with Heterogenous Bidders*, Games Econ. Behav. 3, 467-486 (1991).

[27] The 2010 Horizontal Merger Guidelines, Section 6.2 Bargain and Auctions, Washington, D.C.: U.S. Dept. of Justice (2010).

Extreme value distribution is often used in computing the winning probabilities and price. For example, Froeb, Tschantz and Crooke's paper "*Mergers among asymmetric bidders: a logit second-price auction model*"[28] used extreme value distribution to predict a bidder's winning probability and winning price. It used a logit model to estimate the location parameter $\eta$ of the distribution and a minimum distance estimator to recover the unknown parameters in the distribution, like the scale parameter. With the estimation of the winning probability and price of both merging parties, we can estimate their expected value of profit from the auction. We can also do the same to assess the expected profit of the merged entity. By comparing the price when the maximum expected profit happens before after the merger, we will know the effect of the merger on auction product's price. In "*Mergers among bidders with correlated values*"[29] and "*The effects of mergers in open-auction markets*"[30], accumulative distribution function is also used in analyzing the winning price and probability in auction models.

---

[28] Luke Froeb, Steven Tschantz, & Philip Crooke, *Merger Among Asymmetric Bidders: A Logit Second-Price Auction Model* (1998).

[29] Luke Froeb & Steven Tschantz, *Mergers Among Bidders with Correlated Values*, Contrib. Econ. Analysis 255, 31-44 (2002).

[30] Keith Waehrer & Martin K. Perry, *The Effects of Mergers in Open-Auction Markets*, RAND J. Econ. 287-304 (2003).

For more details about unilateral effect calculation, see Werden and Froeb's 2010 paper, "*Unilateral competitive effects of horizontal mergers I: Basic concepts and models*"[31], and Budzinski and Ruhmer's 2009 paper "*Merger Simulation in Competition Policy: A Survey*"[32]. Both papers provide clear and detailed merger simulation methods and examples.

---

[31] Gregory J. Werden, *Unilateral Competitive Effects of Horizontal Mergers I: Basic Concepts and Models*, Issues in Competition Law and Policy (2010).

[32] Oliver Budzinski & Isabel Ruhmer, *Merger Simulation in Competition Policy: A Survey*, J. Compet. Law Econ. 6, 277-319 (2009).

## 2.2.2 Unilateral Effects and the UPP test

The Upward Pricing Pressure Test (UPP) was first proposed by Farrell and Shaprio in their paper, "*Antitrust evaluation of horizontal mergers: An economic alternative to market definition*"[33]. It is claimed to be "practical, more transparent, and better grounded in economics than are concentration-based methods"[34]. UPP is designed as a simplified screen to find mergers to be further reviewed based on the theory of unilateral effect in a heterogeneous product market. "In the pure form of our test, a merger is flagged for further scrutiny if the net effect of the two forces creates upward pricing pressure."[35]

The UPP model compares two forces caused by a merger. One is the "Upward Pricing Pressure". The other is the "Downward Pricing Pressure". The loss of direct competition between the merging parties will definitely create a unilateral effect and cause price increase if the merging party's products are substitute of each other[36]. Therefore, the merger will create an upward pricing pressure. On the other hand, the merger might also decrease marginal cost as a result of scale effect and other efficiency considerations, therefore there is also a downward pricing pressure. The UPP test assumes the existence of the efficiency. Although there are works in academia questioning the existence of efficiency[37][38]. The existence of efficiency is a generally accepted assumption.

---

[33] Joseph Farrell & Carl Shapiro, *Antitrust Evaluation of Horizontal Mergers: An Economic Alternative to Market Definition*, BE J. Theor. Econ. 10, 1-41 (2010).

[34] id.

[35] id.

[36] Raymond Deneckere & Carl Davidson, *Incentives to Form Coalitions with Bertrand Competition*, RAND J. Econ. 16, 473-486 (1985).

To be specific, in order to calculate the Upward Pricing Pressure, the UPP test uses demand models to calculate diversion ratio. It then uses diversion ratio, product price and gross marginal cost of the none price increasing product to calculate the Upward Pricing Pressure. The diversion ratio is "the fraction of unit sales lost by the first product due to an increase in its price that would be diverted to the second product."[39] In our context, it's the ratio between the sale diverted to another product and the total sale loss of one price-raising product. It was first proposed in the paper "Mergers with differentiated products"[40] and has been widely used in Antitrust premerger investigation models. To calculate the Downward Pricing Pressure, it uses demand elasticity and gross marginal cost of the price increasing product. The net Upward Pricing Pressure is the difference between the Upward Pricing Pressure and the Downward Pricing Pressure.

Set the diversion ratio from product 1 to product 2 as $D_{12}$. Set the price of product 1 as $P_1$ and the price of product 2 as $P_2$. Set the price elasticity of product 1 as $E_1$. In the final function produced by the paper, all the price and cost are price and cost before the merger. The function to calculate net UPP is shown in Equation 5.

---

$$UPP_1 = D_{12}(\overline{P}_2 - \overline{C}_2) - E_1\overline{C}_1$$

*Equation 5*

"A merger between firms selling differentiated products may diminish competition by enabling the merged firm to profit by unilaterally raising the price of one or both products above the pre-merger level."[41] The theory behind UPP test is based on the fact that if the merging companies are selling substitutive product of each other, the price increase of one product will "divert" a part of its sale to the other merging company. If the two products are close substitutes or they together are relatively more different from the other products (hence a big quantity of sale will be diverted to the other merging party), the price increase is likely to be profitable. Therefore the merged company will have an incentive to increase product price.

Regarding the Downward Pricing Pressure, since there are various forms of efficiency effect, the UPP test therefore is not fixed with the Downward Pricing Pressure factor, i.e., the product of $E_1$ and $\overline{C}_1$. This form is determined according to the most usual form of efficiency, the decrease in production cost. However, some of the products have fixed marginal cost. In those cases, the efficiency might result in increase of quantity or product quality. The UPP paper did not discuss much about this.

---

[41] The 2010 Horizontal Merger Guidelines, Section 6.1 Pricing of Differentiated Products, Washington, D.C.: U.S. Dept. of Justice (2010).

Like all the other models calculating unilateral effects, the UPP test is a non-structural way to investigate the anticompetitive effect of a merger. According to the Merger Guidelines, when it comes to the unilateral effect, the Agencies will give more weight to the UPP test than the HHI, which will be discussed later in this chapter and in depth in Chapter 4. Since it is not a structural approach, it does not rely on market definition to quantify market power. It is also less demanding in data than the Hypothetical Monopolist Test. Therefore, the authors claimed that it is better than the structural approach of market definition and market share. "This "structural presumption" drew on the then-dominant structure-conduct-performance paradigm in industrial organization, linking concentration to poor market performance. In recent decades, however, industrial organization scholars and the courts have been more apt to stress that high concentration can be compatible with vigorous competition and efficient market performance. Thus, while *Philadelphia National Bank* has never been overruled, the strength of its structural presumption has weakened over the past 30 years"[42].

However, the wide application of market definition approach is not only due to academic trend. As will be discussed in the next session, market definition is an important base of assessing coordinate effects. The fact that the non-structural approach is more discussed in industrial organizations academia in the past two decades is irrelevant to the choice of legal approach. The fruits of those discussions are all used in assessing the unilateral effects after all.

---

[42] Joseph Farrell & Carl Shapiro, *Antitrust Evaluation of Horizontal Mergers: An Economic Alternative to Market Definition*, BE J. Theor. Econ. 10, 1-41 (2010).

While most of the premerger investigation models are about price and pricing power, market power to a company means more than merely the power of pricing. Merger effect are not limited to power of pricing either, it involves the danger of coordination, suppression of entrance, discouragement to innovation, which are all relevant to the conclusion of market devision. In fact, more of the none quantitative approach done by the court are based on the market definition.[43]

As will be discussed in depth in Chapter 3, there are a lot of problems in the current method of market definition. Since the antitrust academia has widely believed a market definition approach beyond a mere artificial "bright line" is not feasible, the methodology of market definition has not developed since the invention of Critical Loss Analysis[44]. The work after it is mostly discussing and debating about it[45][46][47]. We will discuss this in detail in Chapter 3. However, the shortage of tools is not a reason why the market definition should not be done. As long as there is still a need for such market definition tools in litigation, we should continue improving it.

---

[43] Jonathan B. Baker, *Market Definition: An Analytical Overview*, Antitrust Law J. 74, 129-173 (2007).

[44] Critical Loss Analysis uses a function to calculate the critical loss. When the price increases, the quantity in demand will decrease. In most case, there will be a point where the decrease in sales is so large that the increase in price can no longer maintain profitable. The loss in sales at that point is the critical loss. The critical loss analysis compares the actual loss happens when the price increases by a SSNIP and the critical loss to determine if the company is going to profit from increasing the price.

[45] Kenneth L. Danger & H.E. Frech III, *Critical Thinking about "Critical Loss" in Antitrust*, The Antitrust Bulletin 46, 339-355 (2001).

[46] Malcolm B. Coate & Joseph J. Simons, *Critical Loss vs. Diversion Analysis: Clearing up the Confusion*, The CPI Antitrust Chronicle 1, 1-15 (2009).

[47] Malcolm B. Coate & Joseph J. Simons, *Critical Loss v. Diversion Analysis: Another Attempt at Consensus*, The CPI Antitrust Journal 1, 1-8 (2010).

This paper will propose a new way to perform market definition, which is data oriented and constant between cases. Since the conclusion is not case specific, we can also use the conclusion to set up the threshold of premerger investigation, so as to resolve another concern of the arbitrariness of the threshold and make premerger investigation more efficient.

## 2.2.3 Coordinate effect

Summarizing the Guidelines, coordinate effect means the impact that a merger will cause by reducing the hardship or increasing the probability of collusion between the merged firm and its rivals after the merger. To be specific, according to the 2010 Guidelines, there are some situations where the Agencies need to be more careful about the coordinate effect. They are[48],

1. A market where the "competitively important firm's significant competitive initiatives" can be observed by the competing firms within a relatively short period.

2. A market where the participant's "prospective competitive reward" from the consumers switching from their competitors will be significantly diminished by the competitor's timely response. Meanwhile, the competitor is also likely to conduct a timely response.

3. A market where the participants are having "small and frequent" transactions rather than "big and long-term" ones.

4. A market where the participants are having a big stake in the status quo.

If a market has fewer participants in it, the participants are more likely to observe and react according to each other's action. Also, the participants will also be in relatively large scale, so the chance that they have large scale material purchase and sales is also higher. It is self evident that if a market has fewer participants, their stake in the status quo will be bigger. Therefore a coordinated effect tends to be stronger in a more concentrated market.

---

[48] The 2010 Horizontal Merger Guidelines, Section 7.2 Evidence a Market is Vulnerable to Coordinated Conduct, Washington, D.C.: U.S. Dept. of Justice (2010).

On the other hand, for the market where the competition is marked by "leapfrogging technological innovation" (the response from the other participants cannot significantly affect the acted company's interest), the coordinate effect is smaller while the efficiency brought by the merger might be more prominent. The Agencies will also take this into consideration. Meanwhile, the previous conduct will be considered as well. The Agencies believe that the if a high-market-share firm has previously engaged in express collusion, the firm's merger is assumed to be conducive to coordinated interaction.

The four situations listed all require market delineation to further the analysis. After delineating the market, the market share can be calculated, which is a strong indication of market power. The assumption of previous conduct also involves calculating the market share. Therefore, to assess the coordinate effect, we need to conduct market definition, and find relevant product market. This is the major difference between evaluation of coordinate effect and unilateral effect. In the following session and chapters, we are going to focus on discussion about market definition methodologies.

## 2.2.4 Market Definition and SSNIP test

The concept of lessening competition was first established in 1948, in the landmark case U.S. v. *Columbia Steel Co.*[49] The Supreme Court held the steel fabricator to have not violated the antitrust law, because the acquisition of the assets purchase did not unreasonably lessen competition or attempt to monopolize the market.

The case *Brown Shoe Co. v. U.S*[50] decided in 1962 was the first merger case to emphasize the concept of the relevant market and confirmed its importance in assessing the competition consequence of a merger. The Supreme Court held that "Determination of the relevant market is a necessary predicate to a finding of a violation of the Clayton Act, 15 U.S.C.S. § 18, because the threatened monopoly must be one which will substantially lessen competition within the area of effective competition."[51] The Supreme Court also decided on the standard of defining the relevant market: based on "reasonable interchangeability of use or the cross-elasticity of demand"[52] between the merging product and the other substitutes. These two aspects have heavily influenced the 1968 Horizontal Merger Guidelines, setting up the framework of the relevant market delineation.

---

[49] United States v. Columbia Steel Co., 334 U.S. 495 (1948).

[50] Brown Shoe Co., Inc. v. United States, 370 U.S. 294 (1962).

[51] id.

[52] id.

In Brown Shoe, the Supreme Court also defined submarkets within the product market, which may exist in the relevant market defined by interchangeability and cross-elasticity. The submarket is defined by "practical indicia"[53], such as:

1. industry or public recognition of the submarket as a separate economic entity,

2. the product's peculiar characteristics and uses,

3. unique production facilities,

4. distinct customers,

5. distinct prices,

6. sensitivity to price changes,

7. specialized vendors.

Submarkets will be considered in assessing marketing control and merger consequences. If a merger will be "substantially to lessen competition or tend to create a monopoly"[54] in a submarket, it may also be enjoined. However, significantly controlling one aspect of practical indicia does not indicate the control of submarket. Like in *PepsiCo, Inc. v. Coca-Cola Co.*[55], Coca-Cola's involvement in behavior to control *"distribution of fountain syrup by independent foodservice distributors"* did not constitute a significant control to the relevant market. Therefore, it was allowed. Although defining a submarket is not necessarily quantitative, it is based on the delineation of relevant market. Therefore, market definition is an important and necessary component in merger review, even if there are other factors to be considered as well.

---

[53] id.

[54] id.

[55] PepsiCo, Inc. v. Coca-Cola Co., U.S. District Court for the District of Columbia - 641 F. Supp. 1128 (D.D.C. 1986).

One year after the Brown Shoe, in *U.S. v. Philadelphia National Bank*[56], based on the congressional concern established in the *Celler-Kefauver Amendments*, the Supreme Court announced the illegality of a merger which would have caused a significant increase in market concentration. The Supreme Court held, "[I]n certain cases, with elaborate proof of market structure, market behavior, or probable anticompetitive effects… [A] merger which produces a firm controlling an undue percentage share of the relevant market, and results in a significant increase in the concentration of firms in that market, is so inherently likely to lessen competition substantially that it must be enjoined in the absence of evidence clearly showing that the merger is not likely to have such anticompetitive effects. "[57]

Since Section 7 of the Clayton Act stipulated that "an acquisition which may be substantially to lessen competition or tend to create a monopoly shall be prohibited"[58], the question of how to define an acquisition which fits the description above has been around for decades. This landmark cases developed an approach based on the relevant market theory. Following the cases, the DOJ established the first Horizontal Merger Guidelines. The 1968 Merger Guidelines, establishing the first standard of product market definition based essentially on product distinction and substitution.

---

[56] United States v. Philadelphia Nat'l Bank, 374 U.S. 321 (1963).

[57] id.

[58] Federal Antitrust Act (Clayton Act), 63 P.L. 212, 38 Stat. 730, 63 Cong. Ch. 323 (1914).

The definition of relevant product market in the 1968 Guidelines is a long and descriptive one. "The sales of any product or service which is distinguishable as a matter of commercial practice from other products or services will ordinarily constitute a relevant product market, even though, from the standpoint of most purchasers, other products may be reasonably, but not perfectly, interchangeable with it in terms of price, quality, and use. On the other hand, the sales of two distinct products to a particular group of purchasers can also appropriately be grouped into a single market where the two products are reasonably inter-changeable for that group in terms of price, quality, and use. In this latter case, however, it may be necessary also to include in that market the sales of one or more other products which are equally interchangeable with the two products in terms of price, quality, and use from the standpoint of that group of purchasers for whom the two products are interchangeable."[59] Summarizing from this long definition, we find that the market definition at that time is based on two things:

1. Products in the relevant market should be relatively distinguishable from the products outside the relevant market.

2. Products are reasonably inter-changeable in terms of price, quality, and use from the perspective of a group of purchaser.

The 1968 Guidelines definition was established right after the landmark cases establishing the concept of market definition and market control. Compared to the market definition in the 1982 Guidelines, the one in the 1968 Guidelines is the closest to the definition by the Supreme Court.

---

[59] The 1968 Merger Guidelines, Section 3 Market Definition, Washington, D.C.: U.S. Dept. of Justice (1968).

The idea of "reasonable interchangeability" was first brought up by the Supreme Court in case "*United States v. E. I. Du Pont de Nemours & Co.*" in 1956. The Supreme Court commented, "[t]he market which one must study to determine when a producer has monopoly power will vary with the part of commerce under consideration. The tests are constant. That market is composed of products that have reasonable interchangeability for the purposes for which they are produced -- price, use and qualities considered."[60] *In Brown Shoe Co. v. United States*, the Supreme Court held, "[t]he outer boundaries of a product market are determined by the reasonable interchangeability of use or the cross-elasticity of demand between the product itself and substitutes for it."[61] The definition in 1968 Guidelines is consistent with the precedents with regard to the application of interchangeability in price, quality, and use. Meanwhile, it is also a natural deduction from the concepts of market competition and price mechanism. We are going to discuss about this in Chapter 4.

However, the 1968 definition was widely criticized because of its impossibility in quantification[62]. As a result, the Merger Guidelines of 1982 established the Hypothetical Monopolist Test, i.e., the SSNIP test (the test of a "small but significant and non-transitory increase in price"). The DOJ explained in the reprint of the Guidelines that "[t]he logic of this approach is that such a group of producers, if they are able to coordinate their behavior, have the joint ability and incentive to lessen compensation and raise prices. By contrast, a proper subset of these producers would not have this ability."[63] This "logic" is unfortunately not that logical. In the Hypothetical Monopolist Test, working with a small but significant and non-transitory increase in price can only partially prove that the manufacturers have the incentive to coordinate. This test is irrelevant with whether they are capable of doing so or not. We will discuss more about this in Chapter 3.

---

[60] United States v. E. I. du Pont de Nemours & Co., 351 U.S. 377 (1956).

[61] Brown Shoe Co. v. United States, 370 U.S. 294 (1962).

[62] Thomas E. Kauper, *The Goals of United States Antitrust Policy-The Current Debate*, Zeitschrift für die gesamte Staatswissenschaft/J. Inst. Theor. Econ. H. 3, 408-434 (1980).

Summarizing the Guidelines[64], the common approach of Hypothetical Monopolist Test with a SSNIP is as follows.

1. The Agencies will define an as-small-as-possible relevant market containing at least one product sold by one of the merging firms, then the Agencies assume that there is only one hypothetical firm who is "the only present and future seller of the products" in that relevant market.

2. The Agencies assume that the firm is profit-maximizing, and it will impose a five to ten percent increase on product price (a small but significant and non-transitory increase in price, SSNIP).

3. If after the increase, there are too many sales shifted to the product outside the control of the hypothetical monopolist, the price increase is not profitable, i.e., relevant market is too small.

---

[63] 1984 Department of Justice Merger Guidelines § 2.0. reprinted in 2 Trade Reg. Rep. (CCH) 4491 et seq. (1984).

[64] The 2010 Horizontal Merger Guidelines, Section 4.3 Implementing the Hypothetical Monopolist Test, Washington, D.C.: U.S. Dept. of Justice (2010).

In this scenario, the Agencies will bring the closest substitute product into the relevant market and test again.

4. After adding the new product under the control of the hypothetical firm, the firm will be the only supplier and so conduct a SSNIP again. The Agencies will see whether the new firm's SSNIP will be profitable…

5. This process will be repeated until the price increase is profitable. The products under control of the hypothetical monopolist consists the relevant market that the Agencies will use. The process above is also know as the SSNIP test.

It is worthy to mention that the Guidelines did not specify whether the price increase is imposed on all the product under control of the monopolist or just one or several product under control. This will be discussed in depth in Chapter two.

The SSNIP test is used to identify a set of products that are reasonably interchangeable with a product sold by one of the merging firms. According to the Guidelines, after the market boundary is delineated, the Agencies will analyze the impact of the merger on product price in the relevant market with specific information of each merger, based on the result of the Hypothetical Monopolist Test. The SSNIP test solved the problem in the earlier market definition that the interchangeability is not quantifiable—whether the SSNIP is profitable is highly correlated to interchangeability, and it's quantifiable. However, the SSNIP test is only a theoretical framework. It did not address how to determine whether the SSNIP is profitable quantitatively.

"The Guidelines have been harshly criticized for this omission, with some critics contending that such an omission makes the Guidelines unworkable in practice."[65]

As a resolution of this defect, "[w]hen the necessary data are available, the Agencies also may consider a [']critical loss analysis['] to assess the extent to which it corroborates inferences drawn from the evidence noted above"[66]. When the data is not available, the Agencies will conduct a conceptual analysis to analyze the pertinent evidence and determine whether the merger will substantially lessen competition. Although the wording in the Guidelines is "may consider", the critical loss analysis is actually the only quantitative way available to determine whether the price increase will be profitable for the hypothetical monopolist, i.e., the only way to set up a bright line on market definition and make the Hypothetical Monopolist Test work in a less arbitrary way.

The Critical Loss Analysis asks "whether imposing at least a SSNIP on one or more products in a candidate market would raise or lower the hypothetical monopolist's profits."[67] The theorem of Critical Loss Analysis is based on the idea that a price increase will incur loss of sales, while in a non-perfect competition market (which is usually the case) some customers will stay. Thus, thus if there are more customers staying after the SSNIP, the SSNIP is still profitable. The critical loss is defined as the decrease of the number of the unit sale for which the profit remains unchanged.

---

[65] Barry C. Harris & Joseph J. Simons, *Focusing Market Definition: How Much Substitution is Necessary*, J. Reprints Antitrust L. & Econ. 21, 151-172 (1991).

[66] The 2010 Horizontal Merger Guidelines, Section 4.1.3 Implementing the Hypothetical Monopolist Test, Washington, D.C.: U.S. Dept. of Justice (2010).

[67] id.

The predicted loss is defined as the decrease in the unit sale after the SSNIP. In the paper first proposing the Critical Loss Analysis[68], it is called the "Actual Loss". Therefore, if the Actual Loss is less than the Critical Loss, the SSNIP will be profitable. That's the point where the Hypothetical Monopolist Test stops adding new products into the relevant market.

As was mentioned in the earlier session, Critical Loss Analysis received a lot of criticism in the pasts few decades. Despite the criticism, the Critical Loss Analysis was adopted by the 2010 Merger Guidelines, and become the only quantitative model mentioned in the Guidelines to assess the coordinate effects. We are going to explore the Critical Loss Analysis in great details in the Chapter 3. We are also going to establish new methodologies to define a relevant market in Chapter 4, which can hopefully be a supplement or a substitution of the SSNIP hypothetical monopolist test combined with the Critical Loss Analysis.

It is worth noting that, according to the Guidelines, firms who are considered as "rapid entrants" will be regarded as market participants as well as all the firms currently earn revenues in the relevant market[69]. Whether a firm will be considered as rapid entrants are based on the commitment of entrance, sunk cost of switching production, possibility of switching geographic market to enter, possibility of switching target customers, possession of necessary asset to supply into the relevant market, and possession of idle or "swing" capacity which can produce products in the relevant market. All those aspects will be considered by the Agencies, and their approach cannot be conducted without the market definition. Therefore, again, it is important to delineate the market. Even if the UPP and the merger simulations can predict the post-merger price effect, the analysis in Unilateral Effects cannot replace the analysis which is based on the market definition.

---

[68] Barry C. Harris & Joseph J. Simons, *Focusing Market Definition: How Much Substitution is Necessary*, J. Reprints Antitrust L. & Econ. 21, 151-172 (1991).

[69] The 2010 Horizontal Merger Guidelines, 9. Entry: Timeliness, Washington, D.C.: U.S. Dept. of Justice (2010).

A geographic market means "the geography limits some customer[s'] willingness or ability to substitute to some products, or some supplier[s'] willingness or ability to serve some customers"[70]. If a geographic market exists, the Agencies will examine the market power of the merging firms in a relevant market defined with a geographic dimension and a product dimension. In case *Brown Shoe Co. v. United States*, the Supreme Court held, "[t]he criteria to be used in determining the appropriate geographic market are essentially similar to those used to determine the relevant product market. … Congress prescribed a pragmatic, factual approach to the definition of the relevant market and not a formal, legalistic one. The geographic market selected must, therefore, both correspond to the commercial realities of the industry and be economically significant."[71] Therefore, we believe our discussion in relevant product market definition will also be helpful in defining the geographic market.

### 2.2.5 Market concentration and HHI

After we have achieved the conclusion about the market definition, we can calculate Herfindahl–Hirschman Index (HHI) to measure the level of concentration in the relevant market before and after the merger, so as to assess the effect of the merger on the level of competition. HHI was established by the Agencies in the Horizontal Merger Guidelines of 2010. "Market shares may be based on dollar sales, units sold, capacity, or other measures that reflect the competitive impact of each firm in the market."[72] HHI measures the overall level of competition in the relevant market.

---

[70] The 2010 Horizontal Merger Guidelines, 4.2 Geographic Market Definition, Washington, D.C.: U.S. Dept. of Justice (2010).

[71] Brown Shoe Co. v. United States, 370 U.S. 294, 296, 82 S. Ct. 1502, 1508, 8 L. Ed. 2d 510, 519, 1962 U.S. Dist. LEXIS 2290, *1, 1962 Trade Cas. (CCH) P70,366 (1962).

The HHI is the sum of the squares of the market share of all the current market participants in the relevant market. Therefore, as was mentioned above, HHI is dependent on market definition, i.e., for now it depends on a result of the Hypothetical Monopolist Test. It is worth mentioning that HHI and the standard applied with HHI are not the only analysis done with the relevant market conclusion, even though it is the only further quantitative step following the market definition. The court will also consider factors other than market concentration about the coordinate effects based on the conclusion of the market definition.

The Agencies will consider both the post-merger HHI and the increase in HHI resulting from the merger. Based on the Agencies' experience, they classify market into three types: a market in which HHI is below 1500 is an Unconcentrated Market; a market which HHI is above 1500 and below 2500 is a Moderately Concentrated Market; and a market whose HHI is above 2500 is a Highly Concentrated Market.

---

[72] Federal Trade Commission, Mergers – Competitive Effects, Federal Trade Commission (Dec. 21, 2018, 12:38 AM), https://www.ftc.gov/tips-advice/competition-guidance/guide-antitrust-laws/mergers/competitive-effects, url.

The Agencies use HHI as a preselective method of mergers which require further investigation. A small change in concentration (defined as increase of HHI less than 100 points) or mergers occurring in an unconcentrated market does not require further analysis. A merger occurring in a moderately concentrated market causing an increase of more than 100 points in HHI or a merger occurring in a highly concentrated market causing an increase between 100 and 200 points in HHI "raises significant competitive concerns and often warrant scrutiny"[73]. A merger that occurs in a highly concentrated market causing an increase in HHI of more than 200 points "will be presumed to be likely to enhance market power"[74], while this presumption is rebuttable.

The Agencies use HHI to examine whether a merger is unlikely to raise anticompetitive concerns so the Agencies can focus on the ones do raise that concern. Thus, the HHI is not intended to be a final say of whether a merger will be permitted or not. The HHI is not perfect, especially so since its thresholds are only based on a rule of thumb. In fact, according to the quality of the market shares they can very well fit the logarithmic normal distribution: thus the log of the market shares, if squared, will fit chi-square distribution. Since we have the distribution, we can have a more scientific discussion about the thresholds.

Although the HHI can very well reflect how concentrated the market is, market concentration is only a reflection of the competition level. The Agencies are aware of that so they clarified in the Guidelines that high concentration of the market does not necessarily mean a less competitive market. The Merger Guidelines talk about some qualitative rules in deciding this, while they do not address any model which can directly quantify the competition itself. In academia, there are some other models to quantify the competition of a market, like the Panzar-Rosse Model[75] and Status Based model[76]. On the other hand, HHI does not directly quantify the level of pricing power. Therefore, the threshold of HHI increase set by the merger guidelines is arbitrarily. HHI is purely indicating the general structure of the market.

---

[73]The 2010 Horizontal Merger Guidelines, 5.3 Market Concentration, Washington, D.C.: U.S. Dept. of Justice (2010).

[74] id.

However random as it seems to be, we may regard it as following the Chi-square distribution. There is a paper *A probabilistic approach of Hirschman-Herfindahl Index (HHI) to determine possibility of market power acquisition*[77], which discussed "the probability of any market participant acquiring market power" and "a trend for this probability with the changes of HHI" based on the assumption above.

---

[75] Jacob A. Bikker, Sherrill Shaffer, & Laura Spierdijk, *Assessing Competition with the Panzar-Rosse Model: The Role of Scale, Costs, and Equilibrium*, Rev. Econ. Stat. 94, 1025-1044 (2012).

[76] Joel M. Podolny, *A Status-Based Model of Market Competition*, Am. J. Sociol. 98, 829-872 (1993).

[77] Anil Kanagala et al, *A Probabilistic Approach of Hirschman-Herfindahl Index (HHI) to Determine Possibility of Market Power Acquisition*, IEEE PES Power Systems Conference and Exposition (2004).

## 2.2.6 Limitations of the models in the Horizontal Merger Guidelines

As we have mentioned for several times, none of the models used in premerger investigation is absolutely outcome determinative. Although "throughout the history of U.S. antitrust litigation, the outcome of more cases has surely turned on market definition than on any other substantive issue"[78], neither HHI nor the other aspects in coordinated effects can decide the result of a merger review by itself. With regard to the unilateral effects, the merger simulations are not outcome determinative either. "The Agencies do not treat merger simulation evidence as conclusive in itself, and they place more weight on whether their merger simulations consistently predict substantial price increases than on the precise prediction of any single simulation."[79]

Apart from the coordinated effect and the unilateral effect, the Merger Guidelines also talk about powerful buyers, entry, efficiencies, and other aspects considered about a merger in a permerger investigation. All of them will be considered by the Agencies when deciding whether to submit a merger review to the court or not. The court will also consider all the related aspects to decide whether to approve a merger or not.

The Guidelines also emphasiz that the methodology suggested in the Guideline is not a uniform application[80]. The investigation is based on fact-specific processes, which is decided by the nature of the cases. Therefore, we believe that even though our model cannot replace the traditional approach, our model can still be a supplement to the traditional approach and help the premerger investigation.

---

[78] Jonathan B. Baker, *Market Definition: An Analytical Overview*, Antitrust Law J. 74, 129-173 (2007).

[79] The 2010 Horizontal Merger Guidelines, Section 6.1 Pricing of Differentiated Products, Washington, D.C.: U.S. Dept. of Justice (2010).

[80] The 2010 Horizontal Merger Guidelines, Section 1. Overview, Washington, D.C.: U.S. Dept. of Justice (2010).

Like the Guideline models, the methodologies proposed in this dissertation will not apply to all the cases that can be encountered in mergers either (for example, it cannot be applied in to assess the geographic market). However, the nature of our model is such that it can be applied to a nationwide market without losing its integrity in "artificial" assumptions. This advantage supplements the shortcoming of the SSNIP test that it cannot be extended into a larger market without assuming the nationwide market follows from the pattern of behavior of the market where the SSNIP is performed in (since the cashier data used to calculate the elasticity is collected from some specific states). We will discuss more about the limitation of our models in Chapter 4.

Additionally, as the name suggests, all the Horizontal Merger Guidelines do not give guidance in vertical mergers or any other types of non-horizontal mergers even though horizontal mergers are not the only type of merger which raises a concern of lessened market competition. While this dissertation will only discuss methodology in horizontal mergers, similar technology might also be capable of solving problems in non-horizontal mergers, if proper changes are made accordingly.

As suggested in the Merger Guidelines, the Guidelines are not intended to dictate or exhaust the range of evidence to be introduced in litigation. Thus we believe the new methodology proposed in the following chapters can be helpful in the investigation and litigation. Correspondingly, the new kinds of data required in the proposed methodology can also be acquired by the Agencies from the participants of the market in issue. In fact, the data used by our model can be collected from the general information of the market participants and from surveys. A survey is a traditional method used in merger simulation. The Agencies have enough experience working with data collection companies to get the data we require.

This dissertation is going to discuss the current market definition methodology in details in the second chapter, including the Hypothetical Monopolist Test (SSNIP), and the Critical Loss Analysis model and the demand elasticity models required to calculate the actual loss. In Chapter Three, we are going to establish two new models to define relevant market. In Chapter Four, we are going to discuss about the quantification of the competition and our new method to set up the threshold of HSR premerger review based on the result of the models proposed in Chapter 5. Chapter 5 will conclude this dissertation.

# CHAPTER 3: THE SSNIP TEST AND THE CLA

Ever since the landmark case, Brown Shoe Co. v. United States, market definition has been an essential task in premerger investigation. The Supreme Court held, "[d]etermination of the relevant market is a necessary predicate to a finding of a violation of the Clayton Act, 15 U.S.C.S. § 18, because the threatened monopoly must be one which will substantially lessen competition within the area of effective competition. Substantiality can be determined only in terms of the market affected. The area of effective competition must be determined by reference to a product market and a geographic market."[81] Though the case was decided more than fifty years ago, its holdings are still regularly cited by merger review cases[82][83][84] now, evidencing the importance of market definition in premerger investigations.

As we have mentioned in Chapter 2, market definition is the start of the analysis of coordinated effects, market power, market concentration, potential entrant and entry barrier. Moreover, "[t]he criteria to be used in determining an appropriate geographic market are essentially similar to

[81] Brown Shoe Co. v. United States, 370 U.S. 294 (1962).

[82] United States v. Energy Sols., Inc., 265 F. Supp. 3d 415, 2017 U.S. Dist. LEXIS 109663, 2017-1 Trade Cas. (CCH) P80,050 (2017).

[83] United States v. Anthem, Inc., 236 F. Supp. 3d 171, 2017 U.S. Dist. LEXIS 23613, 2017-1 Trade Cas. (CCH) P79,906, 2017 WL 685563 (2017).

[84] Golden Boy Promotions LLC v. Haymon, 2017 U.S. Dist. LEXIS 29782, 2017-1 Trade Cas. (CCH) P79,884 (2017).

those used to determine a relevant product market"[85] Therefore the market division methodology is also helpful in the case where a geographic market is involved.[86][87][88]

Despite the importance of market division in the legal analysis of merger reviews, the development of its methodology framework has not been advanced much since the Hypothetical Monopolist Test combined with Critical Loss Analysis was established. The academic scholarship of quantitative premerger investigation has been focusing on Unilateral Effects on price caused by merger in the past two decades. Although along with the development of merger simulation models assessing unilateral effects, the demand prediction models used in both merger simulation and Critical Loss Analysis have been advancing, the theoretical framework itself remains static.

Critical Loss Analysis has been receiving criticism for decades[89][90][91][92]. Some economists believe it should be replaced by the merger simulation models. "In some cases, more reliable simulation analyses could be constructed, but if they are it is likely that they could be used to evaluate

---

competitive effects of firm conduct directly, rendering market definition superfluous."[93] However, the reliability of the model is what an industrial organization economist considers, and it cannot replace what the court believes is necessary to decide a merger review case. What we do in premerger investigation research is to provide support to the judge, not to tell the judge what he should or should not need.

Price influence is not the only anticompetitive effect of a merger. Even though the merger simulation models can reliably predict the price, it cannot substitute the market definition to provide a bigger picture of the market competition. In fact, the Critical Loss Analysis was adopted by the 2010 Merger Guidelines despite all the problems and criticisms on the model. This fact is strong evidence of the needs for market definition. On the other hand, we also find out that some of the criticisms are not correct.

In this chapter, we are going to discuss the current approach to market definition. It will be the framework set up by the Hypothetical Monopolist Test (SSNIP test), the quantitative approach based on the Critical Loss Analysis, and the demand models required in the calculation of Actual Loss, an essential part of the Critical Loss Analysis. In addition to providing a detailed introduction, we will also defend the models from the criticisms. We are also going to discuss the pros and cons, limitations, conceptual and methodological problems of those models.

---

[93] Joseph Farrell & Carl Shapiro, *Antitrust Evaluation of Horizontal Mergers: An Economic Alternative to Market Definition*, BE J. Theor. Econ. 10, 1-41 (2010).

# 3.1 THE SSNIP TEST SINCE 1982 MERGER GUIDELINES

The 1982 Merger Guidelines (the 1982 Guidelines), established by the Department of Justice (the Department), was the first official document which announced the application of the Hypothetical Monopolist Test, which is also known as the SSNIP test. As the 1982 Guidelines elucidates, "the market definition used by the Department can be stated formally as follows: 'a market consists of a group of products and an associated geographic area such that (in the absence of new entry) a hypothetical, unregulated firm that made all the sales of those products in that area could increase its profits through a small but significant and non-transitory increase in price (above prevailing or likely future levels).'"[94]

The 1982 Guidelines implemented this definition by assuming that buyers can only switch to another product in response to a price increase if there are enough immediate alternatives. In aggregate, the alternatives will cause enough loss of sales to the company increasing the price, making price increase no longer profitable[95]. Then, the products within the hypothetical monopolist's "control" cannot consist a broad enough market. As a result, the Department will include the substitutions products into the provisional market and test again.

---

[94] The 1982 Horizontal Merger Guidelines, Section 2 Market Definition and Measurement, Footnote 1, Washington, D.C.: U.S. Dept. of Justice (1982).

[95] id.

It is worth noting that the 1982 Guideline do not mention whether the increase of the product price means all the products are under the control of the hypothetical monopolist, or that the hypothetical monopolist can increase some or one of the products' price. This omission makes a significant difference. We will discuss it in later sections.

To make sure that the patterns of the switching between the products won't change too much after the supply and demand changes according to the price increase, the Department will test to make sure there is no significant subsequent switch so that the hypothetical monopolist can stay profitable for a year.[96] This arrangement is a response to the concern from academia that the SSNIP test is unreliable because it is not dynamic to the changes in the market.[97] The Guidelines mentioned that "The potential weakness of such a market based solely on existing patterns of supply and demand is that those patterns might change substantially if the prices of the products included in the provisional market were to increase. For this reason, the Department will test further and, if necessary, expand the provisional market. "[98] This is an important modification that the Department made to assure the accuracy. If this arrangement can be done as intended, it will make the Hypothetical Monopolist Test significantly more accurate. We are going to discuss this in the later sections as well.

---

[96] id.

[97] Douglas H. Ginsburg & Joshua D. Wright, *Dynamic Analysis and the Limits of Antitrust Institutions*, Antitrust L.J. 78, 1 (2012).

[98] The 1982 Horizontal Merger Guidelines, Section 2 Market Definition and Measurement, A, Washington, D.C.: U.S. Dept. of Justice (1982).

The framework of SSNIP set up above remains unchanged since 1982. Although the SSNIP test is criticized in a lot of aspects, it was a revolution in premerger investigation that sets up a framework to replace the previous demand-elasticity test[99][100] and cluster service test[101][102], which are known to be much more arbitrary and without a clear quantitative criterion.[103][104]

Led by the SSNIP trend initiated by the United States, many other countries adopted the SSNIP test into their anti-monopoly legislation. These initiatives include the 1991 Merger Enforcement Guidelines of Canada, the 1993 Market Dominance Guidelines of Australia, 1996 Business Acquisition Guidelines of New Zealand, 1997 Draft Notice on Market Definition of the European Community, and so on.

The major criticism from academia about the SSNIP test in 1982 Guidelines is that it does not address how to determine when the "small but significant and non-transitory increase in price" (will also be noted as "SSNIP" in the following content, while being different from the

---

[99] Example case: United States v. E. I. du Pont de Nemours & Co., 351 U.S. 377 (1956).

[100] Paper commenting on this method: William M. Landes & Richard A. Posner, *Market power in antitrust cases*, Harv. L. Rev. 937-996 (1981).

[101] Example case: Behrend v. Comcast Corp., 264 F.R.D. 150, 2010 U.S. Dist. LEXIS 1049, 2010-1 Trade Cas. (CCH) P76,869 (2010).

[102] Paper commenting on this method: Jonathan B. Baker, *Market definition: An analytical overview*, Antitrust L.J. 74, 129-173 (2007).

[103] Donald F. Turner, *Antitrust Policy and the Cellophane Case*, Harv. L. Rev. 70, 281-318 (1956).

[104] Thomas E. Kauper, *The Goals of United States Antitrust Policy-The Current Debate*, Zeitschrift für die gesamte Staatswissenschaft/J. Inst. Theor. Econ. H. 3, 408-434 (1980).

"SSNIP test") will be profitable.[105][106] "The Guidelines have been harshly criticized for this omission, with some critics contending that such an omission makes the Guidelines unworkable in practice."[107] To be specific, the 1982 Guidelines did not specify how to calculate how much diversion of sales is enough to make the SSNIP unprofitable. The Critical Loss Analysis (CLA) with a benefit of low data requirement was proposed to solve this problem. Later, the CLA became an important model in the antitrust premerger investigation. We are going to discuss the CLA in the following section.

---

[105] Steven C. Salop, *Symposium on mergers and antitrust*, J. Econ. Perspect. 1, 3-12 (1987).

[106] Peter Bronsteen, *A review of the revised Merger Guidelines*, Antitrust Bull. 29, 613 (1984).

[107] Barry C. Harris & Joseph J. Simons, *Focusing Market Definition: How Much Substitution is Necessary*, J. Reprints Antitrust L. & Econ. 21, 151-172 (1991).

# 3.2 THE CRITICAL LOSS ANALYSIS

Before the Critical Loss Analysis, the Hypothetical Monopolist Test was widely regarded as non-operational. *G. Stigler* and *R. Sherwin* commented, "This market definition has one, wholly decisive defect: it is completely non-operational. No method of investigation of data is presented and no data, even those produced by coercive processes, are specified that will allow the market to be determined empirically."[108]

More specifically, *they* questioned about the availability and high requirement of data. "[T]his construct is simply impracticable. It would require an enormous number of calculations of cross-price elasticities between any given provisional product or geographic market definition and all possible additions to it. There will virtually never be adequate data to make such calculations, and we doubt that the Department has the resources to make the calculations if the data were available."[109] The test was also criticized by a federal Circuit Court: "a decision based on these Guidelines remains as inexact as the data gathered to make the assessment … [T]hese Guidelines are more useful for setting prosecutorial policy than delineating judicial standards."[110]

---

[108] G. Stigler & R. Sherwin, *The Extent of the Market*, J. Law and Econ., 28, 555-582 (1985).

[109] R. Harris and T. Jorde, *Market Definition in the Merger Guidelines: Implications for Antitrust Enforcement*, Cal. L. Rev. 71, 464-481 (1983).

[110] Monfort of Colorado. Inc. v. Cargill. Inc., 761 F. 2d 570, 579 (10th Cir. 1985), rev'd, U.S. _ . 107. S. Ct. 484 (Dec. 9, 1986). See also United States v. Virginia National Bank-Shares. Inc., 1982-2 Trade Cas. (CCH $ 64,871 (W.D. Va. 1982).

To address the problems above, *Barry C. Harris* and *Joseph J. Simons* had proposed a new approach, later called, Critical Loss Analysis in paper "*Focusing Market Definition: How Much Substitution is Necessary*"[111] The Critical Loss Analysis (CLA), "has been a standard method of implementation for the market definition algorithm of the Department of Justice and Federal Trade Commission Horizontal Merger Guidelines."[112] Meanwhile, "[i]t was recognized as one of the major developments of the modern Merger Guidelines era."[113]

To better explain SSNIP and CLA, the following section will reiterate the methodology proposed in that paper in great details. We want to demonstrate that most of the criticism about the CLA, even very recent ones[114][115], are in fact caused by bad applications of CLA by the Agencies. They are not problems with the CLA itself.

---

[111] Barry C. Harris & Joseph J. Simons, *Focusing Market Definition: How Much Substitution is Necessary*, J. Reprints Antitrust L. & Econ. 21, 151-172 (1991).

[112] Malcolm B. Coate & Joseph J. Simons, *Critical Loss vs. Diversion Analysis: Clearing up the Confusion*, The CPI Antitrust Chronicle 1, 1-15 (2009).

[113] David Scheffman, Malcolm Coate & Louis Silvia, *Twenty Years of Merger Guidelines Enforcement at the FTC: An Economic Perspective*, Antitrust L.J. 71, 277-285 (2003).

[114] Serge Moresi, Steven C. Salop & John Woodbury, *Market Definition* (2017).

[115] Russell W. Pittman, *Three Economist's Tools for Antitrust Analysis: A Non-Technical Introduction*, (2017).

### 3.2.1 The Critical Loss Analysis Model

The Critical Loss Analysis(CLA) was established to offer the SSNIP test a non-arbitrary threshold. In this section, we are going to introduce the CLA model with its technical details for further discussing.

CLA was proposed to improve the utility of the SSNIP test. The SSNIP test sets the rule under the Merger Guidelines to determine whether the group of products under the control of a hypothetical monopolist can have price increase by 5% without producing less profit than before. Thus, the critical question asked by the Guidelines is how much is the difference between the profit after and before the increase and whether the difference is positive. The Harris and Simons paper chose to get the answer in two steps. The first step is "by examining the relationship of marginal or variable cost to price, it is possible to determine, for any given price increase, the percentage loss in sales necessary to make the specified price increase unprofitable."[116] This percentage of loss was named as "Critical Loss" in the paper. The second step is to determine the actual percentage of sale loss by the hypothetical monopolist.

The paper assumed that all the products on the market are homogeneous and set the initial price and quantity as $P_0$ and $Q_0$. The marginal cost at $Q_0$ is $MC_0$. The profit earned at sales of $Q_0$ is $Profit_0$. Thus the gross revenue of the firm is $(P_0Q_0)$. The paper noted average variable costs as $AVC_0$, which is the part of the cost which goes up as the quantity produced goes up. There are also fixed costs to produce the product, the paper noted it as "Fixed Costs". Thus we have the following equation:

$$Profit_0 = P_0 Q_0 - Q_0 \cdot AVC_0 - Fixed\ Costs$$

---

[116] Barry C. Harris & Joseph J. Simons, *Focusing Market Definition: How Much Substitution is Necessary*, J. Reprints Antitrust L. & Econ. 21, 151-172 (1991).

*Equation 1*

Since $Q_0 \bullet AVC_0$ is equal to the total variable cost, we can regard it as the aggregation of the variable cost of every single product. Thus we have this:

$$Profit_0 = P_0 Q_0 - Q_0 \cdot AVC_0 - Fixed\ Costs$$
$$= P_0 Q_0 - \sum_{i=1}^{Q_0} MC_i - Fixed\ Costs$$

*Equation 2*

Here we need to discuss quantity between the aggregation of the margin at quantity $Q_0$, and the product of $Q_0$ and $AVC_0$. To get the Equation 2, the paper assumed equation 3 and equation 4. We write them out for future use.

$$\sum_{i=1}^{Q_0} MC_i = AVC_0 \cdot Q_0$$

*Equation 3*

$$\sum_{i=1}^{Q_1} MC_i = AVC_1 \cdot Q_1$$

*Equation 4*

The modeling continues as the price increases. The new price and quantity are designated by $P_1$ and $Q_1$. The new marginal cost, or variable cost, is denoted as $MC_1$. The new profit is $Profit_1$. The new average variable cost is $AVC_1$. The fixed cost remains the same even the price increases. So similarly we have:

$$Profit_1 = P_1Q_1 - Q_1 \cdot AVC_1 - Fixed\ Costs$$
$$= P_1Q_1 - \sum_{i=1}^{Q_1} MC_i - Fixed\ Costs$$

*Equation 5*

The paper explained that the Critical Loss is the threshold for which if the sales loss gets bigger than the Critical Loss, $Profit_1$ will be smaller than $Profit_0$. Thus the Critical Loss is the sales loss when $Profit_1$ equals to $Profit_0$. Thus we have:

$$P_0Q_0 - \sum_{i=1}^{Q_0} MC_i - Fixed\ Costs = P_1Q_1 - \sum_{i=1}^{Q_1} MC_i - Fixed\ Costs$$

*Equation 6*

Then the paper set Y as the proportion that the price increases, which was defined as $Y = (P_1 - P_0)/P_0$. For most of the cases in premerger investigation, $Y = 0.05$. Set X as the proportion of quantity that decreases as the price increases, and we have:

$$P_1 = P_0(1 + Y)$$

*Equation 7*

$$Q_1 = Q_0(1 - X)$$

*Equation 8*

Now that, as X and Y are set, we notice that the ratio X/Y is the overall elasticity of demand that the hypothetical monopolist is facing. Substitute Equation 7 and 8 into Equation 6, we get:

$$P_0 Q_0 - \sum_{i=1}^{Q_0} MC_i - Fixed\ Costs = P_0\ (1 + Y)\ Q_0(1 - X) - \sum_{i=1}^{Q_1} MC_i - Fixed\ Costs$$

*Equation 9*

The paper then collected the terms and transformed Equation 9, so we can get an equation for X, i.e., the Critical Loss, which is:

$$X = \frac{Y}{1+Y} + \frac{\sum\limits_{i=1}^{Q_0} MC_i - \sum\limits_{i=1}^{Q_1} MC_i}{P_0 Q_0(1 + Y)}$$

*Equation 10*

If we substitute Equation 8 into Equation 4, we get:

$$\sum_{i=1}^{Q_1} MC_i = AVC_1 \cdot Q_0(1 - X)$$

*Equation 11*

Then we substitute Equation 3 and Equation 11 into Equation 10, we get:

$$X = \frac{Y}{1+Y} + \frac{AVC_0 \cdot Q_0 - AVC_1 \cdot Q_0(1-X)}{P_0 Q_0(1+Y)}$$

*Equation 11*

After collecting terms, we get:

$$X = (YP_0 + AVC_0 - AVC_1)/(P_0 + YP_0 - AVC_1)$$

*Equation 13*

The paper elucidated that in this equation, the $Y$ is the percentage that $P_0$ increased by the hypothetical monopolist, which is decided by Department. $P_0$ is the prevailing price of the production in issue. $AVC_0$ are usually calculated by the companies in their daily accounting records. The only variable unknown is $AVC_1$. The paper suggested using $AVC_0$ to approximate $AVC_1$, and offered two reasons[117]:

1. Many production processes have relatively flat average variable cost curves.

2. In the long run, the competitive equilibrium $Q_0$ is to the right of the minimum point of the AVC curve. A reduction in Q will first lower and then raise the AVC. This change in the direction of AVC attenuates any tendency of $AVC_0$ and $AVC_1$ to differ.

Thus we assume $AVC_0 = AVC_1$, and substitute this into Equation 13:

$$X = YP_0/(P_0 + YP_0 - AVC_0)$$

---

[117] id footnote 17.

*Equation 14*

The paper then divided Equation 14 by $P_0/P_0$, which yields:

$$X = Y \div (P_0/P_0 + Y - AVC_0/P_0) = Y \div [Y + (P_0 - AVC_0)/P_0]$$

*Equation 15*

The paper defined $(P_0 - AVC_0)/P_0$ as the contribution margin, which is denoted as CM. We can regard it as the ratio of gross profit plus fixed costs, and $P_0$. So Equation 14 can be rewritten as X = Y/(Y + CM). The paper later stated X in percentage terms, so Equation 15 becomes:

$$X = [Y/(Y + CM)]*100$$

*Equation 16*

As the paper suggested[118], the merit of this model is that it only requires the contribution margin to calculate the Critical Loss. In addition to that, this model can freely change the percentage of the price increase as the Department wants. The contribution margin can be calculated by the current prevalent price and the average variable cost. In the case of an Hypothetical Monopolist, however, there are actually many different ways to mimic the Hypothetical Monopolist's data and action since the monopolist's data does not exist in real life.

Note that the calculation of CL in the CLA model is only based on 3 assumptions: the products under control of the hypothetical monopolist are homogeneous; the product demand is smooth near $P_0$ (if they want to have the ratio X/Y as the overall elasticity of demand) and $AVC_0$ = $AVC_1$. Besides, although the paper assumed that the products under hypothetical monopolist's control are homogenous, it is not necessarily required in any part in the deduction of the model itself. It only influences the way we get the data required in the analysis, which we will discuss later. Also, if we do not assume that the ratio of X/Y is the overall elasticity of demand, we do not assume that the demand curve is smooth either. "Other than the fact that it involves an estimation of the margin, it is pure arithmetic algebra to be precise."[119] This has actually rebutted the widely spread concern that the CLA is inconsistent with classical economics.[120] Even if sometimes the conclusion is inconsistent with the classical economics, it is at least not caused by the CL calculation and caused by the AL calculation instead.

---

[118] id.

There are also different ways to implement the term "a hypothetical, unregulated firm that made all the sales of those products in that area could increase its profits through a small but significant and non-transitory increase in price." The increase in price can be done with simultaneous and identical price increase on all of the products, simultaneous but different price increase on some of the products, or price increase on an individual product followed by price reset upon the price increase of the next product, etc. There are so many ways to understand this term in the 1982 Guidelines. The agencies need to choose the way of implementing the CLA model as reliably as possible, given the availability of data. We are going to discuss those different understandings in the later sections. For now, we will assume that we have calculated the critical loss and proceed to the calculation of the actual loss.

---

[119] Adriaan Ten Kate & Gunnar Niels, *The Concept of Critical Loss for a Group of Differentiated Products*, J. Compet. Law Econ. 6, 321-333 (2009).

[120] The critical loss is relatively big when the contribution margin is small, and the critical loss is relatively small when the contribution margin is big. Some people believe it is against the classic economics because when the contribution margin is small, there is usually more competition in the market. They believe the critical loss should be smaller so it is easier for the actual loss to be bigger than the critical loss. As a result, the relevant market concluded can be broader as well. This is incorrect. We are going to illustrate why it is incorrect in Section 2.6.

# 3.3 CALCULATION OF ACTUAL LOSS

We cannot find much research discussing the calculation of actual loss. This is probably because the crucial question in the calculation of actual loss is demand estimation. There is much research discussing demand estimation. We will discuss them in the next section. When we get the demand estimation function, the actual loss can be easily calculated by equation 17. Set $D_A$ as the demand function of product A, and $D_B$ as the demand function of product B; they can be estimated as linear demand, Logit demand, or AIDS demand model, etc.

$$Actual\ Loss = (D_A(P_0) - D_A(P_1)) - (D_B(P_1) - D_B(P_0))$$

*Equation 17*

The Merger Guidelines did not set forth any particular method to calculate Actual Loss. The paper *"A Critical Analysis of Critical Loss Analysis"*[121], describes a way to calculate actual loss. The authors of this paper claimed that they are FTC Economist doing premerger investigations. They also accredited many other economists in the footnote of the paper. Therefore, this method might be one of the methods that the Agencies use. We are going to dig deeper into this. Also, since this paper is one of the most cited criticisms to the CLA, exploring the model they use can also help us in finding out whether this important criticism is correct.

---

[121] Daniel P. O'Brien & Abraham L. Wickelgren, *A Critical Analysis of Critical Loss Analysis*, Antitrust L.J. 71, 161 (2003).

The paper claimed that they used a standard economic approach to calculate Actual Loss under a hypothesis of two products A and B, where product A and B are both inside the candidate market. The actual loss was interpreted as the difference between product A's loss of sales when SSNIP occurs and the product B's gain of sales in accordance to the price change of A. Under the same notation of section 2.2, the actual loss was expressed as Equation 18.

$$Actual\ Loss = Y(\ E_{AA} - E_{BA}\ )$$

*Equation 18*

Using the same train of thought as was used in CLA, to maximize the profit, a company should make the benefit of price increase equal to the loss of sales decrease. So if we set $Q_0 - Q_1 = \Delta Q$, $P_0 - P_1 = \Delta P$, we have Equation 19.

$$\Delta P(Q_0 + \Delta Q) = -(P_0 - AVC)\Delta Q$$

*Equation 19*

The author assumed Fixed Costs = 0, so $P_0 - AVC = CM$. Since $E_{AA} = (\Delta Q/Q_0)/Y$, if we divide Equation 19 by $Q_0 * \Delta P/P_0$ and, we get Equation 20.

$$CM = \frac{1}{E_{AA}}\left[1 + \frac{\Delta Q}{Q_1}\right]$$

*Equation 20*

Let $\Delta Q$ approaches 0, and we get,

$$CM = \frac{1}{E_{AA}}$$

*Equation 21*

If we substitute Equation 18 to Equation 20, we get Equation 22 to calculate the actual loss.

$$Actual\ Loss = Y\left[\frac{1}{CM} - \mathrm{E}_{AA}\right]$$

*Equation 22*

We deduced the Actual Loss function above (AL). In contrast with the CL function, this AL function assumed the demand curve being smooth to get equation 21 by letting $\Delta Q$ approaches 0. The criticism in the paper *"Critical loss: Let's tell the whole story"*[122], shows that the demand curve can kink near $P_0$. Therefore the concern on CLA about non-smooth demand curve is in reality not about the CL function but the AL function mentioned in the paper. As we have already mentioned, this is one of the methods proposed in FTC economists, O'Brien, Daniel P., and Abraham L. Wickelgren' paper. If we still use equation 17 to calculate the AL, this concern may not be true because some of the demand forms do not use smooth demand curve. For example, the demand is discrete in the Logit demand model. Therefore the concern about smoothness will be irrelevant, i.e., the smooth demand curve is not a problem of CLA.

We found that the equation has another intrinsic problem aside from the smooth demand curve assumption. Actual Loss in equation 18 is calculated with the difference between $E_{AA}$ and $E_{BA}$. E is the ratio between $\Delta Q$ and Q, so $E_{AA} = \Delta Q_{AA}/Q_A$ and $E_{BA} = \Delta Q_{BA}/Q_B$ . We should be careful about taking the difference in this ratio. $\Delta Q_{AA}$ and $\Delta Q_{BA}$ are related because $\Delta Q_{BA}$ is a part of

---

$\Delta Q_{AA}$. If the purpose of this ratio difference is to calculate the quantity of sales which did not switch to B, it can only stand if $Q_A$ and $Q_B$ are at least at a very similar scale. Otherwise dividing them is a meaningless calculation. For example, if product A loses 1000 sales and it used to have 10000 sales, $E_{AA}$ is equal to 10%. On the other hand, if product B used to have 3000 sales, and 500 of A's 1000 sales loss went to B, $E_{BA}$ is equal to 16.7%. In this situation, equation 18 concludes a negative AL.

There is another problem, which is the fact that linear demand $E_{AA} = (\Delta Q/Q_0)/Y$ was used to produce Equation 20. It was already proven that using a linear demand curve would conclude a narrower relevant market.[123] The reason is that the demand curve is concave near $P_0$. The customers are mixed with different demand flexibilities. The customers with flexible demand or less switching costs are more sensitive to price. They will switch quickly when the price goes higher, making the demand slope stiffer for a small price increase. Meanwhile, there are also customers whose demand is more rigid. Those customers are much less sensitive to a price increase. So the decrease in demand has a steeper slope in the beginning and a flatter slope for a greater price increase, i.e., the demand curve is concave. If the demand curve is concave and we calculate it as linear, we are going to underestimate AL and eventually end up with a narrower market. Therefore, if the actual loss was calculated by equation 22, it will also tend to conclude a

---

narrower market. There are several papers criticizing the CLA for concluding a narrower market than it should have been.[124][125][126]

By going through the CLA model and O'Brien, Daniel P., and Abraham L. Wickelgren's AL model, we can conclude that the CLA function is correct in its deduction and will not incur a narrow market conclusion as it was criticized for. At the same time, the AL function should not be applied for the calculation of AL in CLA; otherwise, it will bring an unreliable result to the market definition. Therefore, we suggest that we should use Equation 17 to calculate the Actual Loss if we are going to conduct a CLA and have estimated demand functions. If we have demand elasticity, like the situation in the discussed paper, we should use equation 23.

$$Actual\ Loss = \Delta Q_A - \Delta Q_B$$
$$Actual\ Loss = Y \cdot E_{AA} \cdot Q_A - Y \cdot E_{BA} \cdot Q_B$$

*Equation 23*

On the other hand, the paper "*A critical analysis of critical loss analysis*"[127] (the AL model paper) also concluded that the CLA is inaccurate, based on a calculation of AL conducted with equation 22. Therefore we can safely say that this criticism is incorrect.

---

Another influential paper, "*Critical Loss: Let's Tell the Whole story*"[128], written by *Michael Katz* and *Carl Shapiro*[129]. The paper proposed a formula to "use aggregate diversion ratio (ADR) to tell the rest of the story"[130] after calculation of CL. Therefore, the ADR is a way to replace the calculation of AL. The ADR is the ratio of the number of sales switching to the products inside the candidate market and the number of sales lost by the price-increase product. According to the paper, we only need to compare ADR and CL to see whether it is profitable to increase the product price. The aggregate diversion ratio, if there are more than two products, is calculated by Equation 24.

$$Diversion\ Ratio = \frac{\Delta Q_B}{\Delta Q_A}$$

*Equation 24*

Let gross margin be M and the elasticity of product A whose price has increased as $E_A$. They claimed $M = 1/E_A$. Therefore, they have Equation 25.

$$\Delta Q_A = \frac{Y}{M}$$

*Equation 25*

---

Let the diversion ratio be D. Based on Equation 24, we know that the proportion of quantity which does not switch to product B is 1 - D. Therefore, we have another actual loss function, which is Equation 26.

$$AL = \frac{Y}{M} \cdot (1 - D)$$

*Equation 26*

By comparing Equation 26 and Equation 16, the paper came up with the conclusion, "[i]f and only if the aggregate diversion ratio is larger than the critical loss, then the actual loss is less than the critical loss and thus a hypothetical monopolist would find a SSNIP profitable."[131], i.e., "AL < CL if and only if D > CL". This is for the case where the product A and B have the same profit ratio. If they do not, the paper concludes that "AL < CL if and only if $(P_B - C_B)/(P_A - C_A) \cdot D > CL$", where C is the total cost of product A and B.

The whole deduction above was based on the equation $M = 1/E_A$. The reason they gave in support of this conclusion was "[a]s any microeconomics textbook demonstrates, an economically rational firm acting unilaterally sets its price so that its gross margin is inversely related to its elasticity of demand: *M = 1/E*, where *E* is the elasticity of demand facing the firm in question."[132] Other papers claimed that $M = 1/E_A$ is based on "Lerner's condition."[133] However, "Lerner index analysis is not applicable when the market exhibits product homogeneity or dynamic differentiation."[134] Another paper "*Critical Loss v. Diversion Analysis: Another Attempt at Consensus*"[135] commented that "[t]hat relationship between the margin and demand responsiveness, however, is a theoretical one, derived from the mathematics underlying the classical monopoly model. It has not been demonstrated to reliably predict demand elasticities, especially in more competitive markets."

---

[131] id.

[132] id page 3.

[133] Malcolm B. Coate & Joseph J. Simons, *Critical Loss vs. Diversion Analysis: Clearing up the Confusion*, The CPI Antitrust Chronicle 1, 1-15 (2009).

Not only is ADR based on Lerner's formula, which may not hold, it also assumes that the fixed cost is zero, which is usually not true, either. The paper "*The SSNIP test and market definition with the aggregate diversion ratio: A reply to Katz and Shapiro*"[136] pointed out an error in the deduction of the ADR formula in 2007, the error also caused a narrower conclusion of relevant market products as well. In the years when ADR was used instead of the actual loss, the CLA tends to conclude a narrower market, which caused CLA to be less trusted by the court. The author of CLA later suggested in his paper "*Critical Loss vs. Diversion Analysis: Clearing up the Confusion*" that we should use AL instead of ADR.[137] Eventually, the 2010 Merger Guidelines adopted the CLA in the market division. During the introduction to CLA, the Guidelines only mentioned AL (as "Predicted Loss"), not ADR.

---

[134] Malcolm Coate & and Joseph Simons, *Critical Loss: Modeling and Application Issues* (2010).

[135] Malcolm B. Coate & Joseph J. Simons, *Critical Loss v. Diversion Analysis: Another Attempt at Consensus*, The CPI Antitrust Journal 1, 1-8 (2010).

[136] Øystein Daljord, Lars Sørgard, & Øyvind Thomassen, *The SSNIP Test and Market Definition with the Aggregate Diversion Ratio: A Reply to Katz and Shapiro*, J. Compet. Law Econ. 4, 263-270 (2007).

[137] Malcolm B. Coate & Joseph J. Simons, *Critical Loss vs. Diversion Analysis: Clearing up the Confusion*, The CPI Antitrust Chronicle 1, 1-15 (2009).

# 3.4 THE DEMAND SYSTEMS TO CALCULATE ACTUAL LOSS

There are several commonly used demand models in premerger investigation. The demand calculation provides crucial input to both CLA and merger simulation models. As was suggested by Figure 2, the choice of demand form will directly affect the estimation of demand even when the price increase is the same. As was mentioned in Chapter 2, the Agencies will conduct multiple demand estimations according to the availability of data and market status, and then consider all the results in merger simulations.

The choice of demand form will also result in a different result of market definition. Regarding the different results of market definition, the 2010 Guidelines committed that, "[t]he hypothetical monopolist test ensures that markets are not defined too narrowly, but it does not lead to a single relevant market. The Agencies may evaluate a merger in any relevant market satisfying the test, guided by the overarching principle that the purpose of defining the market and measuring the market share is to illustrate the evaluation of competitive effect."[138] Although there is discussion on which product to include when there are similarly "substitutable" ones according to the SSNIP test, the essence of this discussion is the attitude towards fuzzy market edges. Therefore, the Agencies will likely consider what roughly constitutes the relevant market more to get the big picture of the competitive effect.

---

[138] The 2010 Horizontal Merger Guidelines, 4.1.1 The Hypothetical Monopolist Test, Washington, D.C.: U.S. Dept. of Justice (2010).

Although the Guidelines indicate we can pick any of the results when there are multiple ones, meaning using either demand form might work fine, there are discussions questioning if the CLA will conclude a too narrow market if a linear demand is used.[139][140] Therefore we believe the antitrust scholars tend to pick the form of demand which is the closest to the circumstance in the market.

In this section, we are going to introduce the most commonly used demand forms and how to calculate demand with them. We are going to introduce the linear and log-linear demand model, the Logit demand model, and the Almost Ideal Demand System model (AIDS). There are plenty of discussions about demand estimation in economics. There are several variations of those demand models as well. Because the theme of our paper is not about the demand forms, we are not going to go through great details of those demand models. We are going to introduce the frameworks, demand functions, assumptions and data required of those models instead.

---

[139] Kai Hüschelrath, *Critical Loss Analysis in Market Definition and Merger Control*, European Competition Journal 5, 757-794 (2009).

[140] Daniel P. O'Brien & Abraham L. Wickelgren, *A Critical Analysis of Critical Loss Analysis*, Antitrust L.J. 71, 161 (2003).

### 3.4.1 Linear Demand Model

The earliest demand forms used are the linear and log-linear demand models. They are usually just linear regressions using the least squares method to assess the parameters. Let product i be the one whose price will increase. There are j products whose data are acquired and used to assess the price-demand relation of i, and $i \in \{1, 2, ..., j\}$. Let P be the product price and $Q_i$ be the quantity product i was sold. Let $V_k$, $k \in \{1, 2, ..., k\}$ be the other demand shifting variables, which will also affect the quantity sold. So we have Equation 27 for linear regression and Equation 28 for log-linear regression. $\alpha, \beta, \gamma$ are the coefficients.

$$Q_i = \alpha_i + \sum_{i=1}^{j} \beta_{ij} \cdot P_i + \sum_{k=1}^{k} \gamma_{ik} \cdot V_k$$

*Equation 27*

$$log\ Q_i = \alpha_i + \sum_{i=1}^{j} \beta_{ij} \cdot log\ P_i + \sum_{k=1}^{k} \gamma_{ik} \cdot V_k$$

*Equation 28*

The merit of the linear demand models is that we can get the elasticity directly from the estimated parameters. When all the $\beta$ are estimated, $\beta_{ij}$ will be the cross elasticity between product i and j, and $\beta_{ii}$ will be the own elasticity of product i. The prediction is not perfect, but it is quite effective and also easy to calculate. When the computational capacity was not very advanced, the linear demand has played a crucial role in the market division and other analysis about the market for decades.

The downside of the linear demand is the constant elasticity as the coefficients. "Own and cross-price elasticity are usually expected to change with the level of prices."[141] "It often results in the prediction of negative quantities for highly asymmetric mergers."[142] As was mentioned earlier, linear demand does not fit the reality of the market. The demand is concave near $P_0$, and using a linear will conclude less actual loss. Therefore, the hypothetical monopolist's profit will be overestimated, and the market will be narrower than it should be. Therefore, the linear demand is scarcely used by itself nowadays. If the linear demand is applied, there will be caution of underestimation of demand loss, and it will mostly be used as a rough indicator in merger screening.[143]

---

[141] Gregory J. Werden, *Demand Elasticites in Antitrust Analysis*, Antitrust L.J. 66, 363 (1997).

[142] Philip Crooke, et al, *Effects of Assumed Demand Form on Simulated Postmerger Equilibria*, Rev. Ind. Organ. 15, 205-217 (1999).

[143] Oliver Budzinski & Isabel Ruhmer, *Merger Simulation in Competition Policy: A Survey*, J. Compet. Law Econ. 6, 277-319 (2009).

### 3.4.2 Logit Demand Model

The most used demand form nowadays is the Logit demand model. Logit demand model is not a single model, but a category of the models where logistic regression is used to estimate the correlation between demand and price. When another input is added to the model, the output will also have an estimation of the correlation between demand and other parameters. It is "a model where the log-odds of the probability of an event is a linear combination of independent or predictor variables."[144] There are many commentaries talking about using Logit demand model in merger simulations.[145][146][147] For example, the Antitrust Logit Model is a merger simulation about the heterogeneous products based on the Bertrand model and the logit demand.[148] The Logit demand model can be used when market-level data on price, quantity, other product characteristic data, and consumer characteristics data are available. In a simpler version, only the demand, price and product characteristics are necessary.

The Logit demand model is a discrete choice model using Logistic regression to estimate the coefficients and assess the price-demand relation. Logistic regression estimates the probability of one certain option chosen among several different options. Therefore, it is used in estimating the probability of one product being chosen among several heterogeneous products. However, the Logistic regression requires the options to be "[i]ndependent and irrelevant alternatives (IIA)."[149]

---

[144] RONALD CHRISTENSEN, LOG-LINEAR MODELS AND LOGISTIC REGRESSION, Springer Science & Business Media (2006).

[145] Gregory J. Werden & Luke M. Froeb, *The Effects of Mergers in Differentiated Products Industries: Logit Demand and Merger Policy*, J. Law Econ. Organ. 407-426 (1994).

[146] Oliver Budzinski & Isabel Ruhmer, *Merger Simulation in Competition Policy: A Survey*, J. Compet. Law Econ. 6, 277-319 (2009).

[147] Roy J. Epstein & Daniel L. Rubinfeld, *Merger Simulation: A Simplified Approach with New Applications*, Antitrust L.J. 69, 883 (2001).

[148] Gregory J. Werden & Luke M. Froeb, *The Antitrust Logit Model for Predicting Unilateral Competitive Effects*, Antitrust L.J. 70, 257 (2002).

The IIA assumption is the biggest obstacle of using Logit demand model because the products are substitutes of each other. Therefore, the probability of one product being chosen is correlated to another product's price and availability, and the strength of correlation depends on the substitutive level between the products. The IIA can only stand if all the products are equally substitutable to each other.[150] It is obviously not true in the context of premerger investigation. Therefore we cannot trust the estimated result. A previous attempt to solve this problem was done using the nested logit model[151][152]. In an application of nested logit model, we need to pre-group the products, so that the products within groups are correlated while the groups are independent from one another. This might work in merger simulation, as long as the products within each group are substitutable with each other in a relatively equal way.

However, we cannot use this method in market definition, since this requires an estimation of substitution level. The traditional way to assess the substitution level is by measuring the demand. If the demand is pre-estimated, it will end up with a circular reasoning. In Chapter 4, we are going to propose a new way to divide the relevant market, which is independent of demand estimation. Our method may be helpful to the pre-grouping in nested logit demand models so that it can further help in merger simulation.

Since the Logit model and its variations [153]are still commonly used in the market definition,[154] we will discuss it further. The Logit model can be used to estimate the probability of an option as long as the data and context theoretically meet the assumption of the Logistic model, regardless of the interpretability of the coefficients. Regardless, it will still be better to be able to interpret all of the coefficients. The paper "*Estimating discrete-choice models of product differentiation*" used an assumption that "consumer utility function depend[s] on observable product characteristics, including price as well as individual-specific coefficients."[155][156] Based on this assumption, the probability of a product being picked is interpreted as the estimation of the average utility of a product divided by the sum of all the average utility of all the products available for picking.

Assume we have j products to be picked. $i \in \{1, 2, ..., j\}$ is one of the products. Let product i's summarized characteristic be variable $C_i$, and price of product i be $P_i$. Let U be the utility function of product i. Let a constant $\alpha$ represent the aggregation of the factors that affect the U in an unobserved way. In here, the error term $\varepsilon$ is interpreted as the consumer-specific taste preference. Let $\varepsilon$ be a universal error term, so in this model, the consumer taste is assumed to follow an independently and identically distribution (i.i.d.) across all the product picking. Therefore we get Equation 29.

$$U_i \ (P_i, \ C_i, \ \alpha, \ \varepsilon) = \alpha - \beta \bullet P_i + \gamma \bullet C_i + \varepsilon$$

*Equation 29*

Therefore, the mean utility level of product i, V, will have an expected value, which is $U_i$ without the error. We put it in Equation 30.

$$V_i \ (P_i, \ C_i, \ \alpha) = \alpha - \beta \bullet P_i + \gamma \bullet C_i$$

*Equation 30*

Based on our assumption, the probability of i being chosen can be represented by Equation 31.

$$S_i \ (V) = \frac{e^{V_i}}{\sum\limits_{k=1}^{j} e^{V_k}}$$

*Equation 31*

Equation 29 - 31 created a visible link between the probability of product being chosen and the relevant qualities of products. This created a theoretical base of application of the model[157]. Since demand models are not the main topic of this dissertation, we are not going to discuss its application in merger simulation models. The paper "*Merger simulation in competition policy: A survey*" gave a good summary of this topic.[158] In this dissertation we are going to use a simple Logit demand model to estimate the required elasticities in Equation 23 to calculate Actual Loss so that we can make a full walk-through in using logit demand model to do CLA.

The data used as the dependent variable in the logit model is a logged odds of whether a certain product will be picked. Following the notation earlier in this section, the product whose price will increase is i, where $i \in \{1, 2, ..., j\}$. Therefore we let the probability that i will be picked be $S_i$. The endogenous variable of this model is $P_i$, the price of the product i. The exogenous variable is the product characteristics $C_i$. Let the intercept be $\alpha$, which indicates the aggregation of the factors that affect the $S_i$ in an unobserved way. Therefore, we have the simplest version of the logistic regression model of demand in Equation 32.

$$ln\left(\frac{S_i}{S_{-i}}\right) = \alpha + \beta \cdot P_i + \gamma \cdot C_i$$

*Equation 32*

Then we can use the maximum likelihood method to estimate the coefficients $\alpha$, $\beta$, and $\gamma$, and get the estimated function of the logged odds ratio. One may notice that a sign change of the coefficient $\beta$. Since the value of $\beta$ will be estimated by the logistic regression anyway, using a negative sign will only create confusion in further analysis. Therefore we will just use a positive sign here. Almost all the regressions are directly done by software today, so we will not look into the details of the estimation process. To get it back to the probability of i being picked, we can transform Equation 32 and get Equation 33. In the context of market demand, the probability of i being picked can be considered at the expectation of market share of product i.

$$S_i = \frac{1}{1 + e^{\alpha + \beta \cdot P_i + \gamma \cdot C_i}}$$

*Equation 33*

Therefore we can substitute $P_0$ and $P_1$ of product i into Equation 33, and use the result value and Equation 34 to estimate Own Elasticity of product i. In there we also need the total demand in the candidate market.

$$Own\ Elasticity = \frac{\Delta Q_i}{\Delta P_i} = \frac{Q \cdot S_i\left(P_{i0}\right) - Q \cdot S_i\left(P_{i1}\right)}{P_{i1} - P_{i0}}$$

*Equation 34*

To calculate the cross elasticity between i and j, we need to do logistic regression again with product j and estimate the expected market share function of product j, $S_j$. Then we can use Equation 35 to calculate Cross Elasticity of Product i with Product j.

$$Cross\ Elasticity = \frac{\Delta Q_j}{\Delta P_i} = \frac{Q \cdot S_j\left(P_{j0}\right) - Q \cdot S_j\left(P_{j1}\right)}{P_{i1} - P_{i0}}$$

*Equation 35*

Now we can use the calculated Cross Elasticity and Own Elasticity to calculate Actual Loss with Equation 23. The merit of this simple model is that it does not make any additional assumptions other than the ones made by the logit demand model itself. In the application of models in premerger investigation, we should be more careful when making assumptions. Even if those assumptions are widely made in economic research, some of those assumptions are still too risky to be made because the conclusion will affect businesses worth billions of dollars. We are going to discuss this in more details in Chapter 4.

### 3.4.3 AIDS demand model

The AIDS demand model was created by Deaton and Muellbauer in their paper "*An almost ideal demand system*"[159]. This paper is based on the theorem of PIGLOG class preference[160][161], which "are represented via the cost or expenditure function which defines the minimum expenditure necessary to attain a specific utility level at given prices."[162] It assumes the representation of market demand is the outcome of decisions made by a rational representative consumer. Therefore, the cost/expenditure function with utility and price as input is constant between different products. The logged cost/expenditure function is the Equation 33.

$$log\ c(u, p) = (1 - u) \cdot log\ \{a(p)\} + u \cdot log\ \{b(p)\}$$

*Equation 33*

$u$ is a specific utility the rational representative consumer gets from the product. $p$ is the minimum price the consumer is willing to pay. $a(p)$ is the cost of subsistence and $b(p)$ is the cost of bliss. The AIDS paper formed *a(p) as Equation 34.*

$$\log a(p) = a_0 + \sum_k \alpha_k \log p_k + \frac{1}{2} \sum_k \sum_j \gamma_{jk} \cdot \log p_k \cdot \log p_j$$

*Equation 34*

j and k are two products among the products considered in the AIDS model. The logged product of j and k's price will later be the non-linear factors representing the interaction of each pair of two products in the regression function; thus it is formed in this way. The bliss cost is formed as Equation 35. It is the subsistence cost plus the product of all the prices to the power of another coefficient. Again, it is formed in this way to serve the need of the non-linear regression which will be performed later.

$$log\ b(p) = log\ a(p) + u \cdot \beta_0 \prod_k p_k^{\beta_k}$$

*Equation 35*

Therefore, we can get the AIDS cost function as Equation 36.

$$log\ c(u,\ p) = a_0 + \sum_k \alpha_k\ log\ p_k + \frac{1}{2} \sum_k \sum_j \gamma_{jk} \cdot log\ p_k \cdot log\ p_j + u \cdot \beta_0 \prod_k p_k^{\beta_k}$$

*Equation 36*

The paper explained that the reason for taking all the prices and their interaction into consideration. "For the resulting cost function to be a flexible function form, it must possess enough parameters, so that at any single point, its derivatives $\partial c/\partial p_i$, $\partial c/\partial u$, $\partial^2 c/\partial p_i \cdot \partial p_j$ , $\partial^2 c/\partial u \partial p_i$ and $\partial^2 c/\partial u^2$ can be set equal to those of an arbitrary cost function."[163] This is saying that since this function is also arbitrarily made to satisfy the interpretation requirement and no one really knows the real correlation of u, p and c to do the regression, we should consider all the possible regressors under the framework of the interpretation. This way, our regression of the demand price relation will have more accurate estimated coefficients. On the other hand, this arbitrary form of Equation 36 also ensures that the derivatives of the AIDS demand Equation 40 and Equation 41 are in the form of $\partial w_i / \partial \log x = \beta_i$ and $\partial w_i / \partial \log p_j = \gamma_{ij} - \beta_i \alpha_j = \beta_i \Sigma \gamma_{jk} \log p_k$, "so that at any point, $\beta$ and $\gamma$ can be chosen so that the derivatives of the AIDS will be identical to those of any true model."[164]

---

[163] id.

The paper used a theorem from the paper "*Intertemporal consumer theory and the demand for durables*" that concluded that "a fundamental property of the cost function is that its price derivative are the quantities demand"[165]. Therefore, we have the Equation 37.

$$\partial c(u, p) / \partial p_i = q_i$$

*Equation 37*

With a mathematical transformation, the partial derivative of the Equation 34 can be transformed as Equation 37, where the $w_i$ is the budget share of good i.

$$\frac{\partial \log c(u, p)}{\partial \log p_i} = \frac{p_i}{c(u, p)} \cdot q_i = w_i$$

*Equation 37*

Then we take the logarithmic differentiation of Equation 36 and get a function of $w_i$ , which is Equation 38.

---

$$w_i = a_i + \sum_j \gamma_{ij}^* \log p_i + \beta_i \cdot u \cdot \beta_0 \prod_k p_k^{\beta_k}$$

*Equation 38*

Where

$$\gamma_{jk}^* = \frac{1}{2}(\gamma_{ij} + \gamma_{ji})$$

*Equation 39*

Then the paper assumes that the consumer is maximizing utility, so his total expenditure $x$ is equal to $c(u, p)$. Therefore we can transform the Equation 36 into a function calculating $u$. At the same time, we can substitute $u$ into Equation 38 to get a function calculating $w_i$ from $p$ and $x$. The function is in Equation 40.

$$w_i = a_i + \sum_j \gamma_{ij} \log p_j + \beta_i \log\left\{\frac{x}{P}\right\}$$

*Equation 40*

Where P is a price index defined by

$$\log(P) = a_0 + \sum_k \alpha_k \log p_k + \frac{1}{2} \sum_j \sum_k \gamma_{jk} \cdot \log p_k \cdot \log p_j$$

*Equation 41*

If we want to use AIDS demand model nowadays, using Equation 40 and cashier data about price and demand, we can have the coefficients $\alpha$, $\beta$ and $\gamma$ be estimated with a maximum likelihood method in statistical software or statistical programming. However, the computational capacity was not as advanced when the paper was published in 1980. Therefore, the author provided some compromised method to eliminate unnecessary parameters. This is done by placing empirically or theoretically plausible restrictions on $\gamma_{ij}$ parameters, which is offered by the paper. However, nowadays, we can apply dimensionality reduction method to the dataset and eliminate the variables which are not contributing much to the prediction.

On the other hand, Equation 40 is a demand function in the form of budget share. The relation between budget share and demand is demonstrated by Equation 42. When we have the demand function, we can calculate the elasticity using $E = \Delta q_i / \Delta p_i$, like what we do to get Equation 34 and 35.

$$q_i = w_i \cdot x = w_i \cdot \sum_j p_j \cdot q_j$$

*Equation 42*

This AIDS model assumes "the demand functions … are first-order approximations to any set of demand functions derived from utility-maximizing behavior." From Equation 36 and the deduction after, we can see that the model assumes the utility that the rational consumer gets from the products is constant among different goods charging a different price. This is in response to the assumption of homogeneous products. However, this assumption is less and less applicable in the product market nowadays. At the same time, AIDS also requires significantly more data than the linear demand or the Logit demand. Given these two conditions, the application of AIDS demand will be limited to some specific circumstances like supermarket goods.[166]

Meanwhile, there is also a conceptual problem in AIDS. The regression mostly depends on the price change of different products. Nevertheless, the model also assumes $u$ to be constant and the consumer who is making decisions in here is totally rational. If so, the consumer will only pick the lowest price item at every time and it is not very meaningful to calculate various coefficients of different products' prices anymore.

However, this AIDS model can still work. As mentioned in the paper, containing as many price variables and their interactions will decrease the importance of the arbitrary function form itself. Therefore, even if we need to make unrealistic assumptions to get the function from the beginning, having enough variables considered in the function will make the assumptions no longer be that outcome determinative as long as we have enough data and computational capacity to do the estimation. This is different from the Logit demand model's IIA problem because the Logit demand model does not require that many regressors.

To solve the problem of AIDS, there are other models proposed, like PCAIDS[167] etc. However, the PCAIDS introduced more assumptions, making its application more restricted too. It assumes the proportionality, homogeneity, and symmetry in the calculation of demand. In fact, those complex deduction of functions and assumptions all aim to simplifying the calculation of the coefficients. However, we no longer need to worry too much about the simplicity of the calculation or the demand function because the computational capacity has increased by many orders of magnitudes since then.

---

[166] Roy J. Epstein & Daniel L. Rubinfeld, *Merger Simulation: A Simplified Approach with New Applications*, Antitrust L.J. 69, 883 (2001).

[167] id.

The most significant example is machine learning, where we do not try to interpret the prediction function at all, but only care about whether the model is making good predictions. In the future, we may use machine learning method to assess the price-demand relation.

It is worth mentioning that almost all the models commonly used in the antitrust premerger investigation, including all the demand models, are written in an R package developed by Charles Taragin and Michael Sandfort.[168] We can easily put in data and parameters into the functions pre-written in the packages to get the model results. The CLA was not written in the package however, probably due to the uncertainty in a correct way of conducting CLA. We are going to discuss this in detail in the next section.

---

[168] Charles Taragin and Michael Sandfort,
Resource from R CRAN: https://CRAN.R-project.org/package=antitrust, url.
Reference manual: https://cran.r-project.org/web/packages/antitrust/antitrust.pdf, PDF.

### 3.4.4 Conclusion about demand models

From the discussions above, we know that all of those demand models have their shortcomings. The linear model is less realistic. The Logit model assumes the IIA. The nested Logit model needs a pre-division of the products. The AIDS can only be applied to products which can be regarded as homogeneous. None of them are perfectly applicable. This will result in unreliability in the final conclusion of the market division.

On the other hand, we noticed from reading historical papers that the demand relation discussed in market definition research is usually continuous. Discrete selection demand models like logit demand model are more often discussed with merger simulation models assessing unilateral effects. The reason is that the discrete models involve more computational calculation. When the models are available for practical use, there are already less discussions about the market division and more discussions about merger simulation. Those models are originally used for merger simulation. Demand as their output is ideal since the merger simulation is only about predicting the price effect on demand.

Hypothetical Monopolist Test also uses the demand estimation as the most important input. This is likely the reason why the market division method has been largely limited within the Hypothetical Monopolist Test framework. Unlike the merger simulation which is focused on price effect however, the market division is focused on product substitution. Using demand change to estimate substitution is the product of the historical limitation in data resource and analytical tools. If the condition permits, we should estimate the substitution directly instead of using demand to approximate it. We are going to discuss it more in Chapter 4.

# 3.5 DISCUSSION ABOUT THE CORRECT APPLICATION OF CRITICAL LOSS ANALYSIS

Although it seems that the calculation of Critical Loss, Actual Loss, and demand are all we need to consider in the CLA model's application in SSNIP, this is not the completed story. How do we conduct the SSNIP as the hypothetical monopolist and how do we get the data required of the hypothetical firm, are the real challenges which have not yet been solved by the academia until now. The CLA paper offered an example to calculated CL of different products, as an illustration of how to apply the CLA model.[169] Nevertheless, the example only brings confusion to the implementation of the model. The paper calculated the CL of the two products in its example. However, all the values in the example are hypothetical so it did not indicate any data acquisition criteria. It did not offer further analysis about how will the CL be applied to implement market definition either. Most importantly, it did not explain how to calculate CM. In the example offered by the paper, the CM was given directly. These are actually the reasons why there are so many confusions and misunderstanding about how to apply the CLA.

The author of the 1989 paper tried to clear up the confusion about the application of CLA in his 2009 paper, *Critical Loss vs. Diversion Analysis: Clearing up the Confusion*[170]. However, it did not address all of the questions asked by the critics. Especially, it did not answer how to calculate

---

[169] Barry C. Harris; Joseph J. Simons, Focusing Market Definition: How Much Substitution is Necessary, page 165 section C, 21 J. Reprints Antitrust L. & Econ. 151, 172 (1991)

[170] Malcolm B. Coate & Joseph J. Simons, *Critical Loss vs. Diversion Analysis: Clearing up the Confusion*, The CPI Antitrust Chronicle 1, 1-15 (2009).

CM and how to apply CL. Later in this section, we will use a few examples to show that it is not the CLA model itself which causes the resulting inconsistency with the classical economics[171], but the wrongful implementation of it.

[171] Daniel P. O'Brien & Abraham L. Wickelgren, *A Critical Analysis of Critical Loss Analysis*, Antitrust L.J. 71, 161 (2003).

### 3.5.1 How do we define the SSNIP — Three Scenarios

According to the Merger Guidelines, we learned that the Hypothetical Monopolist Test is to hypothetically consider all of the products in the candidate market as being produced by one company, and see whether a SSNIP will be profitable. If it is, then the products in the candidate market consists a relevant market. If it is not, add more products and re-calculate. Logically, there are 3 possible strategies to maximize the profit, if we do not consider the price decrease scenario:

1.  If the substitution between the products inside and outside the candidate market is low, the profit-maximizing strategy can be by increasing the price of all the products by a SSNIP.

2.  If the substitution is not that high, the profit-maximizing strategy should be by increasing some of the product prices but not the others. This way, the sales loss can partly go to the other products which price stays the same.

3.  The last strategy is to increase one product price, while keeping the rest of the product prices the same like in scenario 2.

This discussion involves many different scenarios. Chart 1 summarizes the scenarios stating the pros and cons of applying each scenario. This chart is a preview of the discussion to follow in this section.

| | Scenario 1 | Scenario 2 | Scenario 3 |
|---|---|---|---|
| **Pros** | 1. Only need to conduct SSNIP and CLA once. | 1. Conceptually correct, because it is a more plausible profit-maximizing strategy. | 1. Easy to acquire the data which is correct to be used in CLA. This is the easiest way to avoid concluding a too-narrow market by CLA.<br>2. Conceptually less incorrect than the first scenario.<br>3. The number of SSNIP to conduct is equal to the number of products in the relevant market (less than scenario 2). |
| **Cons** | 1. Conceptually incorrect, because it is hard to say it is profit-maximizing strategy.<br>2. Insufficient data to conduct the CLA. Currently this is the reason why CLA will commonly conclude a too-narrow market.<br>3. AIDS is a more suitable model for predicting demand change when there are multiple price increases, while the use of AIDS is limited to specific occasions. | 1. Have to conduct a large number of SSNIP and CLA.<br>2. Insufficient data to conduct the CLA. If we use the current way to get the required parameters, it will impose a risk of concluding narrow market too.<br>3. AIDS is a more suitable model for predicting demand change when there are multiple price increases, while the use of AIDS is limited to specific occasions. | 1. Conceptually less correct than the second scenario. |

*Chart 1*

It is the case that not only the SSNIP is defined differently, but also the approach we do with the calculated CL and AL are different in those scenarios as well.

**The Approach of Scenario One**

If we increase the price of all the products in the candidate market, our biggest challenge is finding a good estimation for the CM of the Hypothetical Monopolist which is hard because we do not know the market's reaction to this hypothetical monopolist at all. However, the steps after finding the CM are relatively simpler than the other scenarios. We need to calculate CL with Equation 15. Then calculate Actual Loss (AL), and compare AL with CL. If AL is smaller than CL, we are done with the test and conclude that all the products in the candidate market form the relevant market. If AL is bigger than CL, we add the product to where the most sales loss goes into the candidate market and calculate a new CM before performing CLA again. If we simply use the product price of the merging firm, we do not need to recalculate CM.

**The Approach of Scenario Three**

Because the third scenario is a relatively simpler version of the second scenario, discussing the third scenario before the second one will enable a clearer introduction. If we only increase the price of one product, we use the CM of that product to calculate CL. Then we see how much sales are diverted to the products outside the candidate market. We regard this part of sales loss as AL, instead of all the sales loss of the products which increased prices. Then we can compare CL with AL; if CL is greater than AL, we are done with the test and all the products within the candidate market form a relevant market. If CL is smaller than AL, we pick another product inside the candidate market and do it again. If all of the products are checked and none of them makes CL greater than AL, we will add the closest substitution outside the candidate market and do the whole process again.

It is worth mentioning that in this scenario, it does not matter whether the product price increase is caused by the merging firms because this test is only for the market division. This test is not for predicting whether the merging firms will make a profit by increasing the price. According to the Guidelines, the Agencies will analyze the possible actions of the merging companies after we define the market.

**The Approach of Scenario Two**

The second scenario also contains two situations. One is to increase some products' prices in the candidate market with the same Y%. The other is to increase some products' prices, with only one  product using SSNIP increase rate of Y% and the rest of the products using SSNIP increase rate less than Y%. The latter one is called "variable SSNIP".[172][173] By saying "some", we mean it contains all and one, so the second scenario is a combination of the first and third one. It will use the first scenario's approach when increasing all of the product prices and use the third scenarios' approach when not increasing all of the products prices. The only difference from scenario three is that it will need to combine different products as we do in the scenario one.

---

[172] Oystein Daljord, Lars Sorgard, & Oyvind Thomasseen, *The SSNIP Test and Market Definition with the Aggregate Diversion Ratio: A Reply to Katz and Shapiro*, J. Compet. Law Econ. 4, 263 (2008).

[173] Malcolm B. Coate & Joseph J. Simons, *Critical Loss vs. Diversion Analysis: Clearing up the Confusion*, The CPI Antitrust Chronicle 1, 1-15 (2009).

Among the three scenarios, which strategy is going to be profit-maximizing for the hypothetical monopolist is decided by the substitution level of the products. If we liberally apply the content of Hypothetical Monopolist Test from the Guidelines, even if we do not consider the situation when the hypothetical monopolist decreases some of the products under his control, we still need to make sure that there is really no profit-maximizing strategy in all the scenarios above, including increasing all, some, and one product price, before we add one more product into the candidate market.

Decades ago, it was believed to be too complicated and impossible to try every scenario.[174] It is no longer difficult using today's technology. For example, with the same price increase of Y%, there are only $\sum_{i=1}^{n} A_n^i = \sum_{i}^{n} \frac{n!}{(n-i)!}$ situations we need to consider for a candidate market with n products. Although it will still take significant amount of work to collect the demand data and calculate it, it is no longer as hard as it was before.

If we want to be consistent with the idea of the hypothetical monopolist test, we also need to consider variable SSNIP. For a variable SSNIP, most of the steps are the same, because we only increase the price by a significant percentage of one product (this product is treated the same as the SSNIP product). The increase of the other product price is not necessarily a significant one (which means it can be a lot smaller than 5%), but as a portfolio of the price increase, it can eventually maximize the profit of the hypothetical monopolist.

---

[174] R. Harris & T. Jorde, *Market Definition in the Merger Guidelines: Implications for Antitrust Enforcement,* 71 Cal. L. Rev. 464, 481 (1983).

So for a variable SSNIP, we apply the sales loss prediction calculation to all the other products which prices also increase, to calculate the sales loss they encounter. For variable SSNIP, we try SSNIP on different products one by one.[175] The major challenge is that the price increase is continuous, so we need to perform some mathematical process on it or simply do the percentage in some certain intervals. It is worth mentioning that, in the CLA author's paper "Critical loss vs. diversion analysis: Clearing up the confusion",[176] only single SSNIP and variable SSNIP are mentioned, but not the scenarios in which we need to combine products and calculate CM. However, the Agencies seem to combine the products and use the first scenario, which is actually less desirable. We are going to talk about this in detail soon.

The variable SSNIP is less often discussed compared with the other strategies because the Agencies and the academia in the 1990s believed that it takes too many calculations. From the perspective of a law practitioner, it is good enough that the standard is clear and simple. That is the reason why in practice, "the market definition test often … takes the form of asking whether the hypothetical monopolist would find it most profitable to raise the prices of all of the products in the candidate market at least 5 percent above prevailing levels."[177][178] It appears that the Agencies usually only use the first scenario.

As we know from the CLA model, CL is calculated with CM and Y, while CM is defined as $(P_0 - AVC_0)/P_0$. From the deduction of the model, we know that $P_0$ is the price of the product whose price has been increased by a SSNIP (SSNIP product). $AVC_0$ is the average variable cost of the SSNIP product. In this scenario, other than the situation where one product forms the relevant market (so we only increase one product's price by a SSNIP), the $P_0$ and $AVC_0$ are actually representing more than one product. As a result, we need to calculate the $P_0$ and $AVC_0$ of the combined products. "In practice, the gross margins of the merging suppliers are typically taken as representative of the industry because the most reliable data on price and cost readily available usually come from the merging parties. In practice, then, the prices and costs of the merging parties serve as the basis for the hypothetical-monopolist calculations"[179]. From here, we can infer that the Agencies uses the CM of the merging companies as the CM of all of the products controlled by the hypothetical monopolist, from which they will then calculate the CL.

Nevertheless, it is very hard to justify this way of approximating the hypothetical monopolist's cost and price except for its convenience in application. We understand that this approach is good enough to non-arbitrarily define the relevant market. However, it does have some serious problems from the perspective of modeling. There are three major problems with increasing all of the product prices controlled by the hypothetical monopolist as the only way to define a SSNIP.

---

[177] M. L. Katz & C. Shapiro, *Critical loss: Let's tell the whole story*, Antitrust, 17: 49 (2002).

[178] The cited paper is written by Michael L. Katz while he was the Deputy Assistant Attorney General for Economic Analysis at the U.S. Department of Justice, so we believe it reflects the fact of antitrust practice.

[179] id.

## 3.5.2 The Discussion about the Three Scenarios

The first scenario is the least desirable approach of SSNIP, which caused most of the criticisms of the CLA. The author of the CLA paper explained in his paper "Critical loss vs. diversion analysis: Clearing up the confusion"[180] that the SSNIP should be the second (variable SSNIP) and third approach (single SSNIP). In this section, we are going to discuss the data resource and the consequence of the first approach, and clear up some wrongful criticism about CLA, which is not due to the model, but due to the application of the first scenario.

1.  **If we apply the first scenario to all the situations of the market, it is against the profit-maximizing assumption of the Hypothetical Monopolist Test.**

The mentioned way of setting SSNIP is contradictory to the question asked by hypothetical monopolist test in the Merger Guidelines. "[T]he test requires that a hypothetical profit-maximizing firm, … likely would impose at least a small but significant and non-transitory increase in price … on at least one product in the market, including at least one product sold by one of the merging firms."[181]

Although whether a hypothetical monopolist's profit-maximizing strategy is to increase all the products in its control or one of the products depends on market features like substitution levels, we at least know that the former one requires lower substitution between products inside and outside the candidate market. As a result, it is harder for the hypothetical monopolist to profit by increasing all of the product prices. Even if the hypothetical monopolist cannot profit under this strategy, it is still possible to profit by increasing at least one product's price in the candidate market. The CLA can be a threshold because it is the critical point between being able and being unable to profit by increasing price. Thus, if CLA is performed in the way stated above, it loses its meaning as a threshold.

---

[180] Malcolm B. Coate & Joseph J. Simons, *Critical loss vs. diversion analysis: Clearing up the confusion*, (2009).

[181] The 2010 Horizontal Merger Guidelines, Section 4.1.1 The Hypothetical Monopolist Test, Washington, D.C.: U.S. Dept. of Justice (2010).

Besides, it is also hard to justify increasing the price of all the products controlled by the hypothetical company as a good approximation of a profit-maximizing strategy. In business, it is a common sense that differentiated products targeting different customers is more profitable than a uniformed strategy. CLA is based on the assumption of profit-maximizing companies. If the method of applying data into the model is based on a contradictory assumption, it is difficult to justify the result as being reliable.

Thus, using SSNIP to increase all the product prices in the candidate market is inaccurate. Even though the 2010 Guidelines stipulated that "merger analysis does not consist of the uniform application of a single methodology",[182] it does not make it acceptable that the model they use adopts a method which is inconsistent with its own logic.

---

[182] The 2010 Horizontal Merger Guidelines, Section 1 Overview, Washington, D.C.: U.S. Dept. of Justice (2010).

2.  **The merging product's AVC is not representative enough to be used as an approximation of $AVC_0$ of the Hypothetical Monopolist, in the first and second scenario.**

CM was defined as $(P_0 - AVC_0)/P_0$. It is the average variable cost of the product whose price has been increased for a small percentage (the SSNIP product). Therefore when we calculate the CM of product i, $P_0$ should be the prevailing price of product i, so it is $P_0i$. At the same time, $AVC_0$ also should be the AVC of product i too. Based on those, the CL calculated can be the amount of sales loss a company can take when it increases $P_0$. For a different product j, CL should be calculated with $P_0j$ and $AVC_0j$.

$AVC_0j$ and $AVC_0i$ are just $AVC_0$ of single products. The data will probably be available for the Agencies since all the market participants have the responsibility to provide data to the Agencies as was required by the Guidelines. As we mentioned above however, the Agencies do not increase the product prices separately. Rather, the Agencies increase all of the products' prices controlled by the hypothetical monopolist simultaneously.

On the other hand, we do not have data from the operation of the manufacturing since the monopolist is hypothetical. We need to define a way to calculate $AVC_0$ of the hypothetical monopolist to produce the products whose price have increased. It is worth mentioning that as long as there is not only one product with an increasing price, we need to find a way to combine the $AVC_0$ of the products which prices have increased. In the first scenario, it is a way to combine the $AVC_0$ of all of the products controlled by the hypothetical monopolist. In the second scenario, when we are not increasing all of the products' prices together, it is a way to combine the $AVC_0$ of the products which prices are increasing. As we have mentioned before, what the Agencies do is use data of the merging firm as data of the hypothetical monopolist.

We believe it is incorrect to use the data of the merging firms to represent the other companies' $AVC_0$. In Figure 1, we show how CL changes as CM increases.

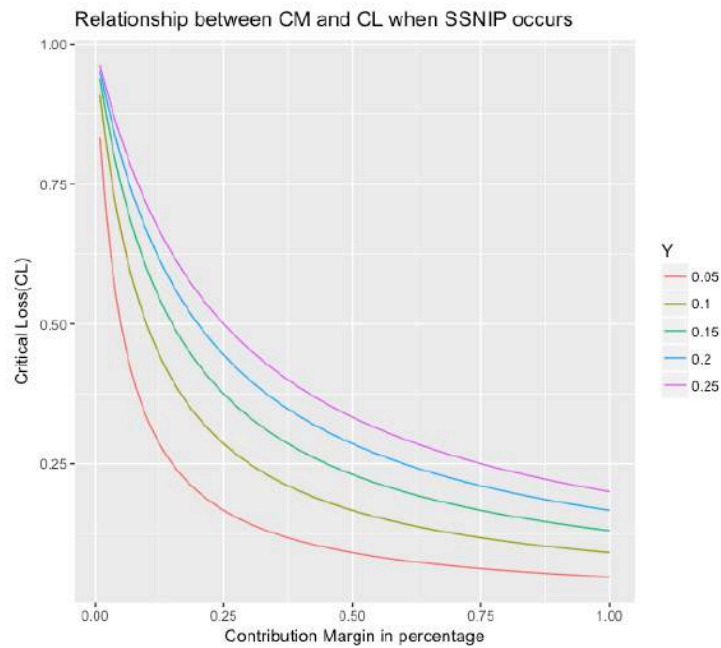

*Figure 1*

As we can see in Figure 1, for a SSNIP of 5%, CL drops from above 80% to below 20% percent when CM changes from close to 0 to 25%. When CM is roughly above 37.5%, CL no longer changes much as CM increases. In reality, CM can have large variations between different companies even when the companies are producing similar products. Unless the data shows that all of the companies which are manufacturing the products in the candidate market have CMs higher than 25%, it is hard to argue that a randomly occurring merger is done by companies which CM are representative enough that it will not produce misleading CLA result.

Apart from the big influence CM can impose on the result of CLA, companies who plan to acquire another company (let's say in half a year) can be well-prepared by intentionally lowering their CM beforehand to get a significantly lower CL. This way, they can mislead the investigation into a larger relevant market. Thus, it is also fundamentally risky to use CM of the merging companies as the CM of the hypothetical monopolist.

About the reason why the Agencies use the merging firms data to represent the other products in the candidate market, the cited paper explained the reason is the information provided by the merging firm is more reliable. This is not a concrete justification.

First, the merging firms have the incentive to manipulate the submitted data. There was actually a case in which the merging firms increased the price to the point of critical loss six months earlier than when they filed the HSR documents in order to make the CLA show that the post-merger SSNIP will be unprofitable for the merging parties. In fact, CLA does not have the mechanism to prevent being tricked by this method as long as the merging companies plan early enough in advance. The model we will later propose in Chapter 4 can perfectly avoid this problem.

Secondly, according to the Guidelines, the Agencies have the power to collect data from market participants other than the merging companies. The other suppliers will not have early notice about the merger. Even if they do have the notice, it is probably too hard and risky to coordinate on this. Therefore, their accounting information is both collectible and reliable. If we collect the data from the other suppliers, the next step will be to find a way to combine the data so that it can be a rational approximation of the AVC and P of the hypothetical monopolist. Then we can calculate a more reliable CM of the hypothetical monopolist.

The first candidate method is to calculate AVC as the average of all the variable costs among all the companies producing the product which is included in the candidate market. This way, we need to collect the data of AVC from the producers whose products are in the candidate market. The quantity to calculate the AVC will just be the current quantity they are producing. Then we average the AVC among them. This is an easy way to approximate the AVC of the hypothetical monopolist, because AVC in many industries today will not change much along the production quantity due to the automated production technology.

The second candidate method is to calculate the average variable costs of a product in different quantities produced, then average with the weight of quantity. If it is an industry where the AVC varies significantly as the quantity of production quantity changes, we need to consider the production quantity while approximating the AVC of the hypothetical monopolist. In fact, we do not know the optimum production quantity of the hypothetical monopolist. To simplify the calculation, we can just assume that the product was produced by the hypothetical monopolist in the same quantity that they are currently producing, although his might not be true since the hypothetical monopolist might have other options of quantity. Therefore, the second way will be using the AVC currently faced by each real-life manufacturers and averaging them with the weight of current production quantity.

The two methods above can improve the accuracy of the current approach of CLA. However, finding a better method to calculate the AVC alone cannot solve the fundamental problem of CLA. The fundamental problem is that when we are calculating CL of more than one product, the CLA forces us to regard them as one product. Otherwise, we cannot insert the parameters into the CLA functions. However, when we combine more than one products, neither the current product price nor the average of the price can be used as the price of the combined product, because none of those prices are the true reflections of the demand anymore.

3. **The merging product's P is not representative as an approximation of $P_0$ of the Hypothetical Monopolist, in the first and second scenario.**

The CLA uses the interaction of product price and demand to estimate the substitution level. In the premerger cases, there will always be a discussion of substitution to check whether the market concluded is consistent to the substitutive level we observe without the CLA model. As we know from the CLA model, CM = $(P_0 - AVC_0)/P_0$. Therefore, CM is Fixed Costs + Profit/$P_0$. For a profit-maximizing company, $P_0$ of a product is directly related to the demand. Thus, CM contains information about profit and demand. Since the demand-price correlation reflects the level of substitution, the information of substitution is also contained in CM.

Since CLA only gives one parameter to fill in CM and the CM function also only has one parameter of $P_0$, too, if we want to increase two products' price together, the two products have to be combined as one. Those two products are not homogeneous in reality. If they are close but different products, as is usually the case for candidate relevant market, their demand will become more rigid after they are combined as one. Similarly, when all of the products controlled by the Hypothetical Monopolist are bundled together, the demand will be significantly more rigid. Therefore, the real price of this combined product should be higher than either of the single product price according to the new demand. The difference between the single product price and the hypothetical product price will be a lot higher when there are a lot of products to be combined. Therefore, the hypothetical product price will not be close to any of the single product price in the candidate market.

If the price we use is not a match with the demand it is correlated to, CLA will either conclude a too narrow or too broad market. That is to say that if we use the first approach in the chart and increase all the prices together, we need to know a $P_0$ which is consistent with the real demand of the bundled "product". However, none of the single product's CM is comparable with a substitution level of the bundled "product". Since we cannot get this $P_0$ from any single price, the $P_0$ has to be calculated from another demand system which is not using P as the independent variable. We cannot know this price unless we find a method other than the price-demand correlation to quantify substitution. We are going to propose another method to quantify substitution in Chapter 4. Based on our model, and some further calculation, we will be able to get an estimation of $P_0$. Though our model can come up with the market definition directly, if the Agencies find the CLA is more desirable, our model can at least be used to calculate the real $P_0$.

On the other hand, this underestimation of the product price will cause an underestimation of CM. This is the fundamental reason why CLA almost always conclude a narrower market. If we increase all of the product prices together while using a single firm's CM, the CM of the bundled "product" will always be underestimated, as the $P_0$ is underestimated. Since $CL = Y/(Y + CM)$, CL will be overestimated. If CL is overestimated, it will be easier to have AL lower than CL. Consequently, the SSNIP is easier to be profitable. The hypothetical monopolist test will end earlier and conclude a narrower market than reality. This is consistent with the most common criticism of CLA. Those criticisms are usually based on evidence that CLA result is contradictory to a more reasonable result from other approaches that the market should have been broader.[183][184]

However, as we can infer from the paper "*Critical loss: Let's tell the whole story*"[185], the Agencies use the first scenario to conduct Hypothetical Monopolist test. The $P_0$ used by the Agencies is the price of the merging firms' products. The paper did not specify how they combined or picked from the merging firms' products. The Agencies are aware of that the CLA

---

tends to conclude a narrower market, but they simply regarded it as a shortcoming of CLA.[186][187][188]

The reason of misunderstanding CLA is probably that we calculate CM as $(P_0 - AVC_0)/P_0$—we do not see profit in the equation. It is easy to disregard considering whether the profit is a good approximation or not, given that $AVC_0$ does not seem to be a problem. For the hypothetical monopolist, AVC cannot be very different from the real companies' AVC, and thus the Agencies might believe there will not be a problem to use the merging companies' AVC to "represent" the hypothetical monopolist's AVC. As we already discussed, it is not safe to use the merging companies' AVC, because CL can vary largely with a small change in AVC.

Until now, we know there are three different scenarios (meaning three possible strategies) that the hypothetical monopolist can possibly take to maximize its profit. In the ideal world, we should try all of the strategies to see whether it is profitable to increase the price. We need to be clear about the Hypothetical Monopolist Test being a way to define the market, and not a prediction of what the merging company will do after the merger. The merging companies' action will be explored after the relevant market is found. Although the Agencies' choice of price increase strategy is inaccurate, the test still serves its purpose of creating a less arbitrary standard to define the relevant market since the purpose of the test is only to set a line.

However, it will still be better to avoid the three problems above by applying the third scenario, which is the single SSNIP as was mentioned by the CLA author's paper "Critical loss vs. diversion analysis: Clearing up the confusion"[189]. However, the author also mentioned variable SSNIP without mentioning the danger of concluding a narrower market. Since a variable SSNIP will also increase product price at the same time as done in the first scenario, the problem in estimating $P_0$ will exist as well. Therefore we recommend using the third scenario. If CL is calculated with a single product SSNIP approach, we can use the actual CM and $P_0$ of the products to calculate CL, so the $P_0$ will be consistent with the real demand.

The third scenario is not perfect. It is not as complete as the second scenario. However, despite being conceptually incomplete, it is the easiest way to avoid the result of a narrow market. It will also be a more reasonable strategy for the hypothetical monopolist to increase the price of one product rather than increasing the price of all of the products since most of the sales loss of the SSNIP product will go to the other products under his control.

The hypothetical monopolist test is in its essence a way to find the level of substitution because the method of quantifying substitution is not available. It is a good idea because it uses the switching of the customer to a new product as a measure of substitution. Substitution is a multi-dimensional concept, which we will discuss in Chapter 4. It is multi-dimensional so it is hard to measure with traditional economic tools. It has to be approximated by its mapping on a one-dimensional value. In this sense, the switch of customers is probably the best choice because customers will automatically consider all the dimensions for us and make a one-dimensional decision—switch to something else or not.

---

[189] Malcolm B. Coate & Joseph J. Simons, *Critical loss vs. diversion analysis: Clearing up the confusion* (2009).

However, to achieve a good approximation of substitution, we need to do the hypothetical monopolist test more liberally. We should at least do the single product process, not just simply increase all the products' prices in the candidate market. Otherwise, we will be too far away from the original correct idea of either CLA or hypothetical monopolist test. Therefore, although the CLA seems to make the hypothetical monopolist test a lot easier, it is in fact not as easy as has been done by the Agencies. It can be very complicated and involve significant calculations.

# 3.6 ANOTHER INCORRECT CRITICISM TO THE CRITICAL LOSS ANALYSIS

The Critical Loss Analysis has been criticized as being partly inconsistent with classical economics.[190][191][192] The reason is based on the shape of CLA curves. If we compare the CL with different CM, we will see the CL drops faster when the CM is low. Those researches blames CLA for concluding a higher CL giving more space of demand decrease for higher contribution margin company (the high-profit company) and lower CL to a lower contribution margin company. They believe this is inconsistent to the classical economics. See Figure 2 as a comparison of the general tendency. On the right is a closer look at SSNIP price change. Therefore, it was concluded that the CLA is more lenient to high CM industry.

---

[190] G J. Werden, *Beyond critical loss: Tailoring applications of the hypothetical monopolist paradigm*, (2002).

[191] M. B. Coate, M. D. Williams, *A critical commentary on the critical comments on critical loss*, The Antitrust Bulletin, 53(4), 987-1025 (2008).

[192] D. P. O'Brien, A. L. Wickelgren, *A critical analysis of critical loss analysis*, Antitrust LJ, 71: 161 (2003).

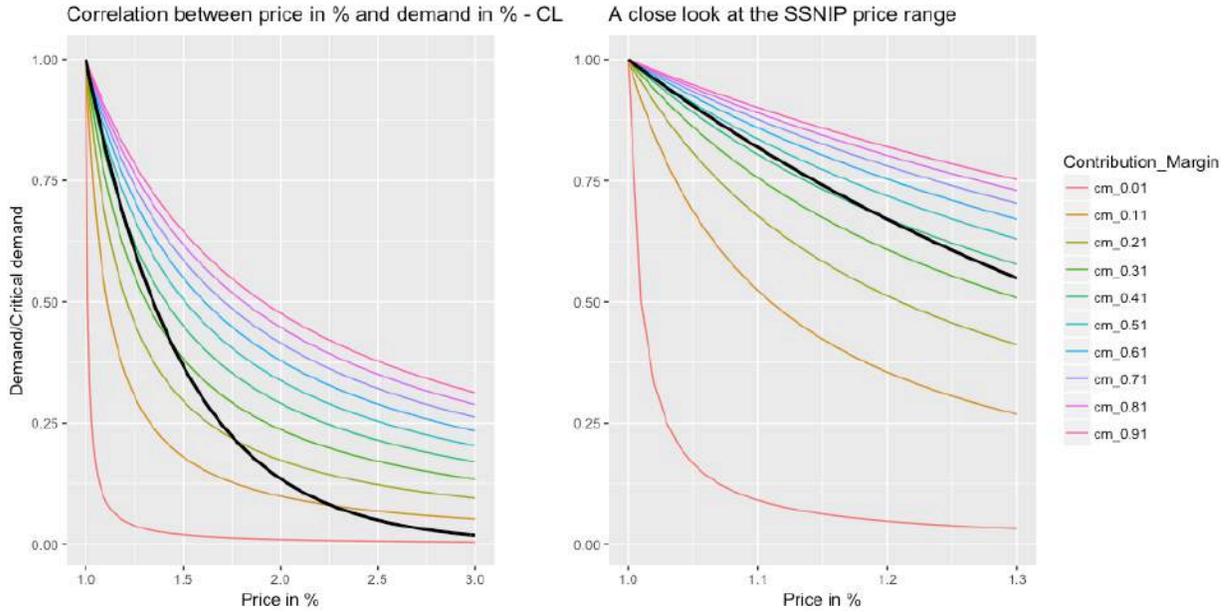

*Figure 2*

Since we do not have specific data to draw the demand, we use a normal-looking random demand curve to represent the tendency that those economists are indicating in Figure 2. From figure 2, we can see that if the CL is compared to a single general tendency of the demand, the CLA tend to conclude that CL is smaller than the demand decrease when CM is small, and CL is larger than the demand decrease when CM is large. Therefore, we see that the tendency of CL is inconsistent with the belief that a company with a higher margin shall be given a more strict standard of CL and not a looser one. However, a company with higher marginal profit will lose more money in each sale they lose. Therefore, it is rational to give a higher CL to a higher CM company.[193]

At the same time, since the function of CLA is just a result of an algebraic deduction, CL and CM in fact will always match each other under the logic embedded in the function. Figure 3 shows a comparison of the change of CL along with the change of CM. To discuss the tendency between CL and CM, we should look at Figure 3 instead of comparing a random demand with CL like Figure 2. In Figure 3, we can see CL does go down as CM goes up, which is consistent with the classical economics.

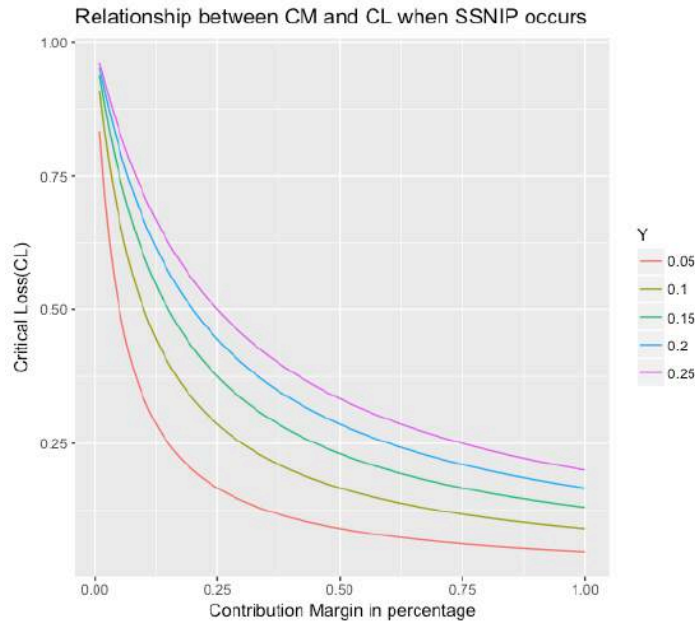

*Figure 3*

---

Also, when we consider the comparison of critical demand and real demand, we should not just consider the general tendency of the demand, but consider it under the context of each CM. When the CM is very small like 1%, it means the company is selling at a price near marginal cost, the competition is fierce and the substitution level is high. It is a market close to the perfect competition market. As we know about the perfect competition market, the reason for such low-profit margin is that the customer will all switch when the price goes up, since the products are homogeneous. Therefore, in the context of CM close to 1%, the demand curve should be like the blue line in Figure 4.

Similarly if a company's CM is as high as 91%, the profit margin is large if this situation is not caused by an extremely high fixed cost This usually happens in a market where the demand is very rigid or where the customers are very insensitive about the price change. If so, the demand curve should be more like the black line in Figure 4.

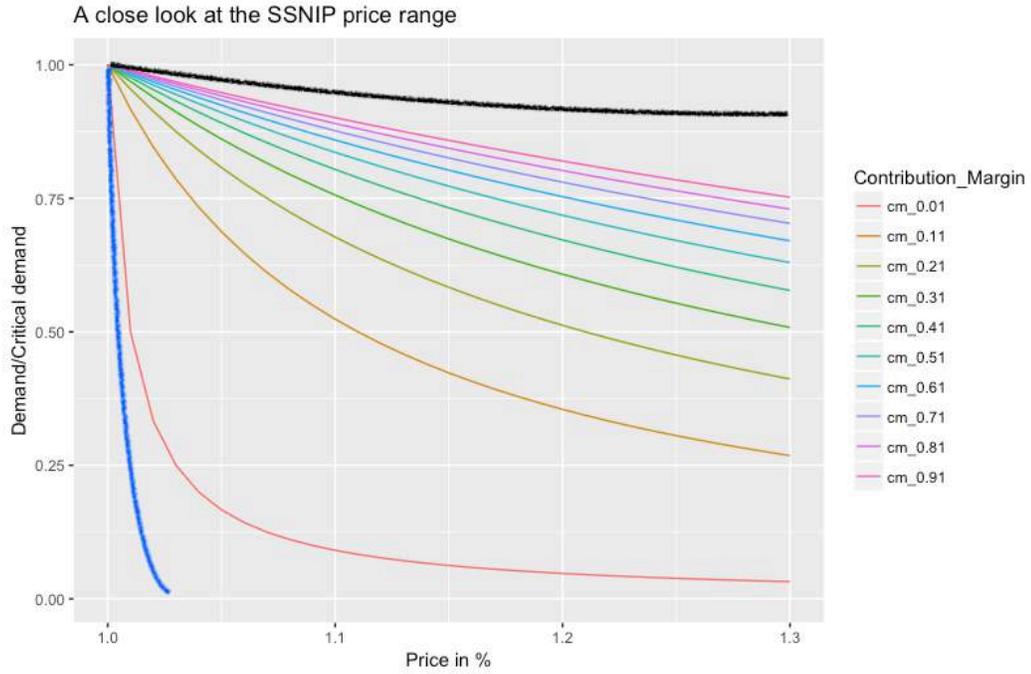

*Figure 4*

Therefore, the criticism of CLA being contradictory to classical economics is incorrect. In fact, since CLA is a pure algebra approach, it is quite accurate as long as the data used is not inaccurate. However, CLA has its limits as well. In the next section, we are going to discuss the limit of CLA and SSNIP test.

# 3.7 THE REAL PROBLEM OF SSNIP AND CLA

Up until now, we have finished the discussion about the correct way to perform a Critical Loss Analysis. As we have mentioned, different ways of performing CLA can produce different results. Some of the results can be quite inaccurate. Although we admit that increasing all of the products' price is one of the possible strategies for maximizing profit, which serves the function of making a bright line of market definition, it is inaccurate under the economic theory. In fact, most criticisms that CLA originated from the specific way of performing CLA. Those criticisms are based on a partial understanding of CLA, causing CLA to be distrusted by the courts.

On the other hand, although it seems that the CLA equation is simple and the only assumption it makes is that $AVC_0$ is equal or approximately equal to $AVC_1$, the application of CLA is not as easy as it was suggested by the author. An approach with fewer steps and calculations will make it more difficult to find the real CM. An approach without the difficulty in calculating the CM will involve a lot of steps and calculations. To do a reliable CLA, we need significantly more calculations and data collection. However these are not the real problem of SSNIP and CLA; as long as we can get a correct estimation of the parameters, these can all be solved. We are going to discuss the real problem of SSNIP and CLA in this section. These problems are harder or impossible to be solved with the tools we currently have.

**1. A conceptual flaw in the foundation of SSNIP approach.**

The Guidelines' Hypothetical Monopolist Test asks whether the whole group of the firms will benefit from the SSNIP, as if they are a hypothetical monopolist. We need to note that the result of the test does not mean that any particular firm can make a profit on the SSNIP. As a result, if there is no conspiracy, those firms will not profit by increasing price with or without the merger by themselves. Therefore, it's not very convincing to use this test. In respond to this concern, the DOJ explained the reason for using this test in the reprint of the Guidelines: "[t]he logic of this approach is that such a group of producers, if they are able to coordinate their behavior, have the joint ability and incentive to lessen compensation and raise prices. By contrast, a proper subset of these producers would not have this ability."[194]

As we have mentioned in Chapter 2, the Hypothetical Monopolist's behavior is relevant to the market definition context because if the firms producing the product "controlled" by the hypothetical monopolist collude in increasing the price at the same time, they will profit, i.e., they have the incentive to collude. However, this conclusion is irrelevant to whether they are capable of doing so or not. Price fixing is illegal. There are also a lot of practical difficulties to collude and unify motion like a "hypothetical monopolist". Based on game theory, if we assume that the products in a group are close substitutes and the manufacturers have a secret deal to increase all of the prices by 5%, the payoff for one manufacturer to breach the deal and sell at a lower price is larger than the payoff for obeying the deal if we do not consider the other cost of the breach and the switching cost. Because of this incentive to breach, it's difficult to conclude that the collusion will happen just because it is theoretically profitable to collude.

---

[194] 1984 Department of Justice Merger Guidelines § 2.0, reprinted in 2 Trade Reg. Rep. (CCH) 4491 et seq. (1984).

Even if the Agencies can use the Hypothetical Monopolist Test to define a relevant market, they have to analysis the merging firms' profit and demand pattern in the relevant market to investigate the effect of a merger because "[d]ifferent firms … may have different cost functions or may face different demand elasticities."[195] We need a company-specific analysis after the market definition to determine whether the coordinated effects are severe enough to bring a risk of collusion. In this case, neither the reason given by the DOJ nor the reason given by the definition of SSNIP provides a foundation to why we should use this specific test.

However, this one-dimensional demand change as a quantification of substitution is continuous. There is no reason provided by SSNIP as why we should draw the line there instead of anywhere else. Despite the arbitrariness becoming more obscure and hidden behind a model, the SSNIP combined with CLA is no less arbitrary than the former demand elasticity approach. After all, there is no solid reason why the line should be there. Meanwhile, a poor application of CLA and SSNIP will make the result even worse as we have discussed in earlier sections.

**2. We cannot rely on prediction outside the range of data.**

Calculating the demand curve is the essential and most evidence-based step. As we have mentioned in earlier sections, none of the demand models perfectly serves our needs. FTC economists mentioned in works that other than linear elasticity, they also use constant elasticity,

---

[195] Barry C. Harris & Joseph J. Simons, *Focusing Market Definition: How Much Substitution is Necessary*, Footnote 12, 21 J. Reprints Antitrust L. & Econ., 151, 172 (1991).

Logit, AIDS, semi-log models to predict demand. However, for all the fitted models we are not sure whether the model makes good prediction outside the range of data. No matter how much data we collect before a merger, none of the models are reliable enough on predicting things after the merger. Although this concern will be partially relieved after we take product features or all of the product price change in the relevant market into consideration, we still do not know the change of production cost after the merger and other resources brought about by a merger. However, all of those have the potential to heavily affect the demand after the merger. Therefore, theoretically we cannot use the predicted result based on premerger data as its exact value to compare with a specific value and get a conclusion, similar to what we do in CLA.

**3. In reality, the consumers do not have perfect information.**

From the history of market division in the Guidelines, we can infer that the hypothetical monopolist test is a way to find the level of substitution. In early the 1980s, the method of quantifying substitution was not available. In essence, SSNIP is to solve the problem of being unable to quantify substitution. SSNIP uses customer reaction as the indicator of substitution so that a multi-dimensional substitution can be mapped in one value and be calculated using traditional economics tools.

Theoretically, this method is more reliable when the customer has perfect information about the product so that they can switch to the next best substitution after the price increase.[196] This is usually true in a material market where the downstream manufacturers usually have enough professional knowledge about the materials that they need to spend a fortune on. However, this is not always true in a consumer goods market. Customers are not able to see the whole picture of all the substitutes in the market. They usually rely on advertisements or personal experiences as a resource of information. Advertisements are usually pro-sales and cannot be fully trusted. Personal experience usually takes time to gain. Thus unless the switch was tracked for long period of time, it won't be a good reflection of the market.

The Department was aware of this problem when making the 1982 Guidelines. Because of this, the SSNIP test has a non-transitory requirement. The non-transitory period is usually one year. We originally believe that this one year period can serve as a social experiment, which makes up for the deficiency of customers' information or the deficiency in predicting things outside the range of the data. However, since SSNIP is only a theoretical framework, the calculation is in fact done by CLA. Meanwhile, CLA relies on demand prediction models to calculate Actual Loss. Although CLA itself does not have a preference in the method for calculating Actual Loss —the choice of method to calculate Actual Loss decides whether CLA takes the imperfect information into consideration.

---

[196] Economic theory assumes consumer sovereignty, in which consumers have perfect information and perfect decision-making capacities. These assumptions are useful in economics, but they are also too far away from reality. Market definition is a legal proceeding, if we have other choices, we should not make an assumption which is too different from the reality.

The use of a year of transactional data before the investigation make sense. Although it is not exactly what the Guidelines say, the one year period will have a similar effect as if we add product feature parameters into the demand prediction model or consider all of the products' prices while making predictions[197]. However, if the product feature is only one of the variables in the model, we will need to single-dimensionalize the product features, after which we will lose a lot of information useful to the regression. If we use more than one variable as the product features, we can actually get to the conclusion of market definition directly since product features are directly related to substitution. There is no necessity to go through the tunnel of demand since the demand has to be one dimensional. This is the basic idea behind the model we are going to propose in Chapter 4.

On the other hand, although the agencies use one year's worth of transaction data, in some industries like durable goods one year might not be enough to explore and switch to other products and build personal experience. For products which are not quickly consumed or products which incur non-minimum switching costs, a one-year period is not enough. For the industries where products are quickly upgraded or the industries where new entries occur frequently, the change of market share is still not representative enough to depict a picture of the real substitution between the goods.

The hypothetical monopolist test combined with CLA is the winner of all the models available now to define the relevant product market. Both the hypothetical monopolist test and the CLA are bright designs. They have been serving the purpose of market definition for decades. However, limited by the technology available at that time, they also have considerable unavoidable problems. Although the Guidelines adopted both of the Hypothetical Monopolist Test and the CLA, the Agencies are also open to other models proposed to serve the same purpose. In the next chapter, we are going to propose a new approach to solve the problem of market definition. Hopefully, it will be useful to the future practice of the premerger investigations.

---

[197] This is the way AIDS uses. By taking all the product price into consideration, the preferences and substitution level can be reflected in the price correlations.

# CHAPTER 4: A NEW APPROACH TO THE MARKET DEFINITION

## 4. 1 BACKGROUND — "SUBSTITUTION" AND ITS SUBSTITUTION

Ever since the market definition and market power test were established by the landmark case "*Brown Shoe Co. v. United States*"[198], they have been the most important factors considered by the court in merger review cases. As Jonathan Baker has commented in his paper "*Market definition: An analytical overview*", "[t]hroughout the history of U.S. antitrust litigation, the outcome of more cases has surely turned on market definition than on any other substantive issue."[199]

We concluded from the discussion in Chapter 3 that the Critical Loss Analysis was wrongfully criticized many times. Meanwhile, there are also many undesirable aspects, which will render the premerger investigation and merger injunction into a risk of unfair results. As we can see from the previous discussion, due to the limitation of the applied models, this risk of unfairness is significantly higher than the risk of unfairness normally expected by the parties in the legal proceedings.

---

[198] Brown Shoe Co. v. United States, 370 U.S. 294 (1962).

[199] Jonathan B. Baker, *Market definition: An analytical overview*, Antitrust Law Journal 74.1, 129-173 (2007).

We also mentioned in Chapter 3 that on the other hand, a hypothetical monopolist's motivation to increase price is not equivalent to an after-acquisition company's motivation to increase the price. Not considering all the compromises made in the models' accuracy for the sake of the calculation convenience, the CLA can theoretically examine whether the Hypothetical Monopolist is able to profit by increasing the product price. However, a positive conclusion on this test is not very pertinent to the prediction of the merged company's post-merger actions. Even though the Hypothetical Monopolist Test equipped with CLA is capable of setting up a bright line on the level of substitutability to decide how far the relevant market should go, the line itself does not have much theoretical significance.[200]

We agree that the CLA under the framework of the Hypothetical Monopolist Test is a breakthrough, and it is "one of the major developments of the modern Merger Guidelines era"[201]. However, based on the background that the 1968 Merger Guidelines' substitution rule was replaced because it cannot be supported by quantitative methods, it is not hard to see that the Hypothetical Monopolist Test and the CLA are both the results of the compromise made due to the limited computing technology. Just as *Farrell* and *Carl* concluded, "[w]hile much has been written in antitrust economics on how best to define markets, the fact is that in many differentiated-product industries, there is no absolute right way to draw boundaries that are inevitably somewhat arbitrary."[202]

---

[200] As we discussed in Chapter 3, the reason given by the DOJ to explain why to apply Hypothetical Monopolist Test is also obscure and illogical.

[201] David Scheffman & Malcolm Coate & Louis Silvia, *Twenty years of merger guidelines enforcement at the FTC: An economic perspective*, *Antitrust LJ* 71, 277 (2003).

Nevertheless, CLA is still a winner in market definition methods. Despite all the downsides and the long-lasting criticisms, the Agencies adopted CLA in the 2010 Horizontal Merger Guidelines[203]. Compared to the Hypothetical Monopolist Test, CLA waited for an extraordinarily long period to be adopted. The adoption is likely due to the fact that there were not better models proposed by the academia to define the relevant market under the theoretical framework of Hypothetical Monopolist Test stipulated by the 1982 Horizontal Merger Guidelines. In the paper "*Antitrust evaluation of horizontal mergers: An economic alternative to market definition*", Farrell and Shapiro commented, "[i]t echoes the difficulty of merger investigation itself by requiring the analyst to predict price changes by a counterfactual firm. In some cases, it can thus fail to provide the desired simple, practical, rapid, and reasonably accurate diagnostic."[204]

Both data science and statistical analysis have been developing fast in the past few decades. After CLA was adopted by the Guidelines, many analysis which was unpractical became feasible under computational power available today. In this chapter, we are going to propose an alternative way to define the relevant market, which is consistent with the spirit of the market definition once upon "*Brown Shoe Co. v. United States*". Our model is not base on any assumptions inconsistent with the principle of merger review or the reality. It was built solely on substitution theory, which we believe is the essence of the market definition theory.

---

[202] Joseph Farrell & Carl Shapiro, *Antitrust evaluation of horizontal mergers: An economic alternative to market definition*, The BE Journal of Theoretical Economics 10.1 (2010).

[203] The 2010 Horizontal Merger Guidelines, Section 4.1.3 Implementing the Hypothetical Monopolist Test, Washington, D.C.: U.S. Dept. of Justice (2010).

[204] Joseph Farrell & Carl Shapiro, *Antitrust evaluation of horizontal mergers: An economic alternative to market definition*, The BE Journal of Theoretical Economics 10.1 (2010).

## 4.1.1 The Substitution Theory in the 1968 Merger Guidelines

The 1968 Merger Guidelines defines the relevant market based on "product dimension (['line of commerce['])" and "geographic dimension (['section of the country['])"[205]. It describes the product dimension as "[t]he sales of any product or service which is distinguishable as a matter of commercial practice from other products or services will ordinarily constitute a relevant product market, even though, from the standpoint of most purchasers, other products may be reasonably, but not perfectly, interchangeable with it in terms of price, quality, and use."[206] In simple words, the relevant product market should be constituted by products which are reasonably inter-changeable considering their price, quality and use.

This definition is consistent with the landmark case "*United States v. E. I. Du Pont de Nemours & Co.*" decided by the Supreme Court in 1956. In that case, the Supreme Court held "that market is composed of products that have reasonable interchangeability for the purposes for which they are produced -- price, use, and qualities considered."[207] At that time, the concept of the relevant product was not invented yet. When the concept of the relevant product market was invented in another landmark case "*Brown Shoe Co. v. United States*", the Supreme Court added "cross-elasticity of demand" into consideration when delineating market boundary. "The outer boundaries of a product market are determined by the reasonable interchangeability of use or the

---

[205] The 1968 Horizontal Merger Guidelines, Section 3 Market Definition, Washington, D.C.: U.S. Dept. of Justice (1968).

[206] id.

[207] United States v. E. I. du Pont de Nemours & Co., 351 U.S. 377 (1956).

cross-elasticity of demand between the product itself and substitutes for it."[208] **The 1968 Guidelines only used "price, use and qualities" when defining the relevant product market.[209]**

The definition above in the 1968 Guidelines is based on the similarity between the products. In regarding each product, the similarity is based on the characteristics of it. The 1968 Guidelines further added, "[o]n the other hand, the sales of two distinct products to a particular group of purchasers can also appropriately be grouped into a single market where the two products are reasonably interchangeable for that group regarding price, quality, and use. In the second case, however, it may be necessary to also include in that market the sales of one or more other products which are equally interchangeable with the two products regarding price, quality, and use from the standpoint of that group of purchasers for whom the two products are interchangeable."

The main damage caused by limiting market competition is limiting the choice consumers have to substitute the product which price has increased due to the market power of the manufacturer. If there are not close substitutions, the consumers will have to choose from not purchasing or purchasing at a higher price. In this case, some consumers will decide not to buy the product.

---

[208] Brown Shoe Co. v. United States, 370 U.S. 294 (1962).

[209] The 1982 Guidelines used "cross-elasticity of demand" instead. The relevant product market in the 1982 Guidelines will be discussed in the next section.

Eventually, there will be fewer people willing to buy and consume the product, meaning there will be less social welfare created and less consumer surplus.[210] In contrast, if consumers have enough close substitute, the price increase will not hurt consumers' interest.

The 1968 Guidelines further explained. "In enforcing Section 7 the Department seeks primarily to prevent mergers which change market structure in a direction likely to create a power to behave non-competitively in the production and sale of any particular product, even though that power will ultimately be limited, though not nullified, by the presence of other similar products that, while reasonably interchangeable, are less than perfect substitutes. "[211] To elucidate the difference between the homogeneous products and heterogeneous but substitutable products, the 1968 Guidelines further stated, "[i]t is in no way inconsistent with this effort also to pursue a policy designed to prohibit mergers between firms selling distinct products, where the result of the merger may be to **create or enhance the companies' market power due to the fact that the products, though not perfectly substitutable by purchasers, are significant enough alternatives to constitute substantial competitive influences on the production, development or sale of each.**"

Since the concern of market power is mostly on insufficient substitution, level of substitutability is a direct and natural basis to define the relevant market. This is probably the reason why both the first landmark case and the first Merger Guidelines chose "reasonable interchangeability" as the standard for defining the relevant market. However, this standard was widely criticized by academia in the 1970s. The concerns are mainly about arbitrariness involved in deciding the case.[212][213] At that time, there was not a good method to quantify substitution. The economists believe there should be a quantitative standard in defining the market. However, the technology limited the complexity of economic models accessible at the time. The available data are also limited to the routine accounting data[214]. Therefore substitution theory was replaced by the Hypothetical Monopolist Test in 1982 due to the technical reasons.

---

[210] N. G. Mankiw, Principles of microeconomics (4th ed. 2006).

[211] The 1968 Horizontal Merger Guidelines, Section 3. Market Definition, Washington, D.C.: U.S. Dept. of Justice (1968).

## 4.1.2 The Hypothetical Monopolist Test—A Substitution of the "Substitution Theory"

In 1981, *Landes* and *Posner* proposed the Hypothetical Monopolist Test in their paper "*Market power in antitrust cases*"[215]. Soon enough, in the 1982 Merger Guidelines, the DOJ accepted the suggestion from the economists and adopted the Hypothetical Monopolist Test. We have discussed the Hypothetical Monopolist Test in great details in Chapter 3. From Chapter 3, we learned that a Hypothetical Monopolist Test which functionally reflects the substitution is not as simple as it seems to be. The Hypothetical Monopolist Test and the CLA should have been more complex and data intensive than the current application. In this section, we are going to compare the Hypothetical Monopolist Test and the concept of substitution.

The 1982 Guidelines stipulated, "[t]aking the product of the merging firm as a beginning point, the Department will establish a provisional product market. The Department will include in the provisional market those products that the merging firm's consumers view as good substitutes at prevailing prices… The Department will add additional products to the market if a significant percentage of the buyers of products already included would be likely to shift to those other products in response to a small but significant and non-transitory increase in price."[216] The Hypothetical Monopolist Test is innovative in using consumer behavior to approximate the

---

substitutive relations. This method is based on a reasonable assumption that consumers will only switch to the products which they believe are reasonably interchangeable with the price-increasing product. This assumption is the reason why we can use the change in sales to reflect the substitutive relations.

However, we can infer from the 1982 Guidelines that compared to the 1968 Guidelines, the Hypothetical Monopolist Test is an approximation of the actual substitution in the market. To be specific, in the 1970s, the multi-dimensional method to quantify and calculate substitutive relations was not available. Products features are multi-dimensional. Consumers will also consider multiple aspects while buying alternatives when the target product price has increased. After considering multiple product features, the consumers will eventually switch to a substitution which is close enough or decide to buy the original product with an increased price. Since the consumers usually do not have much information about the products that they have not used yet, finding the best substitution might take them several switch attemps. Closer substitutions will eventually attract more consumers. The one-dimensional demand change will reflect the multi-dimensional substitutive relations. This is the original intention of the Hypothetical Monopolist Test[217].

Besides the Hypothetical Monopolist Test, the 1982 Merger Guidelines also stipulated the standard of evaluating product substitutability.[218] Since the Hypothetical Monopolist Test is measuring the demand change, a separate evaluation of substitutability should be independent of the Hypothetical Monopolist Test. As we already expounded in Chapter 3, the Hypothetical Monopolist Test was "completely nonoperational"[219] before the Critical Loss Analysis. This was probably the reason why a separate evaluation of substitutability existed.

---

[217] When the high frequency purchaser data was used in estimating demand change, we no longer need to wait for actual sales change to be observed after the price increase, because by taking the historical price-demand relations into consideration, the change of sales quantity can be predicted by the change of price. However, we need to face the common limit to all the predictions based on regressions—the range of data acquired. If the product have not incurred an individual 5% price increase, we cannot assume the demand will maintain the shape and be predicted through the same model at the 5% price increase. If there is a demand theory to explain the correlation, we might be able to justify the prediction outside the range of the data. However, as we have discussed in *Chapter 3: 2.4 The demand systems to calculate Actual Loss*, all the demand models used have some significant problems.

Besides, all the demand models try to find the correlation between demand and price. In those models, product features are not regressors. When more complicated substitutive relations, i.e., the products has many dimensions of features to be considered, the 5% price change may not be significant enough compared to the noise we can get from the real-world.

[218] Besides the Hypothetical Monopolist Test, the DOJ added, a substitution test in the 1982 Guidelines is as follows:

"In evaluating product substitutability, the Department will consider any relevant evidence, but will give particular weight to the following factors:

1. Evidence of buyers' perceptions that the products are or are not substitutes, particularly if those buyers have shifted purchases between the products in response to changes in relative price or other competitive variables;
2. Similarities or differences between the products in customary usage, design, physical composition and other technical characteristics;
3. Similarities or differences in the price movements of the products over a period of years; and
4. Evidence of sellers' perceptions that the products are or are not substitutes, particularly if business decisions have been based on those perceptions. "

The fourth factor can be understood as a mix of intrinsic substitution and extrinsic substitution. When considering the quantity of production, the manufacturer will consider the extrinsic substitution. When considering the production cost, the manufacturer will consider the intrinsic substitution.

[219] G. Stigler & R. Sherwin, *The Extent of the Market*, J. Law and Econ, 28, 555, 582 (1985).

The Critical Loss Analysis was first published in 1989[220]. In the 1992 Merger Guidelines, the level of substitutability test was removed. Although the 1992 Guidelines did not adopt the Critical Loss Analysis, the Hypothetical Monopolist Test along with the observation of consumer switch became the only standard.[221] However, we should not disregard the original purpose of the Hypothetical Monopolist Test—to offer a way to quantify the level of substitutability. Neither should we disregard the original purpose of the Critical Loss Analysis—to offer an operational quantitative model to conduct the Hypothetical Monopolist Test. In the next section, we will discuss the quantification of substitutability. We will also explore the reason why despite the application of the Hypothetical Monopolist Test and the CLA, a non-arbitrary market definition is still believed to be unlikely[222].

---

[220] Barry C. Harris & Joseph J. Simons, *Focusing market definition: How much substitution is necessary*, J. Reprints Antitrust L. & Econ. 21, 151 (1991).
This paper was first published on "Research in Law and Economics 207" in 1989. It was reprinted in 1991.

[221] The 1992 Horizontal Merger Guidelines, Section 1.1 Product Market Definition, 1.11 General Standards, Washington, D.C.: U.S. Dept. of Justice (1992).
Besides the Hypothetical Monopolist Test, the DOJ added the following test to measure the reaction of consumers:
"In considering the likely reaction of buyers to a price increase, the Agency will take into account all relevant evidence, including, but not limited to, the following:
1. evidence that buyers have shifted or have considered shifting purchases between products in response to relative changes in price or other competitive variables;
2. evidence that sellers base business decisions on the prospect of buyer substitution between products in response to relative changes in price or other competitive variables;
3. the influence of downstream competition faced by buyers in their output markets; and
4. the timing and costs of switching products."

[222] Joseph Farrell & Carl Shapiro, *Antitrust evaluation of horizontal mergers: An economic alternative to market definition*, The BE Journal of Theoretical Economics 10.1 (2010).

Meanwhile, the amount switched reflects how close the substitute product is to the SSNIP product[223]. The level of substitutability will affect the profit and the motivation to increase the price. These can be inferred from the 2010 Guidelines: "[w]hen applying the hypothetical monopolist test to define a market around a product offered by one of the merging firms, if the market includes a second product, the Agencies will normally also include a third product if that third product is a closer substitute for the first product than is the second product. The third product is a closer substitute if, in response to a SSNIP on the first product, greater revenues are diverted to the third product than to the second product."[224]

It is worth noting that the Hypothetical Monopolist Test has a non-transitory requirement. The Agencies explained the reason to have the requirement in the 1982 Guidelines. "The potential weakness of such a market based solely on existing patterns of supply and demand is that those patterns might change substantially if the prices of the products included in the provisional market were to increase. For this reason, the Department will test further and, if necessary, expand the provisional market."[225] The test allows a year to modify the result of market definition due to further changes of demand.

Theoretically, this one year period will also allow the consumers to change their mind and try multiple products in their following purchases. It is a practical mitigation to the inaccuracy caused by consumers' limited information about the product that they have not used yet so that the model does not have to make the unrealistic assumption of consumers' perfect information. After a year of trying and switching, the new sales volume will indicate the new substitutive relations. The change in sales volume indicates the level of substitutability between the products in the candidate relevant market, delivering us the result we cannot get by calculation. Therefore, the original idea of "small but significant and non-transitory increase in price"(SSNIP) has the quality of a social science experiment.

---

[223] The product whose price has a small but significant and non-transitory increase in price.

[224] The 2010 Horizontal Merger Guidelines, Section 4.1.1 The Hypothetical Monopolist Test, Washington, D.C.: U.S. Dept. of Justice (2010).

[225] The 1982 Horizontal Merger Guidelines, Section II Market Definition and Measurement, A. Product Market Definition, Washington, D.C.: U.S. Dept. of Justice (1982).

Since the computing technology in the 1970s does not support multi-dimensional substitutability models, the Hypothetical Monopolist Test is a great invention in the history of antitrust premerger investigation. Despite the history of being widely misused[226], the Critical Loss Analysis is also a historical invention in antitrust premerger investigation. It was designed to solve the problem of "how much substitution is necessary" in the Hypothetical Monopolist Test.[227] Learning from the path of the Hypothetical Monopolist Test, we may conclude that the test is serving as a "substitution" of the substitution theory. Therefore, if there is a model for directly evaluating substitutability, it is theoretically acceptable for usage according to the theory of market competition and the theory of the Hypothetical Monopolist Test. Besides, the Hypothetical Monopolist Test and the CLA have many problems as we have discussed in Chapter 3. There is a need to update or supplement the model used in the premerger investigation.

---

[226] See Chapter 3

[227] Barry C. Harris & Simons J. Joseph, *Focusing market definition: How much substitution is necessary*, J. Reprints Antitrust L. & Econ, 21, 151 (1991).

Three decades have passed since then, computational technology has been developing quickly. Data collection and analysis technologies have been predominantly advanced in the past few years. Nowadays, social science academia is equipped with better tools to make inference and predictions. On the other hand, the Guidelines do not exclude the use of other useful models —"[t]hese Guidelines should be read with the awareness that merger analysis does not consist of uniform application of a single methodology. Rather, it is … the Agencies … apply a range of analytical tools to the reasonably available and reliable evidence to evaluate competitive concerns … Where these Guidelines provide examples, they are illustrative and do not exhaust the applications of the relevant principle."

In this Chapter, we are going to propose two unsupervised machine learning method to define the relevant market, which involves fewer assumptions and more data.

# 4.2 QUANTIFICATION OF SUBSTITUTABILITY

## 4.2.1 Utility Is Not Substitutability

From the discussion in Section 3.1, we learned that both the SSNIP test and the CLA use change in product demand to quantify the level of substitutability. In this section, we are going to discuss further about quantification of substitutability. It is easy to associate substitutability with utility. The AIDS demand model we have discussed in Chapter 3 used utility as a quantifiable indicator of consumers' preference. However, although utility has been used in many pieces of antitrust research, we believe utility is not the right quantification of substitutability.

Utility is a commonly used idea in economics. The well-known economics textbook written by Paul Samuelson, *Economics*, defines utility as "… utility denotes satisfaction … it refers to how consumers rank different goods and services"[228] Another famous economics textbook, *Economics*, written by Gregory Mankiw simply defines utility as "level of satisfaction"[229]. There are other definitions of utility which is different from the general ones. In *Kahneman* and *Snell*'s paper "*Predicting Utility*", utility was defined as "a decision maker's anticipation of the hedonic quality of a future experience"[230]. In *Higgins*'s "*Making a good decision: value from fit*", utility was abstractly defined as "worth to some end"[231]. Even though there are so many different definitions, we can still conclude that the idea of utility is to use one abstract value to represent the satisfaction and preference over a product. **Utility is not a fixed number.** It can be quantified in many different ways. The value also changes vis-a-vis different consumers.

---

[228] Paul A. Samuelson & William D Nordhaus, *Economics*, (19th ed 2012).

[229] Gregory N. Mankiw, *Macroeconomics*, (3rd ed 1997).

[230] Daniel Kahneman & Jackie Snell, *Predicting utility*, (1990).

[231] Tory E. Higgins, *Making a good decision: value from fit*, American psychologist, 55.11, 1217 (2000).

Two most common approaches to quantify utility are through ordinal utility and cardinal utility. Ordinal utility is an ordinal scale representing the preferences of a person. The ordinal utility theory believes it is only meaningful to query which option is better than the other. It is not useful to query how much better the option is or how good the option is. In contrast, cardinal utility tries to quantify the difference between the utility of different products. Originally, it uses one product as a unit, and uses the ratio of satisfaction to quantify the utility of the other product. In this method, utility has a tangible meaning. For example, if we use the utility of an orange as a unit, the utility of a meal will be 3 units when the consumer gets three times more satisfied to have a meal than to have an orange. Cardinal utility function is a utility index that preserves preference orderings uniquely up to positive affine transformations.[232] However, "[f]rom most branches of economics, the concept of cardinal utility has been eliminated as redundant since ordinal utility has been found to suffice for doing the job of predicting the choice of consumers. Cardinal utility has been kept only in welfare economics to support the demand for a more equal income distribution … [and] in risk analysis."[233] Utility can be formed in different functions based on the context. There are other functions quantifying the ordinal and the cardinal utilities as well.

---

[232] Daniel Ellsberg, *Classic and Current Notions of 'Measurable Utility'*, Economic Journal, 64 (255), 528–556 (1954)

[233] J. C. Harsanyi, *Cardinal utility in welfare economics and in the theory of risk-taking*, Journal of Political Economy, 61(5): 434-435, (1953)

Besides the two traditional types of utility, there are other ways to quantify utilities. For example, one method is based on an assumption that the amount of money left with a consumer after consumption of goods also creates utility to the consumer, and that the amount of money left is correlated to the utility brought by the consumption of goods. The marginal revenue of the remaining sum of money owned by the consumer was mathematically proven to be capable of representing the marginal utility the consumer get from spending the total cost at the same period of time.[234] This method is called money metric utility. Another example of method will be to infer preference relation from a demand function to calculate utility[235]. This method is to find a sequence of preference rather than to quantify the utility. There are other research on formulating utility functions. However, those methods are not widely discussed or applied.

In economics, utility is usually discussed under the context of describing preference, not quantifying substitution. We do not need to know how much more does one prefer a product to another to describe preference. Besides, there is not a widely acknowledged way to qualify utility. "We only need to know which bunch of product is more preferred… [So] we stick with ordinal utility"[236] Utility is not a fixed number vis-a-vis each product. Ordinal utility only studies the order of choice, so it is not meaningful to compare the quantity of utility. On the other hand, although cardinal utility is not a pure sequence, it is not fixed product-wise. According to the definition of utility, it should vary between different consumers. In contrast, as specified by the

---

previous sections, substitutability is according to the "price, quality and use"[237] of a product. It should not change as we switch between consumers or consumer groups.

There are many pieces of research in quantifying utility, but there are not many pieces of research to quantify substitutability. **In many discussions about antitrust, the difference between utility and substitution is vague. Therefore, before we further discuss defining the relevant market by substitution, we need to clearly differentiate substitution from utility.** For example, the AIDS demand model uses utility as a parameter indicating the difference between products.[238] However, if they aim at describing the quality of a product, substitution should be considered instead.[239]

The current research in utility is not relevant to market definition. Utility is a one-dimensional value quantifying the level of satisfaction. The level of satisfaction is an intangible concept which is interchangeable among different category of products. The utility of a new car and a well-designed course might be the same number. Nevertheless, a car and a course are not substitutable to each other at all. Even if we analyze utility according to different factors, so that we can have a multi-dimensional utility, its intangible nature as the level of satisfaction does not change. Therefore, it is not promising to use utility as a standard of defining a market.

---

[237] The 1968 Horizontal Merger Guidelines, Section 3. Market Definition, Washington, D.C.: U.S. Dept. of Justice (1968).

[238] Angus Deaton & John Muellbauer, *An almost ideal demand system*, The American economic review 70.3 312-326 (1980).

[239] The author of AIDS also realized the idealogical problem that utility varies from person to person. Therefore, it assumes there is a person who is having an average level of satisfaction upon the homogeneous products (AIDS also assumes the products are homogeneous). However, it does not solve the idealogical problem that two totally non-substitutable products can still give people a same "average level of satisfaction".

In contrast, substitutability is a tangible concept. It is based on price, quality and use. Therefore, it is product-specific. The price and quality are fixed and tangible characters of each product. Although the use of products might vary between consumers, we can still find the typical use. Therefore, substitutability is what we should use in market definition. Another merit of using substitutability is that we may form it as a multi-dimensional value.

We will use an example to reinterpret the conclusion that multi-dimensional substitutability based on products' price, quality and use is more suitable in describing substitutional relations between products than utility is. Like many pieces of research in utility, we assume the amount of money that a person is willing to pay directly reflects utility, so that we can compare the tangible individual highest acceptable payments instead of an intangible level of satisfaction that a person gets. To simplify the discussion, we only consider the taste of a drink as the quality considered by a consumer. We also will not compare the use of drinks because we assume they are both used to drink by the majority of the consumers. We also do not consider price in substitutability in this case because we are using highest acceptable payment as an indicator of utility. Although the definitions are different, we chose to not put product price in yet to avoid confusion.

Assume there are only two aspects of a drink's taste — the degree of sweetness and the degree of sourness. There are two drinks, A and B. A is very sweet and not very sour; B is very sour and not very sweet. We quantify the degree of sweet taste and sour taste on the same scale. If a drink is very sweet, we give it 3 points in sweetness. If the drink is a little sweet, we give it 1 point in sweetness. We do the same to the level of sourness. Assume nothing else are considered by consumers when deciding whether a drink can substitute another. In a Cartesian coordinate system, we can quantify the quality of the two drinks as point A(3, 1) and point B(1, 3). These are fixed properties of the products.

Assume a consumer, Jack, likes both sweet taste and sour taste equally, and he prefers strong tastes. The ideal drink for Jack is U(3, 3). As we have introduced above, utility can be inferred from how much money Jack is willing to pay for the drink. We can get ordinal utility from the sequence of the amount Jack is willing to pay. We can also get cardinal utility from setting one drink as the benchmark and compare its amount with others.

A reasonable consumer will decide how much money he is willing to pay according to the distance between A/B and the optimum point U. In Jack's case, the closer the drink is to the point U(3, 3), the more satisfied he can get by consuming the drink. Based on the assumptions, we formed a simple model to calculate how much money Jack is willing to pay for a drink. Assume the ideal taste of sweetness and sourness to a person is $U(a, b)$, and the highest amount that person is willing to pay for his favorite drink is $p$. The decrease in highest acceptable payment is in a linear correlation to the drink's deviation from the ideal taste in sweetness with a slope of $-k$ (so that k is positive). We also add a weight representing the importance of sourness compared to sweetness, which is assumed to be the ratio of the ideal taste of sourness and sweetness. Drink i's level of sweetness is $x_i$. Its level of sourness is $y_i$. The amount of money a person is willing to pay is $M$. We get a model to calculate the highest acceptable payment for a drink in Equation 1:

$$M(x_i, y_i) = p - k \cdot |x_i - a| - \frac{b}{a} k \cdot |y_i - b|$$

*Equation 1*

For Jack, the ideal taste is *U(3, 3)*, so *a = 3*, and *b = 3*. If we substitute *A(3, 1) and B(1, 3)* into the function above, we get *M(A) = M(B) = p - 2k*. Based on our assumption, the utility of A and B are equal.

We add a third and a fourth drink C(2, 1) and D(1, 2) so that C/D is a good substitution of A/B respectively. Among the four, A and B are not good substitutions of each other. Nonetheless, if we calculate the money Jack is willing to pay for C and D, we get *M(C) = M(D) = p - 3k*. Since k is positive, we get *M(A) = M(B) > M(C) = M(D)*. Therefore, C and D are less preferred by Jack. Meanwhile, A and B are equally preferred by Jack. So is C and D. Since Jack's ideal taste is U(3, 3), this sequence of preference is consistent with the assumptions we made. This sequence can be regarded as an ordinal utility. If we set A as the unit product, the cardinal utility that A brings to Jack is 1. The cardinal utility of B is also 1, because A and B has the same distance from U. The cardinal utility of C and D are both *(p - 3k)/(p - 2k)*. Neither the ordinal utility or the cardinal utility we get in this example reflects the substitutional relationship between A, B, C and D.

This example tells us that if the substitutive relationship between A, B, C, and D are measured by the utility based on highest acceptable payment, the result will not reflect the true substitutive relationship between A, B, C and D. In real life, there are more things to be considered by Jack. The highest amount of money Jack is willing to pay is probably not linearly correlated with how far the taste of a drink deviates from Jack's favorite taste. However, this limited example still ideologically shows us how misleading the utility can be in indicating the substitutive relationship between products. Utility may be useful for finding the sequence of preference, but it is not useful in market definition. On the other hand, we can easily find the level of substitutability by calculating the distance between A, B, C and D. The distance between A and B is $2\sqrt{2}$. The distance between A and C is 1. The distance between B and D is also 1. Therefore, in this example, C is a better substitution of A than B; D is a better substation of B than A. This conclusion is consistent with our settings.

From the example, we see that the utility based on the highest acceptable payment cannot be used in quantifying substitutability. There are other methods to quantify utility. We will extend this conclusion in the next section to show that due to an over-deduction of dimensions, other utility models probably cannot work for market definition either. Moreover, the problem of over-deduction of dimensions does not only cause utility to be unsuitable to quantify substitutability. Although demand is currently used in the market definition, it also has a problem of over-deduction of dimensions.

Besides, the example in this section is also intended to preliminarily illustrate the concept of over deduction of dimension while preserving part of the information. The consumer considers multiple aspects when they decide whether to buy a product or not. Equation 1 is an example of getting a one-dimensional value based on multiple aspects of considerations. The decision-making mechanisms will be more complicated in reality, but they are essentially similar to Equation 1 which reaches conclusion which does not sufficiently reflect a complete picture of substitutability. In the next section, we are going to use a more general example to illustrate that even if we use a more complicated and more accurate method to get the value used in market definition, as long as the model has a one-dimensional output, it is hard to avoid the unreliability and arbitrariness of its further application in market definition.

In conclusion, although people mingle the definition of substitution and utility on many occasions, we should be aware of the difference between them. The example above is also a preliminary illustration of quantifying the level of substitutability with Euclidean distance in a space constructed using product quality, price and use. In later sections of this chapter, we are going to propose a method to define the relevant product market with a multi-dimensional substitution.

## 4.2.2 A Method Using Change in Demand Cannot Avoid Arbitrariness in Market definition

One way to quantify the level of substitutability is to find the pattern of sales change when other aspects of the market, typically the product price, is changing. As we have discussed in Section 1, the Hypothetical Monopolist Test decides how much substitution is small enough to define the market. The test uses demand change as a "substitution" of the substitution level. Therefore, the change of demand is essentially a quantification of substitutability.

However, as we have discussed in Chapter 3, a complete Hypothetical Monopolist Test is not as easy as the CL formula indicates. It requires a great effort of data collection and calculation to be performed. Moreover, we cannot solve the problem that the collectible CM data and demand data are not a real match when we include more than one product's price in the candidate market. Ignoring this problem will result in a narrower market definition than the reality. The Hypothetical Monopolist Test is still the best method of market definition up to date. However, as we have discussed, whether a hypothetical monopolist can profit or not is not informative about whether a merger will hurt the competition unless the candidate market only consists of the merging companies. Therefore, the Hypothetical Monopolist Test is just a "bright line" rather than a correct line.

In fact, we can probably conclude that even if we come up with a model better than the Critical Loss Analysis, as long as we use demand to replace the level of substitutability, the result can never be more than a "bright line". When we decide whether to buy a product, almost all of the products have more than one factor to be considered. These factors constitute multiple dimensions of substitutability. Therefore, substitutability is multi-dimensional. If we use a one-dimensional value to quantify or approximate substitutability, it is a dimension reduction. Demand is a one-dimensional value. Using demand change to quantify the level of substitutability is to map multi-dimensional vectors on a line. This incurs a massive loss of information.

Figure 1 is an illustration of two-dimensional data projection on one dimension. The green dots constitute a cluster. The red dots constitute another cluster. To convert them into a one-dimensional value, we need to map them on a line, as we can see in Figure 1.

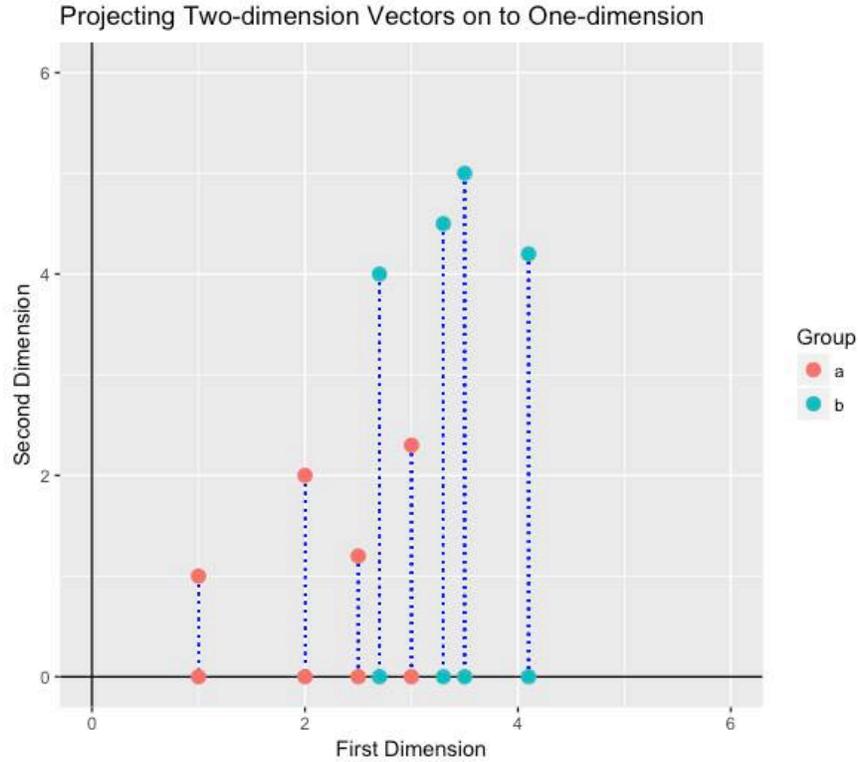

*Figure 1*

After the mapping, we get a set of converted values in Figure 2. As we can see in Figure 2, it is hard to divide Group A and Group B consistent with the correct clusters. This is similar to what happens when we use demand as an indicator of substitutability. Consumers consider multiple product features while all of the considerations will be reflected on the demand change, with which we are unable to divide Group A and Group B in a way consistent to the reality, because of the loss of information.

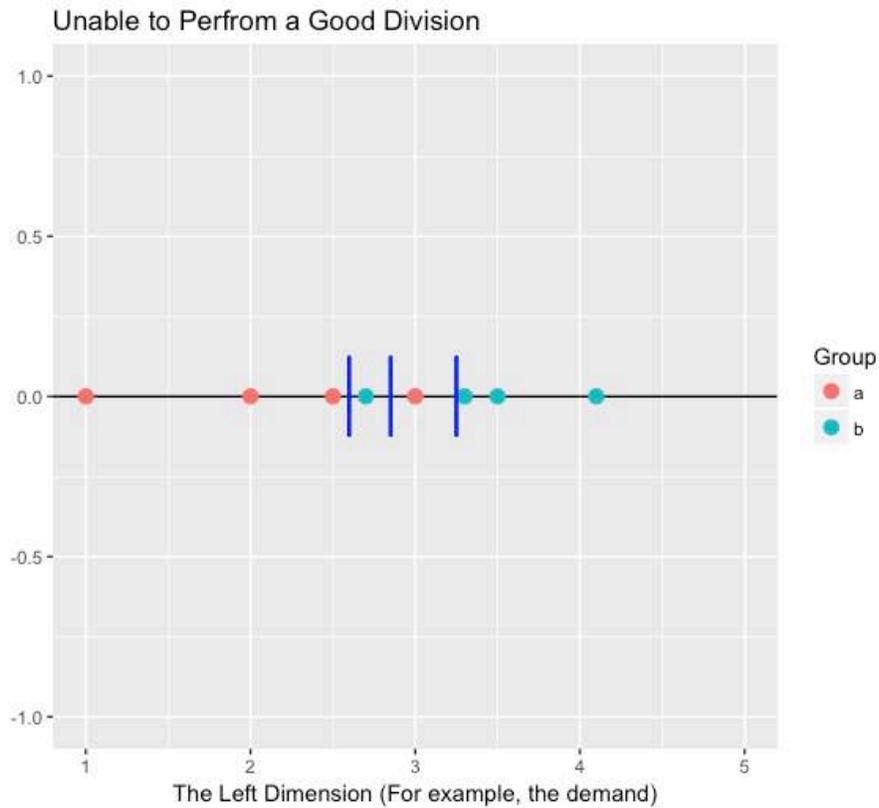

*Figure 2*

In Figure 3, we put the data points back to their original space. We can easily divide them into Group A and Group B with the blue line. Market definition is similar to a classification problem in machine learning. In machine learning, a common trick is to increase the dimension of a dataset using a kernel function. It is a bit different here from manually increasing the dimensions as machine learning does. Instead of using a kernel function, we propose to define the market with a different type of data rather than demand. We believe collecting data of multiple features of the products is a more accurate way to define the relevant market.

As we have discussed in Chapter 3, the Hypothetical Monopolist Test and the CLA are theoretically functional, but the current method used by the Agencies is problematic. Notwithstanding the fact that there is still going to be unsolvable problems as what we have discussed in the last sections of Chapter 3, a correct way of performing the Hypothetical Monopolist Test and the CLA requires significantly more data and calculation than the current application of them. Therefore, even though the model we propose requires collecting another type of data, it can still be desirable for the practice of the premerger investigation.

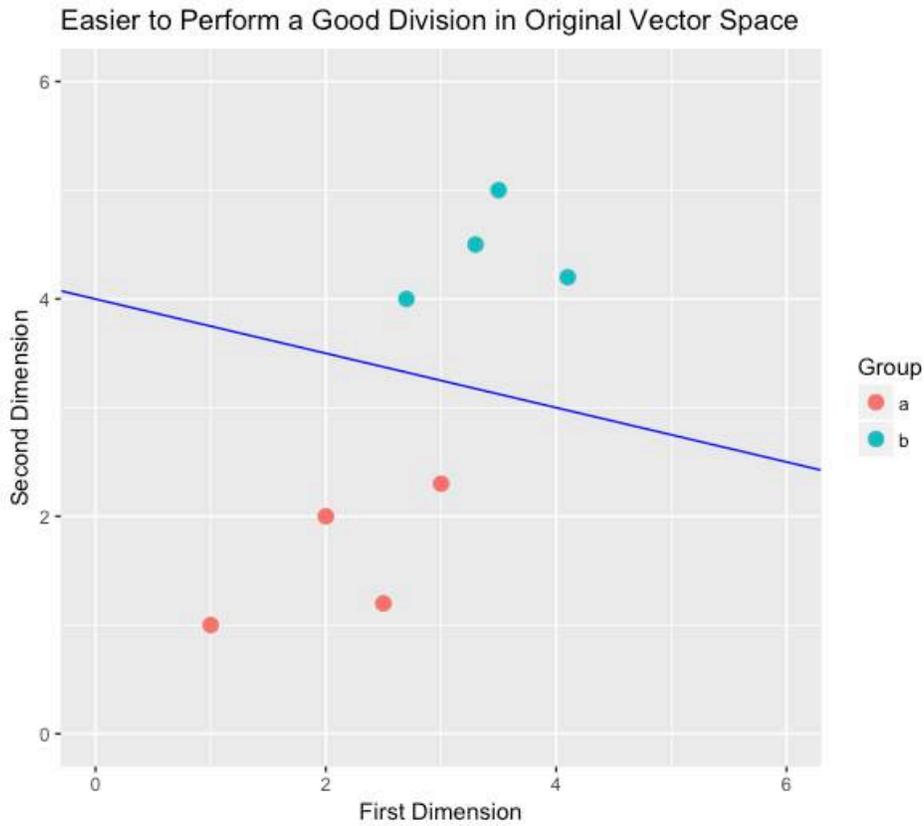

*Figure 3*

On the other hand, the 2010 Horizontal Merger Guidelines stipulates that the level of substitutability depends on the amount of switch in sales, i.e., change in demand. The more sales increase another product has after the SSNIP; the more substitutable the other product is to the SSNIP product. However, this is not necessarily true. For example, in material markets, different materials have different efficiency, a lower efficiency material can incur a larger amount of switched sales. However, if the original product has relatively higher efficiency, a substitutive material with similar efficiency is a better substitution. If we look at the change in sales alone, we will be misled. In fact, according to the HSR annual reports, chemical manufacturing has been the industry group with the highest number of HSR cases filed from 2012 to 2016[240]. It is also an industry where the sales of substitution will be profoundly affected by the material efficiency. Therefore, we cannot just omit the problem of using the switch in sales alone.

On the other hand, the efficiency data might be collected and submitted to the court by the parties anyway, if looking at the demand alone will cause injustice. However, even if the data is submitted, there is no model considering the additional data along with other factors. The court will have to arbitrarily decide how much does each factor matters. Therefore, we need a model capable of considering multiple factors at the same time.

The problem of not considering the efficiency is essentially due to the over-deduction of dimensions. Efficiency is an important aspect of substitutability. It is also a crucial feature of consumable materials. Not considering factors other than demand is therefore problematic. There are other problems caused by over-deduction of dimensions as well — the Hypothetical Monopolist Test cannot model versatile materials.

This problem is also prominent in the chemical material market. We use an example to illustrate this problem. This example is summarized from real chemical material substitutive information in "Solutions for critical raw materials under extreme conditions: a review"[241]. Therefore, the problem in this example is not a theoretical problem. To avoid confusion, we omitted the real names of the materials. We name them material A, B, and C. They are the substitution of each other as additions to alloys. There are two qualities being considered — improving the fluidity and increasing the tensile strength. In each unit of material A, B and C, the effect on fluidity and tensile strength are quantified by numbers. We are using simplified numbers while keeping the essence of the substitutive relations between the materials. Therefore, we assume the qualities of the three materials are A(3, 1), B(1, 4) and C(4, 1).

---

[240] Federal Trade Commission, *Annual Competition Reports*, 34th Report FY2011, 35th Report FY 2012, 36th Report FY 2013, 37th Report FY 2014, 38th Report FY 2015, 39th Report FY 2016 (Aug. 18, 2017, 17:30 PM), https://www.ftc.gov/policy/reports/policy-reports/annual-competition-reports, PDF.

In Hypothetical Monopolist Test, we need to track how much B and C's sales increase when A's price has increased to find the best substitution of A in improving fluidity. Three units of B are required to take A's place, while only 3/4 units of C are required to. Therefore, if all the other conditions are the same, when A's price has increased, B will have more sales increase than C.

However, if we look at the quality directly, it is easy to see that C is a better substitution of A. Using B will cause a twelve-time increase in tensile strength, which is a significant change in the property of the alloy. In the chemical manufacturing industry, there are different products being made while using the same feature of A. The tensile strength is not essential to some of the products. However, it is important to some other products. In the situation where we care about tensile strength, C is the only reasonable substitution. However, this fact may not prevent B from

---

getting the most significant increase in sales and end up being concluded as the best substitution of A in the model currently used by the Agencies.

The industry of chemical manufacturing is different from the industries of consumer goods. It is common that one material has dozens of usage based on many properties of the material. Looking at demand relationship only is unlikely to result in a reasonably realistic result. In this case, the court has to look at the substitutive relations without a model.[242] Chemical manufacturing, and computer and electronic products manufacturing rank the first and the eighth in the number of premerger investigation cases last year[243]. In both of the industries, versatile materials are widely used. Therefore, this problem regarding the versatile materials should not be omitted.

On the other hand, if one material has multiple usages in manufacturing, the Hypothetical Monopolist Test cannot tell us which usage is to be used as the basis for measuring the demand. Facing multiple usage materials, the Hypothetical Monopolist Test only results in multiple candidate markets, while the only solution the Guidelines provides to this situation is to pick any one of those. The essential reason for this problems is still an improper dimension reduction — demand is a one-dimensional value, it cannot correctly reflect all the relevant aspects of multiple-quality objects. The Hypothetical Monopolist Test is based on the assumption that the closer substitution the product is, the more sales will switch to the product. However, when the assumption no longer stands, the result is no longer reliable as well.

---

[242] The other models used in premerger investigation all associates with demand and price as well.

[243] Federal Trade Commission, Annual Competition Reports, 39th Report FY 2016 (Jan 18, 2017, 5:30 PM), https://www.ftc.gov/policy/reports/policy-reports/annual-competition-reports, url.

These two problems of Hypothetical Monopolist also restated our conclusion that products substitute each other on different features. Most of the time, there is more than one feature considered. If we use a one-dimensional value to quantify the level of substitutability, the loss of information will make it difficult to get meaningful segregation other than drawing a "bright line".

To summarize, if we quantify substitutability with a one-dimensional value without information preserving technics, many different data points will be mapped to the same value, losing some essential information. Therefore, the one-dimensional demand may not be desirable to all types of markets. Based on our discussion, what we can do to improve the market definition model is clear—instead of using a number, we should use a vector to quantify substitution. The dimensions of the vector should be what actually constitutes substitutability, which according to the Supreme Court's decision on *United States v. E. I. Du Pont de Nemours & Co.*[244] is the product's price, quality and use.

In later sections of this chapter, we are going to propose a new method to define the relevant product market based directly on substitutability. In our model, the quantified substitutability level will be stable as long as the product features do not significantly change. The result of market definition can be reused as long as the product features do not significantly change and there are no new products available in the relevant market. Even if there is a new product, the previous data is still valid and does not need to be recollected. All we need is to do is get the features of the new product and re-run the model to get a new result of division directly from the model result. Our model is especially good for the material market. It solves the versatile material problem and the efficiency problem, both of which are common in the material market. Data collection in the material market is also easier because personal preference is not crucial in purchasing decisions. Moreover, the products in industries other than consumer goods are relatively stable, especially the chemical material industry. In these industries, market definition results can be valid for long.

---

[244] United States v. E. I. du Pont de Nemours & Co., 351 U.S. 377 (1956).

Hypothetical Monopolist Test only gives out case specific result while it is not able to help in further antitrust practice. In contrast, our method offers a more stable market definition which can be used in future premerger cases and in deciding the annual HSR report threshold. With our method, the threshold for HSR filing no longer has to be a cross-industry rough estimation. Our result can be applied along with HHI or other competition measuring methods. Based on our result, HHI can be used to determine industry-specific standards, and save the efforts of HSR filing of high-target-amount mergers in high-competition markets. This will be discussed further in Chapter 4. In the following sections, we are going to introduce our market definition model which is directly based on a multi-dimensional quantification of substitutability.

# 4.3 K-MEANS CLUSTERING

We have discussed in Section 3.1 that the relevant market should be defined based on the substitutability between the products. We concluded in Section 3.2 that we should quantify the substitutability as a multi-dimensional vector instead of a number. Logically, if two products are close to each other in one dimension, they are more substitutable to each other in that dimension. Therefore, the Euclidean distance of the products in all dimensions measures how much they are substitutable to each other. If two products have small Euclidean distance between each other, they are overall close substitutions. Product within the same market should have similar qualities to be measured. Substitutability measured by Euclidean distance should be universal among the products as long as we use the same qualities. Therefore, we should put the products which are close to each other in Euclidean distance into the same market, and put the products which are far away from each other in a different market.

In statistics, dividing products into different relevant markets is a classification question. It overlaps with some of machine learning algorithms which do cluster analysis. "Clustering (Anderberg, 1973; Jain and Dubes, 1988; Kaufman and Rousseeuw, 1990) is a popular approach to implementing the partitioning operation. Clustering methods partition a set of objects into clusters such that objects in the same cluster are more similar to each other than objects in different clusters according to some defined criteria. "[245]

---

[245] Zhexue Huang, *Extensions to the k-means algorithm for clustering large data sets with categorical values, Data mining and knowledge discovery* 2.3 283-304 (1998).

By definition, cluster analysis is finding objects which are "similar" to each other. Therefore it seems that most of the models doing clustering analysis will suffice for our task of dividing the relevant market. However, this definition only defines the general task of the analysis. We need to look into the algorithm to see what the model actually does and whether what the model does is what we are theoretically looking for. For example, how do they define "close" in the model; how does the model measure the closeness, etc. Market definition is a legal question. The answer provided by the model may have a substantial legal effect on a merger. We need to make sure that the method within the model is also consistent to the market theory and the antitrust purpose.

Cluster analysis is a multi-objective optimization problem in machine learning. "It can be achieved by various algorithms that differ significantly in their notion of what constitutes a cluster and how to efficiently find them."[246] Several decades ago, the computation capacity is still quite limited in most of the industries. Therefore "clustering has been effectively applied in a variety of engineering and scientific disciplines such as psychology, biology, medicine, computer vision, communications, and remote sensing."[247] In the past few decades, the calculation capacity has been developing quickly. Cluster Analysis has been widely used in machine learning, artificial intelligence, advanced computer graphics, and bioinformatics.

---

[246] Cluster analysis, Wikipedia (Jan. 21, 2017 11:10 AM) https://en.wikipedia.org/wiki/Cluster_analysis, url.

[247] K. Krishna & Murty M. Narasimha, *Genetic K-means algorithm*, IEEE Transactions on Systems, Man, and Cybernetics, Part B (Cybernetics) 29.3 433-439 (1999).

Unlike most of the clustering analysis applications and research, efficiency is not essential in our task. Our data size is small compared to big data programs like bioinformatics. Neither do we need to calculate the result fast like it is in computer graphics or smart device. However, we have our own challenges. Since there is no absolute right answer in market definition, our model has to be an unsupervised machine learning model. Unsupervised machine learning models do not have training dataset to compare the result and recalibrate the model. However, the market definition model is influential in the legal progress of mergers with anti-competitive concerns. Therefore, in the context of market definition, the model we use should be not only able to produce a result of market definition but also be consistent to the antitrust theory and the logic of jurisprudence.

Currently, the more advanced clustering technologies are usually uninterpretable. Therefore they are not what we are looking for. Instead, we should use fully interpretable methods, so that we can discuss the rationality of the model result like we always do in normal legal proceedings. The most widely used methods of clustering analysis are grouping with small distances between cluster members; grouping with dense areas of the data space, and grouping with intervals or particular statistical distributions. As we mentioned earlier, the market definition shall put the products which are close to each other in the Euclidean distance into the same market, and the reverse to a different market. Grouping with small distances between cluster members can fit our purpose the best. In the context of premerger investigation of multi-dimensional vector space of substitutability, being dense is similar to being close in distance, Therefore, in some cases, grouping with dense areas of the data space can also be rational under the logic of law. In this Chapter the first algorithm we are proposing is a method of grouping with small distance between cluster members. The second algorithm we are going to suggest is a method of grouping with dense areas of data space.

In this session, we are going to propose to use the K-Means++ Clustering to define the relevant product market. We will explain the algorithm in this session, and apply it with some examples in the next session. When we are doing our clustering analysis, we need to not only consider which algorithm to choose, but also what distance function to use, the number of expected clusters, and whether we need to set a threshold for product density, etc.

On the other hand, we need to figure out which type of data is available or at least collectible under the law, and which data we are going to collect and apply to our model. These will be discussed according to the top 20 most frequent premerger investigation industry groups because products satisfy consumers' needs in different ways across different industries.

K-means ++ clustering algorithm[248] is an improved algorithm of K-means clustering. K-means clustering was initially proposed by Holland in his paper "*Adaptation in Natural and Artificial Systems*".[249] It tried to find a given number of clusters which minimizes the average distance between the data points and the center of each cluster.

---

More specificly, we are given a set of n data points $X \subset R^d$ and the number of clusters, $k$. We use the algorithm to find the centers C, in a number of $k$, which satisfies Equation 2. In Equation 2, $\varnothing$ means the total distance left between data points and the center, after each calculation of the minimization. After finding those centers, we group the data points according to which center is closest to them.

$$\phi = \sum_{x \in \mathcal{X}} \min_{c \in \mathcal{C}} \|x - c\|^2$$

*Equation 2*

We can infer three problems from the description of the algorithm's goal, considering we are aiming to use it in market definition.

The first problem is that the result of K-means clustering is slightly different from the result we are looking for. Our goal is to find a way to group the products so that the difference in each aspect of substitutability between the products are smallest.

In the space of substitutability, we can naturally use the Euclidean distance between the products to represent the overall substitutability between the products. Therefore, there is not a place for centers in our definition. However, all the points being close to the center point of each cluster is equal to all the points within each cluster are closest to each other. It can be proven by the following lemma.

Let $z$ be any arbitrary point in $R^d$, and $S$ be some points in $R^d$ within the group whose center is $c(S)$. We have Equation 3.

$$\sum_{x \in S} \|x - z\|^2 - \sum_{x \in S} \|x - c(S)\|^2 = |S| \cdot \|c(S) - z\|^2$$

*Equation 3*

Therefore, when the distance between $x$ and $c(S)$ and the distance between $z$ and $c(S)$ is minimized, the distance between $x$ and $z$ are also minimized. Thus, if $c(S)$ exists, $z$ and $x$ belong to the same group, in which the distance between $x$ and $z$ is also minimized. Therefore, according to the formation of our vector space $R^d$, the goal stated above is exactly the goal of market definition.

The second problem is what k we are going to use. This is a practical question. In statistics, there are four most common methods used by researchers. Three of them are relevant to our task. They are:

4.  Find the most representative two or three dimensions of the data (which is usually the two or three dimensions which explain the most variations), then graph them in a figure. Observe the figure to tell how many clusters are there. The number of clusters we can see is the $k$ we are looking for. This method is not that useful in high dimensional data, because we can only look up to three dimensions, which will not effetely reflect the whole dataset. Our data is relatively high dimensional. Therefore the usage of this method will be limited.

5. Use a method called the elbow method. $K$ is usually an integer within a range. Therefore, we can run the data multiple times with different $k$. Then calculate the sum of the total distance within each group and compare it with different $k$, and find the best $k$.

For a dataset $\{x_1, \ldots x_N\}$, when we apply K-means clustering on it, we will get $k$ clusters and each cluster $C_k$ has $n_k$ data points. We define the sum of distance within each group as $D_k$, so we have:

$$D_k = \sum_{x_i \in C_k} \sum_{x_j \in C_k} \|x_i - x_j\|^2$$

*Equation 4*

Since $\| \cdot \|$ is a 2-norm, when the number of cluster is $K$, we have the sum of the distance within each group as $W_K$:

$$W_K = \sum_{k=1}^{K} \frac{1}{2n_k} D_k$$

*Equation 5*

We can link the points and calculate the inflection point, where the angle changes fastest, and it will be the best k for our clustering analysis. Or, we can simply graph $k$ as $x$ axis and $W_K$ as $y$ axis, and find the inflection point from the figure directly. For example, in figure 3, we should pick 3 as our $k$.

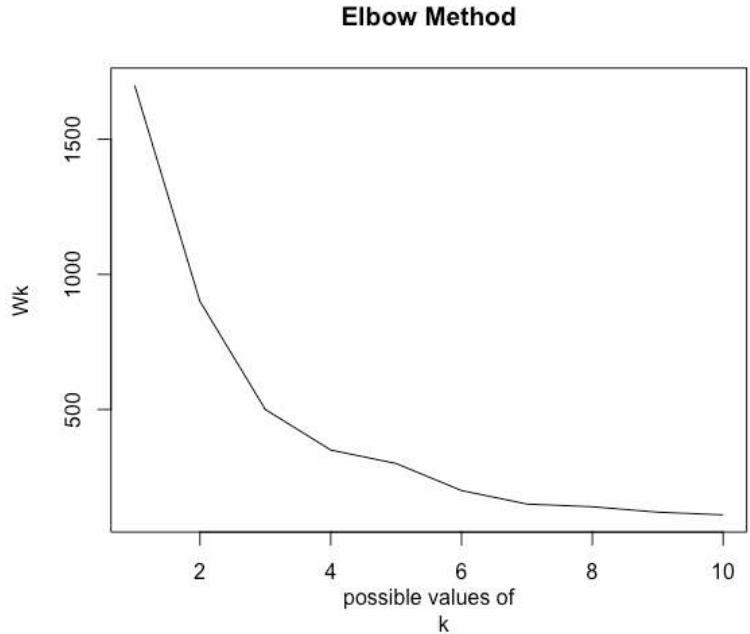

*Figure 3*

This method works because as we increase the number of $k$, $D_k$ will always decrease. At the same time, $W_K$ will also decrease. After a more appropriate k was reached, we will over-divide within an appropriate group. Therefore, the decrease of $W_K$ will significantly slow down when we start to over-divide the data points because the distance between the products within a group is relatively small. Therefore, the $k$ on the inflection point is the $k$ we are looking for. After the inflection point, the change of $W_K$ is no longer significant, meaning we are only doing over-division. In the context of market definition, the elbow method is to make sure that we stop dividing the market when the products are close substitution of each other. The K-means clustering will do the rest of the work to make sure all the closely substitutable products are grouped into the same relevant product market.

6. The gap-statistics method is another method to find an appropriate k. It was first proposed in the paper "*Estimating the number of clusters in a data set via the gap statistic*"[250]. $W_K$ was defined the same as above. The gap-statistics was defined as Equation 4.

$$Gap(K) = E(log(W_K)) - log(W_K)$$

*Equation 6*

In Equation 4, $E(log(W_K))$ is the expected value of $log W_K$. It was usually calculated in a Monte Carlo method. This method is first to simulate data points in the same number (we set it as B) of the data points we need to do clustering analysis. The simulated data points are in a uniform distribution. Each set has simulated uniformly distributed points in a number of B. We make, for example, 20 simulated sets for each k and do K-means clustering for all the simulated sets. Again, *k* is usually between 1 to 10. We do it from k = 1 to k = 10. Therefore, there will be 200 sets, and each set will have simulated points in a number of B. We then average log $W_{Kb}$ calculated from each simulated sets, so we get $E(logW_K)$ in Equation 6. $Log(W_K)$ was calculated as the log of $W_K$ in *Equation 5*. In this way, the *Gap(K)* can be calculated as Equation 7.[251]

$$\text{Gap}(K) = \frac{1}{B} \sum_{b=1}^{B} \log(W_{Kb}) - \log(W_K)$$

*Equation 7*

We then calculate *Gap(k)* with different k and find the maximum of *Gap(k)*. The k which enables the maximum *Gap(k)* is the k we are looking for.

In essence, this method compares the distance between data points, $D_k$, within each group, with the $D_k$ in the situated groups where all the data points evenly spread in the vector space. Since our data is clustered and the evenly spread data is not, the clustered data will have a smaller $D_k$. On the other hand, if we over-divide the data into even small groups, i.e., the k is too large, we are going to make the $D_k$ of the evenly spread simulated points smaller as well. At the same time, the $D_k$ of the real data will not decrease significantly anymore. Therefore, finding the maximum *Gap(k)* will help us find the best number of clusters we need.

---

[251] Equation 7 is not the only way to calculate $E(logW_K)$.

A more intuitive way to explain gap statistic is that it ensures the largest "gap" between groups. When we are not dividing data into enough clusters, the data which should have belonged to other clusters are put together enlarging the range of data points within each cluster. Therefore, the "gap" between clusters will be smaller. Also, when we are over-dividing the data, the data points which belong to the same group will be divided. Therefore, the "gap" between them must be small. As a result, the number of clusters ensures the largest "gap" will be the best number of clusters.

Gap-statistics method is widely used in high volume dataset analysis. We also like to use it in our approach because it is not subjective. Besides, a wider "gap" means a more significant difference between products, i.e., we are making sure that the products in different groups are not closely substitutable to each other.

Although K-means clustering has been around for a long time, the methods of finding k are still evolving in research. There are plenty of papers discussing this topic. The mentioned four methods to find the k are the most widely used by statisticians and scientist. We can find meaningful rationale to use the third and fourth method in market definition context. Therefore, we suggest to use them in the context of market definition. We will have some examples of applying these methods in the next section.

The paper "Integration of self-organizing feature map and K-means algorithm for market segmentation" proposed a two-step approach to do K-means clustering in market segmentation.

Market segmentation has a similar task with market definition. "Market segmentation is the activity of dividing a broad consumer or business market … into sub-groups of consumers based on some type of shared characteristics. In dividing or segmenting markets, researchers typically look for common characteristics such as shared needs, common interests, similar lifestyles or even similar demographic profiles. The overall aim of segmentation is to identify high yield segments so that these can be selected for special attention." Unlike the market definition, market segmentation aims at grouping consumers and finding a target market. In the commercial process, market segmentation is closely related to the product features, because the products were designed according to the need of the target market. Therefore, market definition may also learn from the market segmentation. As a result, we will also discuss the two-steps approach in the context of market definition in this and the next section.

In the two-steps approach, the first step is to use hierarchical clustering to figure out the number of k. The second step is to do the K-means clustering with the k found in the first step. The paper used two examples as experiments to illustrate the merit of the two-step approach. One example used simulated data and the other used real data. The examples show the two-step approach produces a better result than using hierarchical clustering only. Since the examples show K-means clustering produces a better result than the hierarchical clustering, even if we can already get a clustering result in the first step of the two-step approach, it is still meaningful to perform the k-means clustering as the second step. Moreover, hierarchical clustering uses the same data structure as K-means clustering does. Therefore it will not take too much time or resource to do the second step. This might be another reason why the two-steps approach is popular.

It is worth mentioning that since we are discussing this under the context of legal proceeding, we need to consider the rationale of using an algorithm. The algorithm of hierarchical clustering produces a tree-structured result. It does not calculate the distance between two products which are not within the same branch.[252] Therefore, even if the hierarchical clustering eventually produces a similar result as K-means clustering does, theoretically we still have to perform the two-step approach and limit the function of hierarchical clustering in finding a proper k.

There are even more methods available to find k. The approaches we mentioned above can help us to solve the technical problem of finding k. They do not affect the rationality of using K-means clustering to the market definition as we discussed. However, we have one more problem to solve before we conclude that we can use the k-means clustering in market definition.

The third problem of using K-means clustering is that the algorithm only "tries" to find the result which satisfies Equation 2. There is no guarantee of getting the best result. As we can see later from the steps of the algorithm, the start point is random in the genetic k-means clustering method. Since the algorithm will only end up with a local optimum result, if the random start points are not in the range of a theoretically preferable area, the result may not be the one we are looking for. Meanwhile, we do not know where the global optimum is. Furthermore, finding the universal optimal result using K-means clustering is proven to be NP-hard[253] in the paper "The planar k-means problem is NP-hard." At the same time, the k-means clustering has been improved by many papers in computer science, including the paper proposing k-means++ clustering.

---

[252]Lior Rokach & Maimon Oded, *Clustering Methods*, Data mining and knowledge discovery handbook (2005)

K-means++ is one of the most popular modified k-means clustering algorithms. The method we are proposing to use in market definition also includes K-means++ clustering. It can "yield a much better performing algorithm and consistently finds a better cluttering with a lower potential than K-means."[254]

| | Average $\phi$ | | Minimum $\phi$ | | Average $T$ | |
|---|---|---|---|---|---|---|
| k | k-means | k-means++ | k-means | k-means++ | k-means | k-means++ |
| 10 | 10898 | 5.122 | 2526.9 | 5.122 | 0.48 | 0.05 |
| 25 | 787.992 | 4.46809 | 4.40205 | 4.41158 | 1.34 | 1.59 |
| 50 | 3.47662 | 3.35897 | 3.40053 | 3.26072 | 2.67 | 2.84 |

Table 1: Experimental results on the *Norm-10* dataset (n = 10000, d = 5)

---

[253]In computational complexity theory, NP (for nondeterministic polynomial time) is a complexity class used to describe certain types of decision problems. Informally, NP is the set of all decision problems for which the instances where the answer is "yes" have efficiently verifiable proofs. More precisely, these proofs have to be verifiable by deterministic computations that can be performed in polynomial time.
—— Complexity, Wikipedia (Mar. 12, 2018, 15:30 PM), https://en.wikipedia.org/wiki/NP_(complexity), url

NP-hardness (non-deterministic polynomial-time hardness), in computational complexity theory, is the defining property of a class of problems that are, informally, "at least as hard as the hardest problems in NP".

Therefore, saying K-means clustering is an NP-hard question means that solving it is as hard as solving an NP complexing question, which means, in a simple language, one has to calculate all the possible outcomes to see whether one outcome is the answer we are looking for.

[254] David Arthur & Vassilvitskii Sergei, *k-means++: The advantages of careful seeding, Proceedings of the eighteenth annual ACM-SIAM symposium on Discrete algorithms*, Society for Industrial and Applied Mathematics (2007)

Table 1 is the comparison of k-means and k-means++ methods provided by the paper proposing it.[255] In fact, we only need to look at the situation when k is equal to 10 in table 1. Since we are facing a relatively small volume of data, and over-dividing the data will always end up minimizing ⌀, we are almost sure that a k which is more than 10 can end up with a small ⌀ , no matter which variation of K-means clustering we are choosing. We can see from table 1 that the minimum ⌀ decrease significantly. The T value also decreased significantly, meaning the result produced by the k-means++ is a lot better.

On the other hand, even though finding the optimum result of k-means clustering is NP-hard in computer science, it might not be a problem in the context of premerger investigation. Unlike most of the NP-hard" problems concerns, our dataset is relatively small, we can run the model for many times to make sure we are actually getting the smallest ⌀. Therefore, having k-means++ can suffice our need.

Since we are discussing a model's application in a legal context, we do not need to know the underlying calculation done by the computer. Rather than that, we need to scrutinize the main steps done in the computation and what type of data we are using. We are going to look into the main steps in this section and talk about the data in the next section.

---

[255] Id.

A k-means++ clustering algorithm will also do the same steps as a genetic k-means clustering does. The genetic algorithm does several steps of work. Since this is a law dissertation, we would like to put it in more intuitive words.

1. Randomly choose k points from the space $R^d$ as centers of clustering. Name the centers in a set of $C = \{C_1, C_2, \ldots, C_k\}$

2. Calculate the distance between the centers $C_i$ . If a point is closer to one center $C_i$ then that point belongs to $C_i$'s cluster. Repeat this process with all the data points in $X$ between all the candidate centers $C_i$ for each $i \in \{1, 2, \ldots, k\}$.

3. Change the candidate center $C_i$ from the original spot in the average of all the points in each dimension within the cluster $C_i$. Repeat this process with all the candidate centers $C_i$ for each $i \in \{1, 2, \ldots, k\}$. In mathematical terms, $C_i$ is

$$c_i = \frac{1}{|C_i|} \sum_{x \in C_i} x$$

*Equation 5*

4. Repeat step 2 and step 3 until all the centers $C_i$ no longer changes.

As we have mentioned, the reason the model does not come up with a universal optimum is due to the randomness of choosing the starting point. The K-means++ clustering algorithm uses the steps above to find better start points, which replaced the first step of the genetic K-means clustering algorithm. Denote D(x) as the shortest distance from a data point to the closest center we have chosen. The steps are:

1. Choose one data point from X as our first center $C_1$. When choosing, each data point has the same probability of being chosen.

2. Calculate the shortest distance between all the data points left and the chosen centers, which we denoted as D(x). (For now, we only chose $C_1$, so we do not need to compare and get the shorter distance.) Then we give each data point left a probability of being chosen, which is calculated by Equation 6. Then we use a roulette wheel selection method[256] to pick one data point as the next center.

$$\frac{D(x)^2}{\sum_{x \in \mathcal{X}} D(x)^2}$$

*Equation 6*

3. Repeat step 2 until selected the number of centers required.

The essence of the three steps above is to spread the starting points as far away from each other as possible. We will illustrate this with an example below. In the example, we assume there are nine data points in our data frame as is in Figure 4.[257] Assure the point we picked in step one is 3, the relevant values were calculated is in table 2. Meanwhile, instead of choosing random points in the space as the centers, K-means++ chooses data points as the centers.

---

[256] The roulette wheel selection, which is also known as the fitness proportionate selection, is an algorithms for selecting potentially useful solutions for recombination. — Adam Lipowski & Dorota Lipowska, Roulette-wheel selection via stochastic acceptance, *Physica A: Statistical Mechanics and its Applications,* 391.6, 2193-2196 (2012).

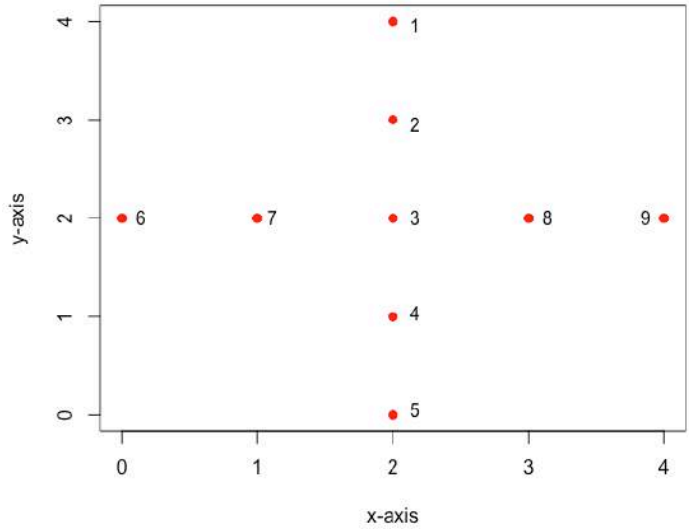

*Figure 4*

| Point number | 1 | 2 | 3 | 4 | 5 | 6 | 7 | 8 | 9 |
|---|---|---|---|---|---|---|---|---|---|
| D(x) | 2 | 1 | 0 | 1 | 2 | 2 | 1 | 1 | 2 |
| D(x)² | 4 | 1 | 0 | 1 | 4 | 4 | 1 | 1 | 4 |
| P(x) | 0.2 | 0.05 | 0 | 0.05 | 0.2 | 0.2 | 0.05 | 0.05 | 0.2 |
| Sum(P) | 0.2 | 0.25 | 0.25 | 0.3 | 0.5 | 0.7 | 0.75 | 0.8 | 1 |

*Table 2*

---

[257] To make it easier to calculate the distance and see how the probability of data point being chosen is calculated, we did not use a clustered example. The read data will not be unclustered like the Figure 4.

As we have mentioned earlier, K-means++ clustering algorithm was proven to help ensure that we have a much higher probability to get the universal optimum result, which ensures that all the products are close substitutions of each other inside the relevant market. In practice, K-means++ can get us to the optimum result almost for sure if we have enough repetition, which does not take much time with a small data size like ours. In contrast, as we have mentioned before, the Hypothetical Monopolist Test does not have a safeguard to make sure the bright line they draw is indeed meaningful to separate the closely substitutable products into relevant markets in one way but not another. Again, the Guidelines does not provide us any better suggestion than picking randomly from similar results.

Since our goal is to find the substitutive relations of the target product[258], we should better make sure the target product is in the center of the substitutability comparison. Therefore, one more merit of using the K-means++ is that instead of randomly picking the first data point as $C_1$, we can simply pick the target product as $C_1$, in which case, we can ensure that in one of the clusters, the target product is in the middle of the cluster. Meanwhile, all the other centers are not to close to the target product, so that we can have one more safeguard that we are not over-dividing the products which are close substitutions to the target products into different relevant markets.

On the other hand, although the Hypothetical Monopolist Test starts from the target product, and add the other products in the sequence of the closeness of the substitution (which is measured by the change of demand), after adding the first product other than the target product into the group which will increase product price, the center of the substitutive relations will be somewhere between the two products, because the sales of the second product is considered as well. In this case, if the first several products happen to be on a similar "direction", the center will be "dragged" far away from the target product by the other products. Therefore, even if in the beginning we are finding the substitution according to the demand of the target product, we will no long do so after the second round of analysis, and what we eventually find is no longer the sequence of closeness to the target product. This is incorrect because in reality, only the target product and the other merging company's product is facing a real danger of price increase.

---

[258] According to the 2010 Horizontal Merger Guidelines, the target product is a product of either of the merging parties.

Another merit of using the K-means++ is avoiding the risk of bad execution of the model.[259] When the start points are by chance badly chosen in a random process, and the people conducting it does not do enough repeat to ensure a better result, the result we get from k-means clustering might be very different from the universal optimum. When we use k-means++, the repetition needed decreases significantly.[260] Therefore, the risk of not conducting enough repetition will significantly decrease as well. Another risk will be choosing a k which is too large. With k-means++, even if the k is too large, the model will not over-divide the cluster near C1, which means at least the relevant market we need in the premerger investigation is not over-divided.

---

[259] We concern bad conduct of the model because the Agencies has conducted the CLA in a very unreasonable way in regarding the data used and the definition of SSNIP as we have mentioned in Chapter 3.

[260] We get this conclusion from the examples in the next section. The results of twenty repetitions of k-means++ have many repetitive values of $\emptyset$ in a small value, which indicates that even twenty repetitions are redundant. However, the standard k-means gives less repetitive results in twenty repetitions, which means in the specific example there is a risk of not repeating the experiment for enough time to get the optimum result.

There are additional points necessary to address about using K-means++ clustering algorithm to do market definition based on substitutability.

The first point is that the Euclidean distance is not the only distance we can use with K-means clustering. Different definitions of distance are widely used in clustering algorithms in similarity measurement. K-means clustering works with all the commonly used types of distance. However, according to the property of the vector space we defined — the space $R^d$ with d dimensions which are the major aspects that the consumers look for from the product, Euclidean distance serves our purpose best.

In our method, we use Euclidean distance in the multi-dimensional space to measure substitutability. There is a paper—A measure of utility levels by Euclidean distance[261], which talked about using Euclidean distance to measure utility. This paper has a similar but different idea from ours based on money metric utility. As we have discussed, utility is different from substitutability in many features. Like the majority of the papers talking about utility, the purpose of this paper is getting the preference sequence, not doing clustering analysis or defining the relevant market. The data used to calculate the distance is also different from ours.

One other thing we must discuss is the input of data. We can form a high dimensional space of substitutability with data collected from product features and well-designed surveys to run the

---

model and perform market definition. Detailed instructions of data collection will be discussed in the next section. On the other hand, K-means clustering takes numeric input. There is no upper limit of the dimensions of data. Technically, the more dimensions we use, the more accurate the result will usually be. However, a too flat data input might introduce too much noise to the model, which might badly affect the performance of the model too. Besides, collecting too much data will also waste the resource. Although data mining and dimension reduction are both important and hot topics in machine learning, it is not that relevant to the purpose of market definition. In session 3.4 and section 3.5, we will talk about the data required to perform the model proposed to define the relevant market and how we are going to collect it.

# 4.4 INTRINSIC SUBSTITUTABILITY AND EXTRINSIC SUBSTITUTABILITY

We would like to address another issue about the substitutability before we discuss the data collection in the next section. As we have discussed in Section 3.1, the 1968 Guidelines following the Supreme Court case use *United States v. E. I. Du Pont de Nemours & Co*[262]. "price, use, and qualities" as the standard of "reasonable interchangeability". On the other hand, the 1968 Guidelines also stipulated a second standard -- "the sales of two distinct products to a particular group of purchasers can also appropriately be grouped into a single market where the two products are reasonably interchangeable for that group regarding price, quality, and use. In the second case, however, it may be necessary to also include in that market the sales of one or more other products which are equally interchangeable with the two products regarding price, quality, and use from the standpoint of that group of purchasers for whom the two products are interchangeable."

Although the second standard seems to be simply setting up the standard of interchangeability from the perspective of consumers, it is essentially different from the first definition — the second standard excludes the characteristics which the consumers do not appreciate. This distinction is important to the data collection in our model. For the convenience of discussion, we call the first one, the substitutive relation in price, quality, and use between the products itself

---

[262] United States v. E. I. du Pont de Nemours & Co., 351 U.S. 377 (1956)

"intrinsic substitutability". We further define it as the objective difference between products in product features. We call the second one "extrinsic substitutability", which means the product features from the perspective of consumers. We define the extrinsic substitutability as the subjective difference between products in product features.

The 1968 Guidelines do not explicitly differentiate the intrinsic and extrinsic substitutability., but there is some content describing how it considers the difference between products. "In enforcing Section 7 the Department seeks primarily to prevent mergers which change market structure in a direction likely to create a power to behave non-competitively in the production and sale of any particular product, even though that power will ultimately be limited, though not nullified, by the presence of other similar products that, while reasonably interchangeable, are less than perfect substitutes. "[263] To elucidate the difference between the homogeneous products and heterogeneous but substitutable products, the 1968 Guidelines further stated, "[i]t is in no way inconsistent with this effort also to pursue a policy designed to prohibit mergers between firms selling distinct products, where the result of the merger may be to create or enhance the companies' market power due to the fact that the products, though not perfectly substitutable by purchasers, are significant enough alternatives to constitute substantial competitive influences on the production, development or sale of each."[264] From these explanations, we may infer when defining the relevant market, the difference between products is considered from the perspective of consumers, i.e., the extrinsic substitutability is more likely to be used.

---

[263] The 1968 Horizontal Merger Guidelines, Section 3 Market Definition, Washington, D.C.: U.S. Dept. of Justice (1968)

[264] id

However, since the Guidelines did not explicitly differentiate the two concepts, intrinsic substitutability and extrinsic substitutability were mingled while defining the relevant market in practice. In most of the case, the two concepts are so highly correlated that they are sometimes interchangeable. However, when they are not interchangeable, their difference will affect the data collection, which will further affect the result of most models.

We will use three examples to illustrate where the intrinsic and extrinsic substitutability should be distinguished. The first example will be evaluating the substitutability of two cellphones which both have high definition digital camera, while one has voice assistant, and the other does not. The voice assistant technology is currently not mature yet. Most of the voice assistants, like Siri in iPhone and Alexa in Amazon devices, have very limited functions. Even if in most of the time they are featured by the manufacturers, they are not considered important by most of the consumers. If we use intrinsic substitutability, both the camera and the voice assistant should be considered. If we use extrinsic substitutability, the voice assistant will probably not be considered. In the context of premerger investigation, since we care more about whether the consumers have enough choice, we should use extrinsic substitutability.

The second example is the ingredients of the food. Different people's sensitivities to taste are different. The same type of materials, like sugar, have different tastes as well. Using the exact amount of ingredients added, which is intrinsic substitutability, is less meaningful because the difference in the number might not produce an exact or a sensible difference in substitutability.

Even if there is a difference in the number we collected, the difference might not affect consumer's decisions at all. Moreover, the taste of the food will be affected by the other ingredients, for example, the amount of sugar added is not always in direct ratio with how sweet the food is. A bitter taste can reduce the sweetness of the food, while a vanilla taste will not. In this situation, we should also use extrinsic substitutability and take a survey to have the consumers rate the taste of sweet by a one-to-ten scale. An averaged value of data collected with extreme values taken away may reasonably be regarded as one variable of the extrinsic substitutability of one product of food.

In contrast, there are also situations where the intrinsic substitutability is appropriate to describe a product. When we deal with product features which are essential to the products and hard to judge without expertise and parameters, the CPU frequency of a laptop or the photosensitivity of a camera for example, we should use intrinsic substitutability because the difference cannot be sensed at the beginning of using the product. Most of the products are designed to accommodate the current usage. The difference will be sensed in the long term or in some extreme cases. To protect the consumers and the good competitor who does not only aim at selling the product, we should consider what is actually inside of the products.

To summarize, we should differentiate the situations in the three examples. In the first case, the standard is whether this feature is influential to the consumers' choice. In the second case, the standard is whether the data on intrinsic substitutability we collect can scientifically reflect the reality of the substitutive relations. The third example established the standard of whether consumers can reasonably sense the reality of the substitutive relations. From these examples, we can also summarize some rules to guide our choice of data:

1.  If there are unimportant features to the consumers, we pick extrinsic substitutability and collect data on what the consumers believe to be important.

2.  If there are non-essential differences which cannot be measured meaningfully in a standard of intrinsic substitutability, we should use extrinsic substitutability.

3.  If there are essential differences which is hard to be noticed by the consumers, we should use intrinsic substitutability.

We can also find support for the selection rules above in the theory of market competition. The primary damage caused by limiting market competition is limiting the choice consumers have to substitute the product which price has increased due to the market power of the manufacturer. If there is not close substitutions, the consumers will have to choose from not purchasing or purchasing at a higher price. In this case, some consumers will decide not to buy the product. Eventually, there will be fewer people willing to buy and consume the product, meaning there will be less social welfare created and less consumer surplus[265]. In contrast, if consumers have enough close substitute, the price increase will not hurt consumers' interest. Since what we care about is the choice available to the consumer, product features which are less considered by the consumers or does not produce an essential difference will not have a strong impact on consumers' choice. On the other hand, we may also consider the fact that there are things that consumers value, but cannot tell the difference without product features. In this case, we will consider the intrinsic substitutability directly.

---

[265] N. G. Mankiw, *Principles of microeconomics* (4th ed.) (2006)

It is worth mentioning that although we tried to differentiate the two concepts and use them on different occasions when they are not interchangeable, there are still a lot of situations where these two concepts are interchangeable with respect to a market definition process. Since the extrinsic substitutability is consumers' opinion towards the products, we can only collect the data from a survey. As a result, in most cases, the intrinsic substitutability, i.e., the product feature data is easier to collect. Therefore, when the two types of substitutability are interchangeable, we will use intrinsic substitutability.

To summarize, differentiating the two concepts of intrinsic substitutability and the extrinsic substitutability can help us make sure the multi-dimensional space we form to do the clustering analysis reflects the true substitutional relations the consumers face when they make purchase decisions. It can also scientifically help us save resources during data collection when it is appropriate to use intrinsic substitutability instead of extrinsic substitutability. In the next section, we are going to discuss the data collection process for the model we propose.

# 4.5 THE DATA TO BE COLLECTED TO FORM A VECTOR SPACE OF SUBSTITUTABILITY

Data collection is the cornerstone of the data analysis. No model can work without proper data. Data mining and data cleaning are also important topics in statistical modeling and machine learning. As the technology advances, there are more types of data available in larger sizes too. However, what we look for from the data is different from what data analysts usually do. As we have emphasized many times in this dissertation, all the steps in defining the market should have a factual and legal basis. Assumptions, especially non-realistic ones, should be avoided in the market division.

The problem of data collection had been a problem in the market definition between the 1960s and the 1980s. This is also a period when antitrust models were a widely pursued topic in academia. However, at that time, antitrust models had to set many unrealistic assumptions to simplify the calculation and accommodate the availability of data. Models which were theoretically good could be severely criticized as unrealistic if the data required were not available or involved a large amount of computation. One of the reasons why the hypothetical monopolist test along with CLA can stand out from the models proposed in the two decades after the establishment of 1968 Merger Guidelines is that it is a method with a low requirement of data quantity and computation.

However, the situation has changed significantly over the past three decades. Both computational and statistical methods have been developing fast. It is a good time now to review the conclusion made three decades ago that it is impossible to quantify substitutability. As we have discussed in section 3.1 and section 3.2, the most logical way to divide the relevant product market is using a standard based on the level of substitutability. The most reasonable way to quantify substitutability is to use a multi-dimensional vector to represent each product's features because substitutability itself is a multi-dimensional concept.

On the other hand, the data for substitutability in our market definition model is collected as intrinsic substitutability or extrinsic substitutability. The intrinsic substitutability data is collected from the parameters of products features. The extrinsic substitutability data is collected by carefully designed surveys. We need first to figure out what do consumers consider when they are making purchase decisions. Most of the consumer goods manufacturers nowadays collect this information on a regular basis. Therefore, we can probably get it from the merging companies or the other competitors in the candidate market. According to section 2.2 of the 2010 Guidelines, a premerger investigation can collect information from the source of the merging parties, consumers, and other industry participants and observers. Therefore, other competitors also have the duty to provide information to the Agencies upon request, although the easiest way is still collecting the data from the merging parties. In a rare case where this information is not available, we can find the information from commercial data company or collect it with a survey. If we use a survey, we need to make sure the sample is random enough and find the features which, for example, 95% of consumers care about when purchasing the product.

When we get to know the feature that the consumers care about, we will decide if we need to use intrinsic or extrinsic substitutability on each feature based on the standard we established in the last section. We will use some examples to explain how to collect the data and form our multi-dimensional substitutability space. There are several important aspects of data collection generally important to all the models. Instead of discussing them all together abstractly then raise examples, we discuss them along with the examples so that it can be more intuitive. The first example is the features of printers. Table 3, 4, and 5 are some printer features comparisons we get from underline amazon.com.

| | HP 3830 | HP 4650 | HP 5255 | HP 6968 |
|---|---|---|---|---|
| **Functions** | Print, Scan, Copy, Fax | Print, Scan, Copy, Fax | Print, Scan, Copy, Fax | Print, Scan, Copy, Fax |
| **Print speed** | 8.5 ppm black, 6 ppm color | 9.5 ppm black, 6.8 ppm color | 10 ppm black, 7 ppm color | 18 ppm black, 10 ppm color |
| **Auto 2-sided printing** | No | Yes | Yes | Yes |
| **Auto document feeder** | 35-sheet | 35-sheet | 35-sheet | 35-sheet |
| **Wired/wireless networking** | 802.11b/g/n | 802.11b/g/n | 802.11b/g/n | 802.11b/g/n, Ethernet |
| **Display(inches)** | 2.2 mono touchscreen | 2.2 mono touchscreen | 2.2 mono touchscreen | 2.65 color touchscreen |
| **Dimensions(inches)** | 17.72x14.33x8.54 | 17.53x14.53x7.5 | 17.52x14.45x7.52 | 18.26x15.35x9.0 |
| **Input/output capacity** | 60 sheets, 25 sheets | 100 sheets, 25 sheets | 100 sheets, 25 sheets | 225 sheets, 25 sheets |
| **Cartridge yield** | black 190, color 165 | black 190, color 166 | black 190, color 167 | black 480, color 330 |

*Table 3: Jet-ink printer features*

|  | ET-2750 | ET-3700 | ET-3750 | ET-4750 |
|---|---|---|---|---|
| **Functions** | Print, Scan, Copy | Print, Scan, Copy | Print, Scan, Copy | Print, Scan, Copy, Fax |
| **Print speed** | 10.5 ppm black, 5 ppm color | 15 ppm black, 8 ppm color | 15 ppm black, 8 ppm color | 15 ppm black, 8 ppm color |
| **Auto 2-sided printing** | Yes | Yes | Yes | Yes |
| **Auto document feeder** | No | No | 30 sheets | 30 sheets |
| **Display(inches)** | 1.44 color | 2.4 color | 2.4 color | 2.4 color touch |
| **Input capacity** | 100 sheets | 150 sheets | 150 sheets | 250 sheets |
| **Connectivity technology** | USB, WiFi, WiFi Direct | USB, WiFi, WiFi Direct, Ethernet | USB, WiFi, WiFi Direct, Ethernet | USB, WiFi, WiFi Direct, Ethernet |
| **Cartridge yield** | 6500 black, 5200 color (Bottles) | 14000 black, 11200 color (Bottles) | 14000 black, 11200 color (Bottles) | 14000 black, 11200 color (Bottles) |

*Table 4: colored laser printer features*

|  | DCP-L2550DW | MFC-L2710DW | MFC-L2750DW | MFC-L2710DWXL |
|---|---|---|---|---|
| **Functions** | Print, Scan, Copy | Print, Scan, Copy, Fax | Print, Scan, Copy, Fax | Print, Scan, Copy, Fax |
| **Print speed** | 36 ppm | 32 ppm | 36 ppm | 36 ppm |
| **Auto 2-sided printing** | Print | Print | Print, Scan, Copy, Fax (all two-sided) | Print, Scan, Copy, Fax (all two-sided) |
| **Auto document feeder** | 50 sheets | 50 sheets | 50 sheets | 50 sheets |
| **Display(inches)** | 2-line LCD | 2-line LCD | 2.7 color touchscreen | 2.7 color touchscreen |
| **Input capacity** | 250 sheets | 250 sheets | 250 sheets | 250 sheets |
| **Connectivity technology** | WiFi, Ethernet, Hi-speed USB 2.0 | WiFi, Ethernet, Hi-speed USB 2.1 | WiFi, Ethernet, Hi-speed USB 2.2 | WiFi, Ethernet, Hi-speed USB 2.3 |

| | DCP-L2550DW | MFC-L2710DW | MFC-L2750DW | MFC-L2710DWXL |
|---|---|---|---|---|
| **Cartridge yield** | 1200 | 1200 | 1200 | 1200 |

*Table 5: monochrome laser printer features*

Some product features like the LCD display and the input/output capacity are probably non-essential to most consumer's choice. The print speed is essential to office users but less important to home users. The color print is essential to some but not all of the users. However, product features such as how many papers each cartridge can print and the price of the cartridge are probably important to all users because they directly affect the cost of printing. Most of the printers have the scan and copy function and wireless print function, so we do not need to consider these features. Printers do not give much perceptual feelings to consumers. The look of the printers is usually not considered important either. Therefore, data for intrinsic substitutability is good enough, and we can form the substitutability space only using the product features data.[266]

There is another point worth raising about using the consumer preference to decide the variable features we consider. The k-means clustering model does not have a limit on the number of dimensions. Generally, the more dimensions we use, the more accurate the result we can usually get. However, a too-flat data input might introduce too much noise, which in extreme cases will lead to a bad performance as well. Besides, collecting too much data will cost extra resources.

---

[266] The weight and size might still have significance to some home users. However, we can quantify the size easily. A non-significant difference in size is perceptible as well. Therefore it is still more appropriate to use intrinsic substitutability.

Although data mining and dimension reduction are both important and widely pursued topics in machine learning, the application of those advanced methods cannot be supported by a legal basis in the context of premerger investigation. They can also result in less interpretability. As we have previously mentioned, our model incurs legal consequences, so the requirement on interpretability should be higher than in other fields. Therefore it is more useful and straightforward for us to use consumers' preference to limit the dimensions of our substitutability space. However, we might still want to use a principal component analysis (PCA)[267] or another method to eliminate the correlations between the features because the k-means clustering will have a more accurate result if we assure the independence of each variable used.

Another important aspect of data collection is the form of data. K-means clustering and its variances take numeric input only. There are product features which are naturally numeric and ready to be compared, like the maximum print speed and the number of paper each replacement toner yields, and the printer/toner price. Those product feature data is numeric but on different scales and units. To avoid the result being affected by the scales and unites, we need to normalize the data. On the other hand, some of the features that the consumers care about are binary. Product features such as whether the printer has double-sided scanning and whether the printer can print from a USB storage directly are some of such examples. For these binary variables, we put having the function as 1 and not having the function as 0, then we normalized them, so the binary value is on a reasonable scale. There are other ways to deal with the binary values, but we believe this one serves our purpose better because we can consider the variables in the same dimensions just like what the consumers do.

---

[267]"Principal component analysis (PCA) is a statistical procedure that uses an orthogonal transformation to convert a set of observations of possibly correlated variables into a set of values of linearly uncorrelated variables called principal components. If there are n observations with p variables, then the number of distinct principal components is $\min(n-1,p)$. This transformation is defined in such a way that the first principal component has the largest possible variance (that is, accounts for as much of the variability in the data as possible), and each succeeding component in turn has the highest variance possible under the constraint that it is orthogonal to the preceding components. The resulting vectors are an uncorrelated orthogonal basis set. PCA is sensitive to the relative scaling of the original variables." — Mahmood, Deeman. (2018). Principal Component Analysis (PCA).

In premerger cases, the court will also consider some of the product features to investigate the substitutive relations between the products, as well as to make decisions on submarkets, as we have mentioned earlier in this chapter. The SSNIP model only provides information about the demand. Therefore, the court still has to weigh different features manually based on the specific case fact. Our model is a quantitative way to consider all the product features together. The data processed for the usage in our model can also scientifically filter out the distracting part of the data, and be presented to the court. Therefore, not only our model result but also our processed data can help the court get a better picture of the substitutive relations in the candidate market.

We also use the example of printers to explain why our model can help the court in deciding the substitutional rations of the products. Table 3 is the features of jet printers. Table 4 is the features of color laser printers. Table 5 is the features of the monochrome laser printer. As we can roughly see from the tables, when it comes to the maximum print speed, the color laser printers and the jet printers are similar. However, the pages yielded per cartridge is very different between the color laser printers and the jet printers. The pages yielded per cartridge is closer between the color laser printers and monochrome laser printers, but there is still a significant difference. The prices of those three types of printers are also very different. As we have discussed in Chapter 3, when there are multiple products being sold by one manufacturer, the SSNIP product is hard to define. To do a complete and reliable SSNIP, we need to perform CLA on multiple combinations of products. In each combination, there are different ways of increasing the price by 5%. If we manage to collect the data and run the model for all the situations we need to consider, the result of the relevant market will no longer be abnormally narrow as we have usually observed. However, it will take a lot of effort in both data collection and calculation. Nevertheless, we can simply put all the product features of all the products available in the market and get a market definition. Then analyze the market share and the effect of the merger based on which products are produced by the merging parties. The data collection is especially easy with non-perceptual products because only intrinsic substitutability is used to form the substitutability space.

Another merit of doing so is that we can avoid the arbitrariness in the decision. As we can see, the substitutability relations are complex between different products. If we simply define the market based on commercial types or our impression, we will probably define the jet printer, monochrome laser printer, and color laser printer as different relevant markets. We might miss the fact that some of the cheaper color laser printers may be highly substitutable with a jet printer, and eventually falls into the same cluster with the jet printers. The clustering analysis based on substitutability can avoid this type of error and define the market in a quantitative way where substitutability is the only standard.

| Insurance plans | Relative value | Ordinary visit fee | Deductible | Coinsurance | Out-of-pocket maximum | Rx. | Out-of-network coverage |
|---|---|---|---|---|---|---|---|
| Silver plan 1 | 0.81 | $45 | $2500 | 20% | $7150 | 15/65/50% Rx Ded. applies Tiers 2-3 | None |

| Insurance plans | Relative value | Ordinary visit fee | Deductible | Coinsurance | Out-of-pocket maximum | Rx. | Out-of-network coverage |
|---|---|---|---|---|---|---|---|
| Silver plan 2 | 0.84 | Ded. 10% | $2800 | 10% | $6000 | 15/65/50% Rx Ded. applies to all tiers | None |
| Silver plan 3 | 0.82 | $45 | $3000 | 30% | $7350 | 15/65/50% Rx Ded. applies Tiers 2-3 | None |
| Silver plan 4 | 0.84 | $45 | $2500 | 30% | $7350 | 15/65/50% Rx Ded. applies Tiers 2-3 | None |
| Silver plan 5 | 0.88 | Ded. 10% | $2800 | 10% | $6550 | 15/65/50% Rx Ded. applies to all tiers | None |
| Bronze 1 | 0.61 | Ded. 50% | $5400 | 50% | $6550 | 15/65/50% Rx Ded. applies Tiers 2-3 | None |
| Bronze 2 | 0.69 | Ded. 50% | $3750 | 50% | $7350 | 15/65/50% Rx Ded. applies Tiers 2-3 | None |
| Bronze 3 | 0.75 | Ded. 30% | $4500 | 30% | $7150 | 15/65/50% Rx Ded. applies Tiers 2-3 | None |
| Gold 1 | 0.97 | $30 | $1000 | 10% | $3500 | 15/65/50% Rx Ded. applies Tiers 2-3 | None |
| Gold 2 | 1 | $30 | $1000 | 10% | $6000 | 15/65/50% Rx Ded. applies Tiers 2-3 | None |

*Table 6: Health insurance plan benefits*

As we have mentioned before, the choice of a printer is usually not perceptual. There are other types of products or services like this, for example, health insurance programs. Table 6 is a fragment of a health insurance company's benefit table. Again, we need to figure out what the consumers or employers care about when making health insurance decisions. The most important features of the health insurance programs are listed. We can use the monthly premium, deductible amount, ordinary visit rate, coinsurance rate, out-of-pocket maximum amount and the reimbursement rate of each tire of medication as the numeric variables we use to form our multi-dimensional substitutability space and apply them to our model after being normalized. There are some binary variables like whether the out-of-network service will be reimbursed, whether the reference from the ordinary visit doctor is required, whether service covers out of state visit, etc. Those binary variables can be normalized and be analyzed with the other numeric variables as well. After collecting and processing the data, we can define the relevant product market directly based on product substitutability.

We can use intrinsic substitutability in all the aspects important to the consumers when we deal with non-perceptual products or services. In fact, the vast majority of the top 20 industry groups which file the most HSR premerger investigation cases each year in the past six years are industries producing non-perceptual products or services. We collected the data of the industry groups which have high percentages in the total number of HSR premerger investigation cases filed in Table 7. The data is collected from the *HSR annual report*[268] from 2011 to 2016. We then predicted the expected percentages of the fiscal year 2017 and found the top 20 industry groups which have the highest predicted percentage in Table 8.

---

[268] Federal Trade Commission, Annual Competition Reports, 34th Report FY2011, 35th Report FY 2012, 36th Report FY 2013, 37th Report FY 2014, 38th Report FY 2015, 39th Report FY 2016 (Aug. 18, 2017, 17:30 PM), https://www.ftc.gov/policy/reports/policy-reports/annual-competition-reports, PDF.

| Industry Group | 2011 | 2012 | 2013 | 2014 | 2015 | 2016 |
|---|---|---|---|---|---|---|
| Oil and Extraction | 1.7 | 1.8 | 2.4 | 2.2 | 1.5 | 2 |
| Utility | 3.5 | 2.6 | 2.6 | 2.6 | 2.4 | 3 |
| Food and Kindred Products | 2.2 | 1.9 | 2.3 | 3.2 | 3.1 | 2.2 |
| Chemical Manufacturing | 5.4 | 4.4 | 6.1 | 6.6 | 6.5 | 5.6 |
| Machinery Manufacturing | 2.8 | 3.4 | 2.3 | 2.3 | 2.2 | 2.2 |
| Computer and Electronic Product Manufacturing | 3.5 | 4.3 | 3.8 | 3.3 | 2.5 | 3.7 |
| Transportation Equipment Manufacturing | 2.6 | 3.0 | 2.9 | 2.5 | 2.7 | 1.9 |
| Merchant Wholesales durable goods | 6.9 | 5.3 | 5.3 | 6.0 | 5 | 4.7 |
| Merchant Wholesales nondurable goods | 5.2 | 5.4 | 4.5 | 5.2 | 5 | 6.4 |
| Publishing Industry Except for Internet | 4.2 | 4.7 | 3.7 | 4.4 | 3.9 | 4.7 |
| Telecommunications | 1.9 | 2.0 | 2.3 | 1.7 | 1.7 | 1.5 |
| Internet Service Providers, Web Search Protals, and Data Processing Services | 2.4 | 2.4 | 1.9 | 2.6 | 3.2 | 3.4 |
| Credit Intermediation and Related Activities | 1.7 | 1.6 | 2.8 | 1.5 | 1.5 | 2.5 |
| Securities, Commodity Contracts, and Other Financial Investments and Related Activities | 3.0 | 2.4 | 2.5 | 2.7 | 1.4 | 2.6 |
| Insurance Carriers and Related Activities | 3.6 | 3.1 | 3.2 | 3.1 | 3.7 | 2.5 |
| Funds Trusts and Other Fiancial Vehicles | 0.1 | 0.1 | 0 | 0 | 0.1 | 0.2 |
| Professional,Scientific, and Technical Serivces | 9.0 | 7.9 | 7 | 7.6 | 8.7 | 8.7 |
| Administrative and Support Service | 2.2 | 2.5 | 2.4 | 1.7 | 2.9 | 2.4 |
| Ambulatory Health Care Services | 3.2 | 1.6 | 1.5 | 2.0 | 2.7 | 2.5 |
| Hospitals | 2.0 | 2.1 | 3.8 | 1.7 | 2.2 | 1.8 |

*Table 7: Percentage of HSR case filed (entity standard)*

| Rank | Industry Group | Prediction of FY2017 |
|---|---|---|
| 1 | Chemical Manufacturing | 7.39 |
| 2 | Professional,Scientific, and Technical Serivces | 7.17 |

| Rank | Industry Group | Prediction of FY2017 |
|------|----------------|----------------------|
| 3 | Securities, Commodity Contracts, and Other Financial Investments and Related Activities | 6.74 |
| 4 | Merchant Wholesales nondurable goods | 5.18 |
| 5 | Merchant Wholesales durable goods | 3.74 |
| 6 | Publishing Industry Except for Internet | 3.18 |
| 7 | Insurance Carriers and Related Activities | 3.15 |
| 8 | Computer and Electronic Product Manufacturing | 2.95 |
| 9 | Food and Kindred Products | 2.68 |
| 10 | Administrative and Support Service | 2.51 |
| 11 | Utility | 2.31 |
| 12 | Internet Service Providers, Web Search Protals, and Data Processing Services | 2.21 |
| 13 | Transportation Equipment Manufacturing | 2.14 |
| 14 | Hospitals | 2.04 |
| 15 | Funds Trusts and Other Fiancial Vehicles | 2.04 |
| 16 | Credit Intermediation and Related Activities | 1.9 |
| 17 | Machinery Manufacturing | 1.89 |
| 18 | Telecommunications | 1.77 |
| 19 | Ambulatory Health Care Services | 1.71 |
| 20 | Oil and Extraction | 1.48 |

*Table 8: Predicted top 20 in HSR case filed*

However, there are still industry groups producing products which are considered in a perceptual standard by the consumers. The most typical example will be food, which is among the top 10 industries groups which file the most HSR cases. As we have discussed earlier in this chapter, we should not use intrinsic substitutability on products which are weighed perceptually because some differences in intrinsic substitutability are not significant enough to be sensed by the consumers. For example, in the food industry, the amount of sugar added is not always in direct ratio with how sweet the food is. The sweet taste will be affected by the other ingredients and the kind of sugar used. Moreover, the ingredients do not have a uniformed quality even if they are technically the same. Therefore, it will be more reasonable to use extrinsic substitutability.

In this dissertation, we propose to design surveys for consumers to rate the product's perceptual functions in different features by numbers. For example, in the food and kindred product industry group, the products usually have two purposes -- the functional purpose and the recreational purpose.

We will analyze cereal as an example to show how we form a space of substitutability for perceptual products like food. Consumers of cereal usually have both purposes. There are some features in the nutrient fact which many consumers will consider. They are calories, protein, carbohydrates, fat, and fiber. These nutrition facts are numeric. Therefore, we can use them directly. These functional features can be measured by intrinsic substitutability as we usually do.

It is worth mentioning that we need to consider the unit price as a necessary numeric factor. Products are not substitutable to each other if their price is very different. Therefore, the price should also be a factor in our substitutability space. There are also binary features which might either affect the taste or affect the health aspects, like whether the cereal is deeply processed, whether the cereal has raisins or almond in it, and whether the cereal is organic. These features are also essentially intrinsic substitutability. They should be processed in the same way we do the other binary factors.

However, the recreational function of cereal should be measured by extrinsic substitutability. The data is collected by survey. The survey should be sent to the consumer who has purchased and used the cereal. The survey should contain questions to let the consumers rate the taste they experienced. In the case of cereal, we should typically have the level of crunchiness, creaminess, chewiness, sweetness, bitterness, sourness, and saltiness, diversity in taste. The consumer can rate from, for example, 0 to 10 on each factor. We will not use the data which is more extreme than the first and last 5% in distribution to avoid the effect of extreme samples.

The sample size does not have to be large to be sufficient for our purpose. According to central limit theorem[269] and the law of large numbers[270], the sample size needed is decided by the type one error rate allowed and the estimated standard deviation of the population. The standard deviation of the population is estimated by sample means. When estimating the population mean, we need to make sure the data points in the sample(s) are independent of each other, and the data points are from identical distributions. Let the sample variance be $\sigma^2$. The standard error of the sample means will be Function 15.

---

[269] In probability theory, the central limit theorem (CLT) establishes that, in most situations, when independent random variables are added, their properly normalized sum tends toward a normal distribution (informally a "bell curve") even if the original variables themselves are not normally distributed.

[270] In probability theory, the law of large numbers (LLN) is a theorem that describes the result of performing the same experiment a large number of times. According to the law, the average of the results obtained from a large number of trials should be close to the expected value, and will tend to become closer as more trials are performed.

$$\frac{\sigma}{\sqrt{n}}$$

*Function 15*

The sample means will follow a normal distribution. The mean of the distribution will be the population mean. Therefore, we have Function 16 to calculate the sample size.

$$n = 16\sigma^2/W^2$$

*Function 16*

The most commonly applied confidence interval is 95%. What interval to choose will be decided by the Guidelines or by the judge. Then the sample size can be estimated based on an experiential estimation of sample standard deviation and the inferential interval. Since we are estimating the same population which is all the consumers in the market who is interested in certain kind of product, we can get a good estimation of the population after conducting the first survey for each type of product. Moreover, our estimation will get better as we conduct more and more premerger investigation cases about the type of product.

On the other hand, the cereal manufacturers including the merging company might have already collected the data of the cereal taste for the purpose of marketing. According to the 2010 Merger Guidelines, the merging companies have the responsibility to truthfully provide information required by the agencies. Therefore, we can use their data to find the population standard deviation with regard to the cereal taste.[271]

There might be a concern regarding the cost of the survey. In fact, the merger simulation models in the premerger investigation also need consumer survey to calibrate the consumer behavior parameters. Although we do not know precisely what is the sample size required in the cases, a rule of thumb in research is that 40 data points will usually be good enough. Therefore, our estimation is the survey will not cost more than a couple of thousand dollars. Compared the target amount of the premerger investigation cases, and how much more precise our model is compared to the SSNIP, the cost of the survey is trivial. Additionally, as we can see from Table 8, the top 20 industry groups are mostly producing non-perceptual products. Collecting the data of product features is a lot cheaper than the survey since the merger parties and the other competitors all have the responsibility to provide the data for their product.

According to the common practice of data collection, there are two ways to collect data from costumers. One is to assemble volunteers, who are going to be paid by a proper amount for a long survey. We can provide them with samples of all the products we might put in the relevant market and have them to rate all the aspects of the taste. In this way, the criteria hold by one person will be the same, so the products are rated more fairly. We need to pay attention to the sample randomness of this method because the volunteers might get the information of the survey from a certain path which will affect the randomness of the people surveyed. The people who are willing to spend the time on the survey are also limited to the groups of people who have more free time. The fortunate aspect of this case is the taste of human beings might not be affected by their occupation. Therefore, as long as we can manage to get people of different age and gender groups to do the survey, the data will not be affected.

---

[271] Since we never really know the population standard deviation, it is always estimated by sample standard deviation when it is theoretically required. Therefore, this is not an approximation without basis. Besides this is just the sample size, with correspondence to how many people will be surveyed. This approximation will not affect the result of the model as long as we collect enough data.

Another way to collect perceptual data is through online data collection companies. Those companies usually have websites and programs to track consumers, and sent questionnaires giving out a small amount of money or benefit to incentivize people to participate. In this method, we send the survey of different products to different consumers. Each product has a certain amount of data, and different products' data comes from different people. In this way, the randomness aspect of the survey answers might still create a problem. Nowadays using online behavior to track the purchaser's information is common. The data companies also have their way to make sure that we collect the survey data from people of different age and gender.

We introduced the cereal example to illustrate how to collect extrinsic substitutability data on perceptual products. For the industry group: Other food and kindred products, a similar method can be used. A survey tailored to a specific product can help us get the data on perceptual aspects. Similarly, there are also non-perceptual factors we need to consider if it is something consumers generally believe as being important. For food in general, examples of non-perceptual factors will be the amount of preservatives added and the amount of fibers provided. Since the data collected via surveys is numeric as well, we can simply put the data together as parallel dimensions in the substitutability space. There are also non-perceptual binary factors, for example, whether the food is organic, whether the food is artificial-hormone-free, etc. Again, we can normalize the data and put it together with the other dimensions.

There are other products which involve perceptual aspects. For example, the industry group "Publishing Industry Except for Internet" which is also among the top 10 HSR cases files, involves perceptual products like newspapers and magazines. Similarly, there are also perceptual and non-perceptual features in newspapers and magazines. The non-perceptual features will be the price, the number of topics they cover, the number of articles in each issue, typical length of the articles, the number of days between each issue's release, the percentage of advertisements in each issue, the age of their target readers, the income of their target readers, etc. There are also binary non-perceptual features like the popular topics covered (each topic will be one variable, so the data is binary), whether the content is colored, whether the content is for academic or other professions, etc. The perceptual data will also be collected by surveys, similar to our previous example on cereals. The survey will ask rating for features such as in impartialness, factual report, being interesting to read, being useful to the reader, vision of the articles, depth of the analysis, etc.

It is worth mentioning that intrinsic substitutability is not as absolute as the number shows in some rare cases. For example, some high-quality CPU with a lower frequency might have a better performance than a low-quality CPU with a higher frequency. In this case, since the market definition is not outcome-decisive in court, the parties can submit evidence suggesting a different conclusion. The court may weigh the fact accordingly, or change the data accordingly based on expert opinions. There might also be some non-perceptual factors which are hard to be measured by its actual parameters. In this case, we can take a survey among the experts and get the relevant data.

Additionally, we did not discuss the geographic market in this chapter, while the geographic market is more crucial than the relevant product market in some specific industry groups, like merchant wholesalers, internet service providers, hospitals, and ambulatory health care services. The Agencies have other methods to define the geographic market. Since our purpose is to find a supplementary model to assist the SSNIP and the CLA, we will limit our discussion in the product market. In any case, according to the Merger Guidelines Section 4.2 Geographic Market Definition, the geographic market will be considered as a premise of product market when the Agencies detects the existence of it. Our approaches can be used based on the conclusion of the geographic market. The agencies could apply our model to the products available in each geographic market. The rest will be the same.

In summary, we can collect the data of substitutability with our method, as long as we stay aware of the fact that the substitutability is a combination of product features and the opinion of the consumers. We need to make sure the data collected about the substitutability is an accurate reflection of what the product can offer and what the consumers care about.

# 4.6 EXAMPLES OF APPLYING K-MEANS CLUSTERING

Due to limited funding, we are going to use simulated dataset in our desired format for the first example. For the second example, we will use real data, even though it is not in the exact format that we desire. Nevertheless, we find it close enough to illustrate our purpose in market definition.

The industry group of Merchant Wholesales on durable and non-durable goods are both among the top 5 in the number of HSR cases filed. To give more illustrations of how the data will look like, we simulated our first example as the features of wholesalers.

It is less intuitive to consider collecting data about what consumers are looking for from services, and the industry group Insurance Carriers and Related Activities is among the top 10 in the number of HSR cases filed. Our second example will be about industry group of car insurance. We picked this example also because there are only three major aspects being considered in a car insurance service. The three aspects can all be measured by the premium. The premium is numeric, and there are only three dimensions. Therefore, we can visualize it in a 3D figure, and offer a direct view of how the clustering works.

### 4.6.1 Example 1, Merchant Wholesales durable goods

In this example, we simulated 30 motor vehicle and motor vehicle parts and supplies merchant wholesalers. The choice of the wholesaler is usually non-perceptual. Therefore, we only simulated the features of the wholesalers and used them as the intrinsic substitutability. After simulating the features of the wholesalers, we standardized the data to make sure the dimensions are on the same scale. Since it is simulated data, what we are looking for from the model result is whether k-means clustering can find the pattern we embedded in the simulated data.

When we simulate the data, we imagine that the motor vehicle and motor vehicle parts and supplies merchant wholesalers varies in three different scales: big franchised wholesalers, medium-sized local wholesalers, and small local wholesalers. The patterns in the simulation will be built based on this setting.

The simulation and its code are below. For the sake of typeset, we have knit the instruction, code and the figures with a markdown file. The following are segments of the code. For more details of this simulation and analysis, see the appendix or my post on Rpubs website.[272]

---

[272] Yan Yang, K-means examples, Rpubs (May 11, 2018, 8:30 AM), http://rpubs.com/AnnYang/388225, url.

The first variable we simulate is the number of categories of products available in each wholesaler. We put 3 patterns in the data. In each pattern, we put 10 simulated values. The code is:

```
ctgr <- c(sample(1:50, 10, replace = TRUE),
          sample(100:200, 15, replace = TRUE),
          sample(250:300, 5, replace = TRUE))
ctgrscore <- (ctgr - mean(ctgr))/sd(ctgr)
```

The second variable we simulate is average number of products in each category available in each wholesaler. We put 2 patterns in the data. In each pattern, we put 15 simulated values. The code is:

```
nCtgr <- c(sample(1:10, 15, replace = TRUE),
           sample(20:30, 15, replace = TRUE))
nCtgrscore <- (nCtgr - mean(nCtgr))/sd(nCtgr)
```

The third variable we simulate is the number of brands usually used by customers. We put 2 patterns in the data. In the first pattern, we put 20 simulated values. In the second pattern we put 10 simuated values. The code is:

```
nBrnd <- c(sample(10:30, 20, replace = TRUE),
           sample(1:10, 10, replace = TRUE))
nBrndscore <- (nBrnd - mean(nBrnd))/sd(nBrnd)
```

The forth variable we simulate is the number of models used by customers. We put 2 patterns in the data. In the first pattern, we put 20 simulated values. In the second pattern we put 10 simuated values. The code is:

```
nMdl <- c(sample(50:100, 20, replace = TRUE),
          sample(1:50, 10, replace = TRUE))
nMdlscore <- (nMdl - mean(nMdl))/sd(nMdl)
```

The fifth variable we simulate is the price deviation from average of a product which has an around-average number of sales. We put only 1 pattern in the data. The code is:

```
pPrdct <- sample(1:300, size = 30, replace = TRUE)
pPrdctscore <- (pPrdct - mean(pPrdct))/sd(pPrdct)
```

The sixth variable we simulate is price deviation from average of a category which has an around-average number of sales. We put only 1 pattern in the data. The code is:

```
pCtgr <- sample(1:100, size = 30, replace = TRUE)
pCtgrscore <- (pCtgr - mean(pCtgr))/sd(pCtgr)
```

The seventh variable we simulate is the percentage of non-parts sold. We put 2 patterns in the data. In the first pattern, we put 20 simulated values. In the second pattern, we put 10 simulated values. The code is:

```
pNp <- c(sample(30:70, size = 20, replace = TRUE),
         sample(1:100, size = 10, replace = TRUE))
pNpscore <- (pNp - mean(pNp))/sd(pNp)
```

The eighth variable we simulate is percentage of uncommon transportation products & parts. We put 3 patterns in the data. In each pattern, we put 10 simulated values. The code is:

```
larget <- sample(c(rep(0, times = 5),
                   sample(10:30, size = 5, replace = TRUE)), size = 10)
mediant <- sample(c(rep(0, times = 6),
                    rep(100, times = 2), 20, 10), size = 10)
smallt <- sample(c(rep(0, times = 6), rep(100, times = 4)), size = 10)
pUt <- c(larget, mediant, smallt)
pUtscore <- (pUt - mean(pUt))/sd(pUt)
```

The ninth variable we simulate is percentage of used products. We put 3 patterns in the data. In each pattern, we put 10 simulated values. The code is:

```
largeu <- sample(c(rep(0, times = 8), 3, 5), size = 10)
medianu <- sample(c(rep(0, times = 7), 3, 5, 8), size = 10)
smallu <- sample(c(rep(0, times = 6),
                   sample(1:20, size = 4, replace = TRUE)), size = 10)
pUp <- c(largeu, medianu, smallu)
pUpscore <- (pUp - mean(pUp))/sd(pUp)
```

Now we build our simulated data into a data frame, so it can be later used in clustering analysis.

```
simMV <- cbind(ctgrscore, nCtgrscore, nBrndscore, nMdlscore,
               pPrdctscore, pCtgrscore, pNpscore, pUtscore, pUpscore)
set.seed(Sys.time())
```

After we get the data frame,  we will conduct k-means clustering in different methods. As we discussed in the earlier section, there are several methods to find the k. There are different techniques to avoid the unwanted effects caused by the random start points as well. We are going to try all of them on our simulated data. Since we do not know how many clusters are there in real life, we will apply the clustering model as if we do not know how many patterns are hidden in our simulated data, even though we know how the data was generated.

The first method we use to find the k is the two-step method. After using the two steps to find the possible k, we are going to conduct the k-means clustering for one hundred times with each possible k we find in the first step. During the one hundred times of running the model, we will use different sets of random numbers provided by R to make sure the start points of K-means clustering are all different each time. In this way, we can make sure that we have tried different start points to cover our multi-dimensional space.

After the two-step analysis, we are going to use the Elbow Method and the Gap Statistic Method to find the best k. If they come up with results different from the k(s) concluded by the hierarchical clustering, we will also run the model with that k with the same one-hundred sets we used earlier, so we can avoid having the comparison be affected by the randomness of the start points. We will not use k-means++ in this stage, so we can make sure the comparison is not affected by the better choice of start point provided by k-mean++.

Now, we perform the clustering and see what happens. First, we need to do the first step of the two-steps analysis — hierarchical clustering to find out the k.

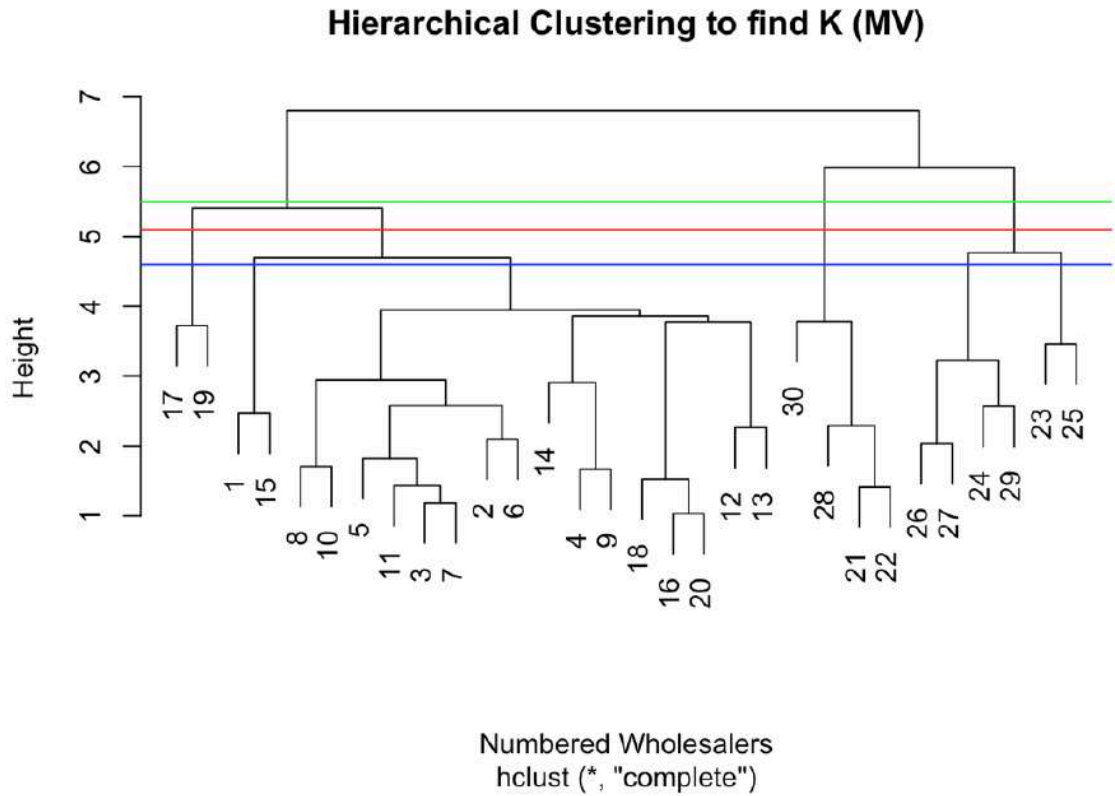

*Figure 5*

From Figure 5, we can see the hierarchical clustering result is not very good. The difference between the three, four and five groups are not very significant. We have discussed in previous sections that the consumer products and services are usually clustered. Therefore, in real life, the hierarchical clustering will probably produce a better result. However, for natural materials, the features are not necessarily clustered. In this case, it is common to get a bad result from hierarchical clustering as well. In a typical desired result of hierarchical clustering, there will be some clusters which are relatively more significant than others. That number of cluster is the k we are going for. In some case, there might be one level which is significantly wider than others.

If so, we should pick that level, and use the number of clusters as our k.

Since there is not a significant difference between having three, four and five groups, we will set our k as 3, 4 and 5. We are going to conduct k-means with k equals to 3 from 100 sets of start points. Then we find the result with the smallest within-group sum of squares and plot the result of the clustering with the smallest within-group sum of squares. The plot is shown in Figure 6.

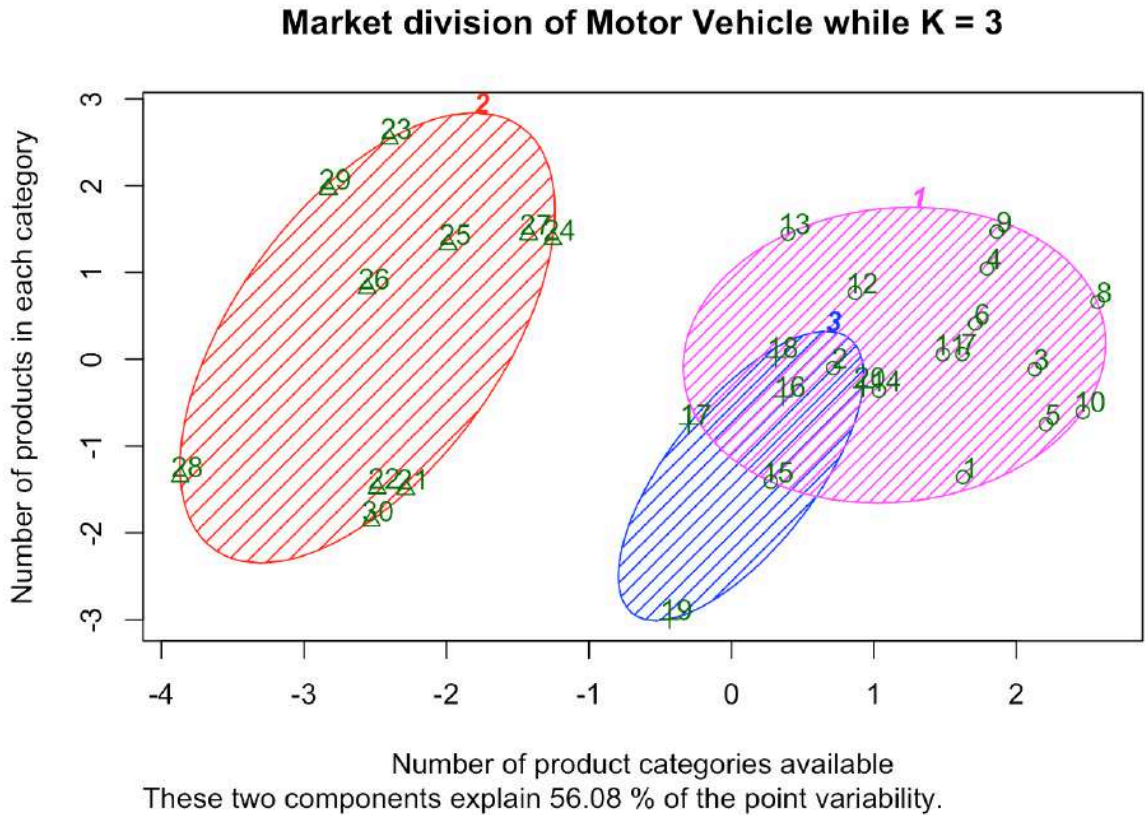

*Figure 6*

In Figure 6, since we are looking at the clustering results of nine-dimensional objects from a two-dimensional picture, there are overlaps of the clusters in Figure 6. In fact, the 30 wholesalers are divided into three groups without overlap. We can see this from the other relevant aspects of the result below. From the tags given to each data points, we can see that if we divide the wholesalers into three groups, it is roughly close to the sequence we intentionally put in the simulation (10 points for each type of data), except that the second group has largely shrunk and many the points go to the first group.

## Other relevant aspects of the result

```
## label of clusters
kmMV3$cluster
```

```
##  [1] 1 1 1 1 1 1 1 1 1 1 1 1 1 1 1 1 3 3 3 3 2 2 2 2 2 2 2 2 2 2
```

```
## cordinates of centers
kmMV3$centers
```

```
##     ctgrscore nCtgrscore nBrndscore  nMdlscore pPrdctscore pCtgrscore
## 1 -0.7217317 -0.9487876  0.4389097  0.6538996 -0.10778829 -0.2323299
## 2  0.8701004  0.8935854 -1.1928434 -1.2187335 -0.07775567  0.6471109
## 3  0.4249945  1.0591919  1.0689576  0.4757683  0.47887620 -0.5972322
##      pNpscore    pUtscore    pUpscore
## 1  0.3242067 -0.3780789 -0.27590332
## 2 -0.6220715  0.3780789  0.38459250
## 3  0.2715231  0.3780789  0.05852495
```

```
## total squared distance from data points to their centers
## it is equal to sum of squares because the center is the mean
## in each dimensions
kmMV3$withinss
```

```
## [1] 56.57548 73.02529 21.88084
```

```
## total within-cluster sum of distances
kmMV3$tot.withinss
```

```
## [1] 151.4816
```

We did not change the sequence of the intended level in generating the data points. The first ten wholesalers are big franchised wholesalers, the second ten wholesalers are medium-sized local wholesalers, and the third ten wholesalers are small local wholesalers. However we also added two variables in which the first twenty data points belong to the same distribution and the last ten data points belong to another. These two variables interfered with the first three patterns while not destroying it. The result of our k-means clustering reflected the interference. Therefore, some data points were grouped into group 3 while it was intended to be in group 2.

This is a good reflection of real-life conditions. Since we used a more random way to generate the data compared to generating certain-shaped data and add errors to it, we have less control over the actual pattern. This is similar to what happens in real life. We intuitively divide the wholesalers based on the scale of them while not knowing that there are more complicated patterns underneath the substitution. There are other imbedded factors which might affect the reality of the substitutive relationship in a stronger way. The computation helped us to find the truth of the market by considering a broader range of things simultaneously and doing it quantitatively. As we have mentioned earlier in this chapter, the court will manually consider the substitutive relations between the products. Some of the underneath substitutive relations are hard to see. Just like us, the judges may not be able to find the pattern synthesized under the surface. Therefore, our model can help the court in making more informed decisions on the substitutive relations.

On the contrary, we can also see that the result of applying hierarchical clustering alone was totally fooled by the 2 interfering variables. When dividing by 3 groups, hierarchical clustering put the first twenty data points into the first group altogether, which means it failed to recognize the pattern of grouping we are actually intending. Therefore, we believe k-means clustering is more reliable to define relevant product market.

Now We look at the clustering when k is equal to 4. Again, we conduct k-means with k equals to 4 from 100 different start points for 100 times and plot the result with the smallest within-group sum of squares. The figure of the clustering is below.

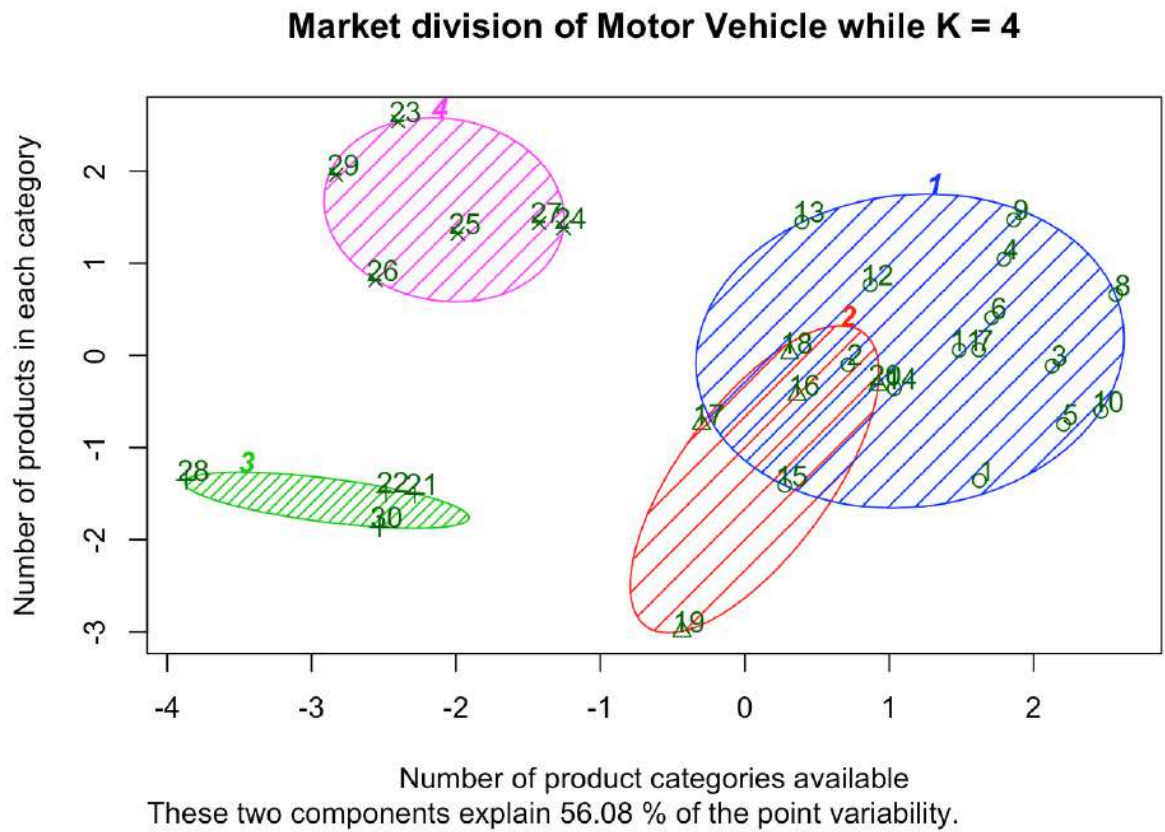

*Figure 7*

The result of grouping and other relevant aspects of the results are as follows:

```
## label of clusters
kmMV4$cluster
```

```
## [1] 1 1 1 1 1 1 1 1 1 1 1 1 1 1 1 1 2 2 2 2 2 3 3 4 4 4 4 3 4 3
```

```
## cordinates of centers
kmMV4$centers
```

```
##     ctgrscore nCtgrscore nBrndscore  nMdlscore pPrdctscore pCtgrscore
## 1 -0.7217317 -0.9487876  0.4389097  0.6538996  -0.1077883 -0.2323299
## 2  0.4249945  1.0591919  1.0689576  0.4757683   0.4788762 -0.5972322
## 3  0.9389722  1.0747176 -1.2406279 -0.9498455   0.8065608  1.0074191
## 4  0.8241858  0.7728306 -1.1609870 -1.3979922  -0.6673000  0.4069054
##      pNpscore   pUtscore    pUpscore
## 1  0.3242067 -0.3780789 -0.27590332
## 2  0.2715231  0.3780789  0.05852495
## 3 -0.8165955  1.9039576 -0.49328169
## 4 -0.4923888 -0.6391737  0.96984197
```

```
## total squared distance from data points to their centers
## it is equal to sum of squares because the center is the mean
## in each dimensions
kmMV4$withinss
```

```
## [1] 56.57548 21.88084 12.33399 32.95276
```

```
## total within-cluster sum of distances
kmMV4$tot.withinss
```

```
## [1] 123.7431
```

As we can see from the clustering, the first one third still roughly belongs to the first group. The rest are incorrectly clustered. It is due to the fact that there were only three real patterns, so the fourth group comes from the two interfering variables.

We can see that the two groups on the right of Figure 6 and Figure 7 are the same regardless of the difference in k. By adding one more group to divide, the algorithm only divided a less desirable group without destroying the integrity of a relatively accurate clustering.  It means the clustering of the two groups on the right is reliable.

As we can see from the simulation, the first two variables are simulated in the three patterns we are intending. Therefore it makes sense that they explain more variations (56.08%) within the grouping result. We can also see from Figure 7 that the green group was relatively slim when looking from the first two dimensions. That is because the first two dimensions do not have much influence in differentiating the fourth group from the other three. It is reasonable because the fourth group is "created" from the interference in other variables, so the difference cannot be reflected from the first two dimensions.

Now we conduct k-means with k equals to 5 from 100 different starting points for 100 times, and plot the result with the smallest within-group sum of squares. The plot is below.

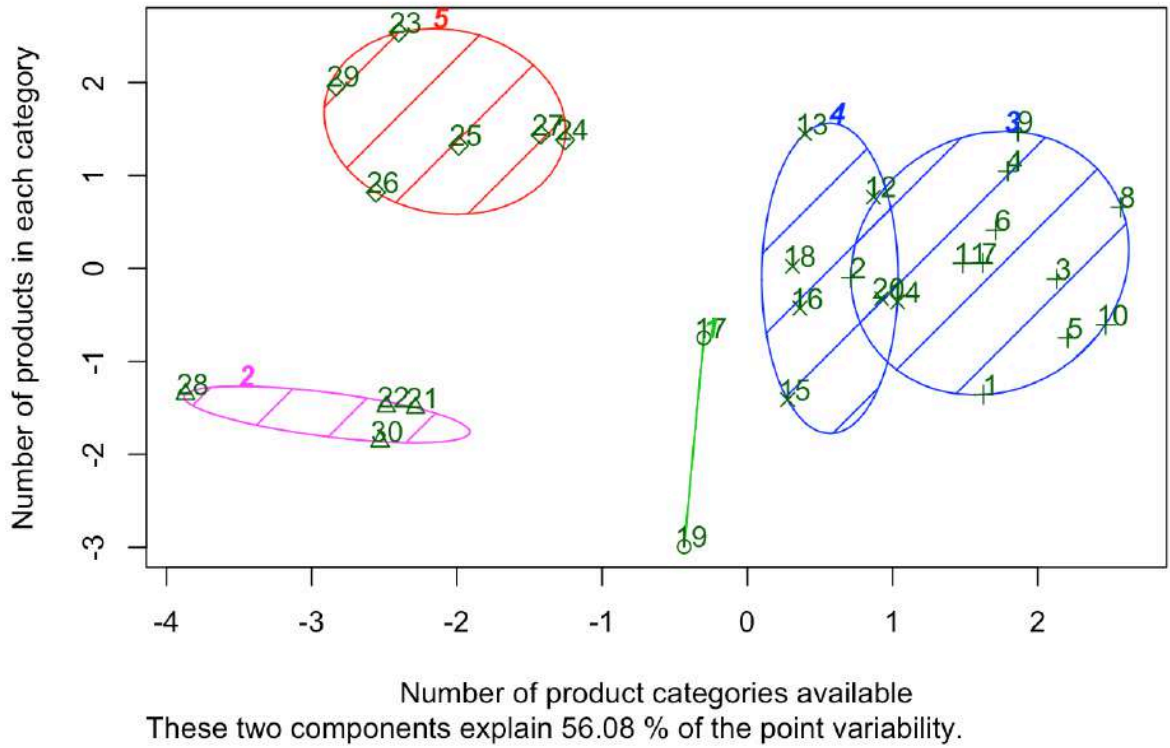

*Figure 8*

The other relevant aspects of the clustering result when k = 5 is as follows.

```
## label of clusters
kmMV5$cluster
```

```
##  [1] 3 3 3 3 3 3 3 3 3 3 3 4 4 4 4 4 1 4 1 4 2 2 5 5 5 5 5 2 5 2
```

```
## cordinates of centers
kmMV5$centers
```

```
##      ctgrscore   nCtgrscore nBrndscore   nMdlscore pPrdctscore pCtgrscore
## 1  0.2815589  0.893585379  1.1751454  0.5806177  0.98860783 -0.5184763
## 2  0.9389722  1.074717551 -1.2406279 -0.9498455  0.80656082  1.0074191
## 3 -1.0669018 -0.969488385  0.4318305  0.6590242 -0.03186524 -0.6366101
## 4  0.3531141 -0.008378903  0.6897153  0.5395477 -0.12130590  0.2240793
## 5  0.8241858  0.772830598 -1.1609870 -1.3979922 -0.66730000  0.4069054
##       pNpscore    pUtscore   pUpscore
## 1 -0.5126518  1.9039576  0.8862349
## 2 -0.8165955  1.9039576 -0.4932817
## 3  0.4406263 -0.3524934 -0.3108663
## 4  0.3427327 -0.5301823 -0.3141237
## 5 -0.4923888 -0.6391737  0.9698420
```

```
## total squared distance from data points to their centers
## it is equal to sum of squares because the center is the mean
## in each dimensions
kmMV5$withinss
```

```
## [1]  6.943531 12.333986 28.328260 31.108963 32.952755
```

```
## total within-cluster sum of distances
kmMV5$tot.withinss
```

```
## [1] 111.6675
```

From Figure 8, we can see that the two groups on the left are the same as the two on the left in Figure 7. However, when the model had to divide the clusters further, it divided the larger groups on the right. Again, we can indicate from Figure 8 that the two clusters on the left are relatively stable. Which means the solution of minimizing the in-group distance is probably a universal optimum. On the other hand, Figure 8 also suggests a similar conclusion as Figure 7 does that the first two dimensions do not provide much information about the green group[273].

The result also indicates that the five clusters are probably too many because the clustering is no longer reflecting the patterns embedded in the simulated data frame. Therefore, it is crucial to find the most appropriate k. The result also tells us that the two-step clustering method is not desirable in the context of premerger investigation, even though it is a useful method in market segmentation. The market segmentation helps the manufacturers and other service providers to find the target consumer and design their products accordingly. It does not have to be as accurate as the market definition. In contrast, the market definition has legal power. It impacts the result of the premerger investigation and the fate of a merger.

In the following content, we are going to use the Elbow Method to find the k. We wrote a function according to the Equation 4 and Equation 5 of Chapter 4 to calculate Wk in the elbow method with the input of a dataset, and the largest meaningful k. The code can be found in the appendix. The output of the elbow function is the Wk we want to calculate. The k-means clustering runs within the elbow function with each k with random start points. Therefore, it will be better if we run the model multiple times. Since our data size is small, it will not take long. We calculate Wk from 100 different start points when k changes from 1 to 20 to avoid the effect of randomness.

---

[273] The green group is label as group 1 this time.

Then we plot it with k as the x-axis and the mean of the 100 Wk calculated as the y-axis. Our code is published on Rpubs to make sure the other researchers can check its reproducibility[274]. The code can also be found in the appendix of this dissertation.

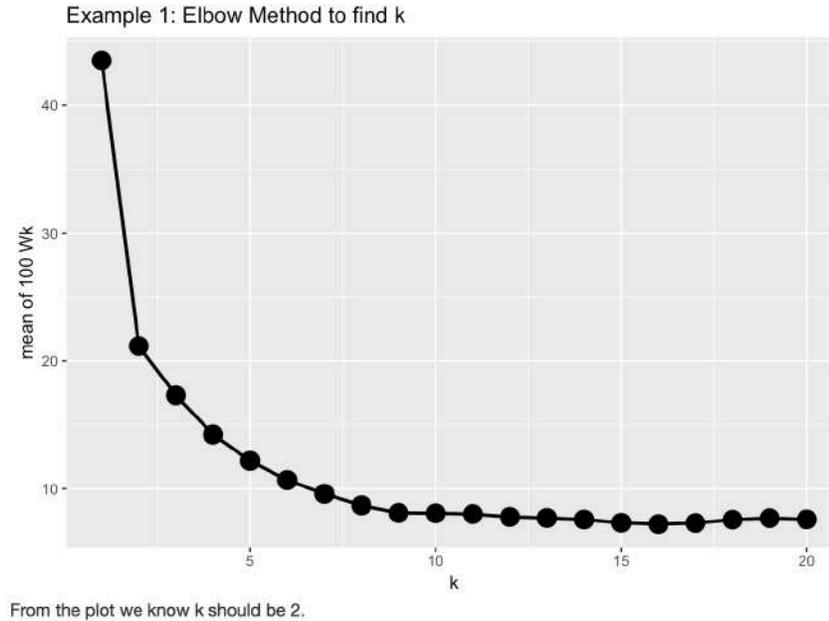

*Figure 9*

We know from the plot that 3 and 4 are desirable k because the change of Wk is still larger than the ones after the 4th k. However, starting from the 3rd point, the slope is already relatively flat compared to the decrease after the data was divided into two groups. Therefore, according to the elbow method, 2, 3, and 4 are all relatively good k, but the best k suggested by the Elbow Method is 2.

---

[274] Yan Yang, K-means examples, Rpubs (May 11, 2018, 8:30 AM), http://rpubs.com/AnnYang/388225, url.

Then we are going to use the Gap Statistic to determine the best k. Since R already has a package to calculate Gap Statistic, we will use the function from package cluster directly. The code is in the Appendix and the Rpubs. We set the maximum k as 20 and the result of the Gap Statistic suggested the best k is 2, which is the same as the result of Elbow Method.

The Elbow Method tries to make sure we are not over-dividing the dataset. The Gap Statistic tries to leave an as-large-as-possible gap between each cluster. Both of the models tell that the best k we should use is 2. As we have mentioned earlier, the Elbow Method is based on the distance between products, and the Gap Statistic is based on the space (gap) between each product. They both fit our purpose of finding the close substitution of the target product without wrongfully dividing less substitutable products inside of the market. Therefore, we should trust the result of these two methods, even more so because they both produce the same result.

Therefore, we are going to conduct the K-means clustering with repeating it for 100 times with different initial points again, using K = 2 as the input of clusters, and plot the result with the smallest within-group sum of squares. The plot of the result is shown in Figure 10.

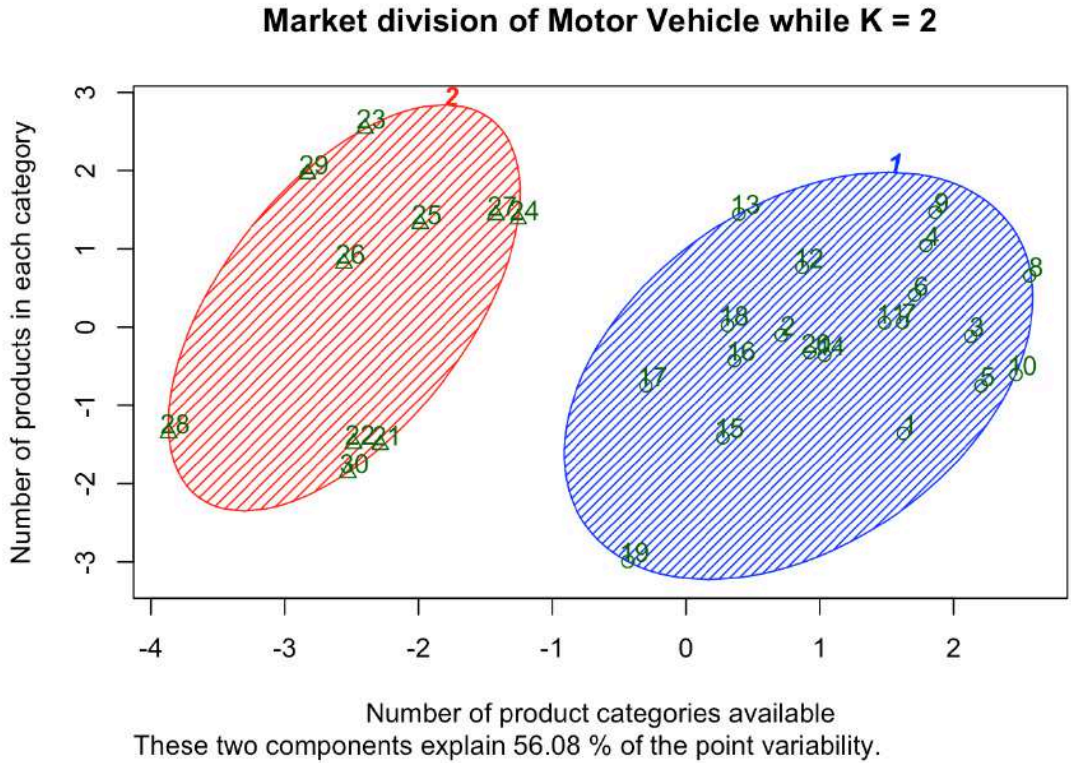

*Figure 10*

The other relevant aspects of the result are as follows.

```
## label of clusters
kmMV2$cluster
```

```
##  [1] 1 1 1 1 1 1 1 1 1 1 1 1 1 1 1 1 1 1 1 1 1 2 2 2 2 2 2 2 2 2 2
```

```
## cordinates of centers
kmMV2$centers
```

```
##    ctgrscore nCtgrscore nBrndscore  nMdlscore pPrdctscore pCtgrscore
## 1 -0.4350502 -0.4467927  0.5964217  0.6093667  0.03887784 -0.3235554
## 2  0.8701004  0.8935854 -1.1928434 -1.2187335 -0.07775567  0.6471109
##      pNpscore   pUtscore   pUpscore
## 1  0.3110358 -0.1890394 -0.1922963
## 2 -0.6220715  0.3780789  0.3845925
```

```
## total squared distance from data points to their centers
## it is equal to sum of squares because the center is the mean
## in each dimensions
kmMV2$withinss
```

```
## [1] 104.47897  73.02529
```

```
## total within-cluster sum of distances
kmMV2$tot.withinss
```

```
## [1] 177.5043
```

We noticed from the labels that this method detected the fact that the last ten elements are by all means following the same pattern. The result of the clustering is the most stable if we divide the data into two groups. Therefore, we should consider using the best k suggested by the two methods which both comply with our context of market definition. If the two methods do not produce the same result, the Gap-statistic is more reliable, because it does not involve subjective opinion.

On the other hand, we should be aware that if the market requires a higher level of substitutability, for example, the chemical material market, we should pick among relatively large k and choose a relatively large k using the Elbow Method and the Gap Statistic. In this case, according to the Elbow Method, either 4 or 5 can work. According to the Gap Statistic, 4 should work. Therefore, we should pick 4 as our k.

## 4.6.2 Example 2, Insurance Carriers and Related Activities

The second example uses the data from Auto Insurance Database Annual Report 2013-2014. We know from the report that there are three major concerned aspects of car insurance, Liability Premium, Collision Premium, and Comprehensive Premium. "Liability insurance is a part of the general insurance system of risk financing to protect the purchaser from the risks of liabilities imposed by lawsuits and similar claims. It protects the insured in the event he or she is sued for claims that come within the coverage of the insurance policy."[275] Collision insurance is the money that the insured can get when his vehicle is involved in a crash with another vehicle or rammed into a fixed structure, he can rely on collision insurance to offer you coverage. "Comprehensive car insurance covers damages from an "act of God," or events that are not caused by a car driving into something else. An "act of God" can include things like damage from a heavy tree branch falling on your car. Since you have no control over when or why a tree branch would fall on your car, this kind of accident would be covered under your comprehensive policy."[276] Premiums were defines by Function 7.

*Average Premium = Written Premiums/Written Exposures*

*Function 7*

---

[275] National Association of Insurance Commissioners, Auto Insurance Database Annual Report 2013/2014, (Dec. 20, 2017, 10:50 PM) https://www.google.com/url?sa=t&rct=j&q=&esrc=s&source=web&cd=1&cad=rja&uact=8&ved=0ahUKEwj8gP-J9f3aAhUL4YMKHa32BgAQFggxMAA&url=http%3A%2F%2Fwww.naic.org%2Fprod_serv%2FAUT-PB-13_2016.pdf&usg=AOvVaw12fsK6pdOcy0A1qW-0Wtwm, PDF.

[276] id.

We averaged the premiums in five years from 2010 to 2014. In the perfect scenario, the data should be the three types of premiums of different companies. We cannot find this data. However, our data is sufficient for the purpose of illustrating the clustering model. Our data is about the average of the three types of premiums in different states. We can treat states as "insurance providers" so that we can define the relevant market with k-means clustering among states.

We averaged the premiums between years so that we can get an overall tendency among states. Here we also standardized the data. It can help to avoid the effect of the difference in the mean among different variables, which can cause unbalanced effect between the features on the grouping. We will do the two-step approach first. After processing the data, we do perform a hierarchical clustering to find k.

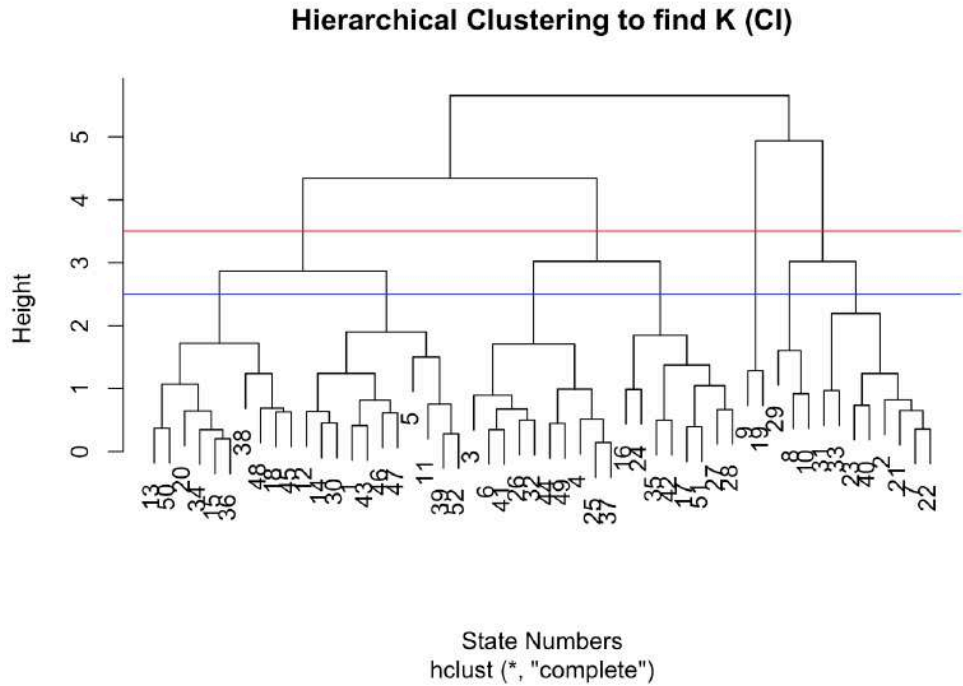

*Figure 11*

As we can see in Figure 11, the hierarchical clustering result is a lot better than the one we get from simulated data, because the real data of substitutability is usually more clustered. From the result of hierarchical clustering, we know that we should either pick 4 or 7 as our k in the next step of the analysis.

As a result, we will first conduct k-means clustering with k equal to 4 from 100 different initial points for 100 times, and plot the result with the smallest within-group sum of squares. As we have mentioned earlier, we should regard the result as the market definition among 52 insurance providers, not 52 states. The code and plot are below.

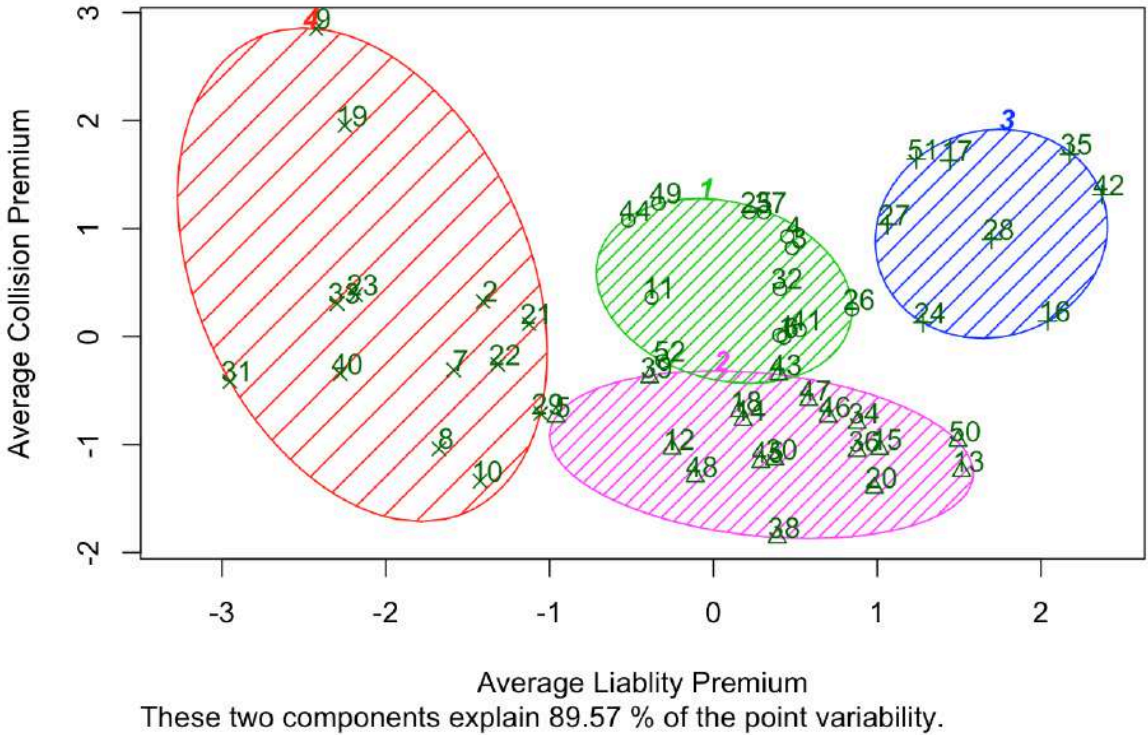

*Figure 12*

Other relevant aspects of the result are included below.

```
## label of clusters
kmCI4$cluster
```

```
##  [1] 1 4 1 1 2 1 4 4 4 4 1 2 2 2 2 3 3 2 4 2 4 4 4 3 1 1 3 3 4 2 4 1 4 2 3
## [36] 2 1 2 2 4 1 3 2 1 2 2 2 2 1 2 3 1
```

```
## cordinates of centers
kmCI4$centers
```

```
##           lap         cap        cpap
## 1 -0.2162251   0.0671289   0.5416765
## 2 -0.4250923  -0.4167178  -0.9077268
## 3 -1.0233950  -1.0358827   1.3526659
## 4  1.4345960   1.1473312  -0.1172338
```

```
## total squared distance from data points to their centers
## it is equal to sum of squares because the center is the mean
## in each dimensions
kmCI4$withinss
```

```
## [1]  7.330818 13.474020  4.984122 25.873355
```

```
## total within-cluster sum of distances
kmCI4$tot.withinss
```

```
## [1] 51.66232
```

We are dealing with real data, so we do not know much about the true pattern beneath it. Therefore, we cannot discuss the result from the perspective of generating data. We are going to conduct the analysis with different methods and compare the results and find the most reliable conclusion of market definition. Again, the court can use legal reasoning and other facts in the case to make a final decision, if there are multiple results.

Now We conduct k-means with k equals to 7 from 100 different initial points for 100 times, and plot the result with the smallest within-group sum of squares. The plot is shown in Figure 13.

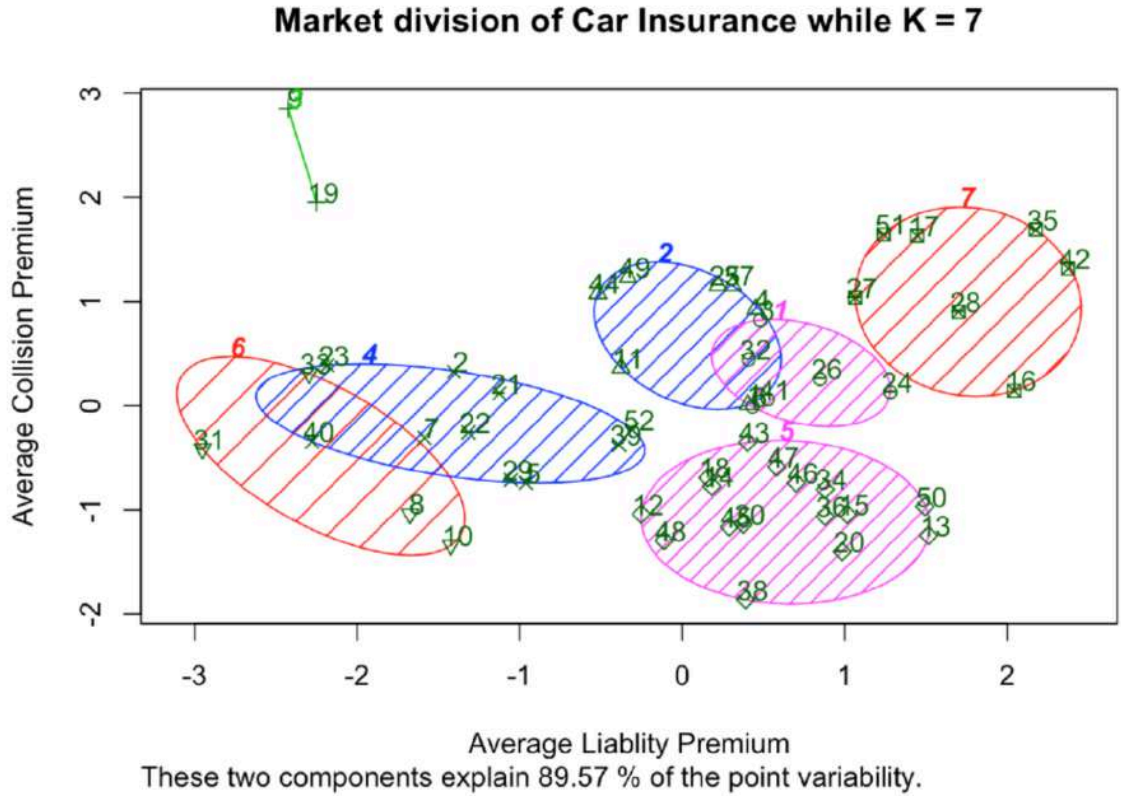

*Figure 13*

The relevant aspects of the result is below.

```
## label of clusters
kmCI7$cluster
```

```
##  [1] 2 4 1 2 4 1 4 6 3 6 2 5 5 5 5 7 7 5 3 5 4 4 4 1 2 1 7 7 4 5 6 1 6 5 7
## [36] 5 2 5 4 4 1 7 5 2 5 5 5 5 2 5 7 4
```

```
## cordinates of centers
kmCI7$centers
```

```
##         lap         cap        cpap
## 1 -0.2043081 -0.6370889  0.4854048
## 2 -0.3011256  0.4464633  0.7098097
## 3  1.3140978  2.4377592  1.8389374
## 4  0.7083975  0.9803677 -0.4775267
## 5 -0.4727496 -0.5727570 -0.9136164
## 6  2.2576671  0.5235082 -0.6085935
## 7 -1.1207312 -0.9874039  1.4669332
```

```
## total squared distance from data points to their centers
## it is equal to sum of squares because the center is the mean
## in each dimensions
kmCI7$withinss
```

```
## [1] 1.5108276 2.8056016 0.8263802 7.0045401 8.8901085 3.8771411 3.5907569
```

```
## total within-cluster sum of distances
kmCI7$tot.withinss
```

```
## [1] 28.50536
```

Now we conduct the Elbow Method to find the K. We used the function written earlier to calculate Wk and the plot we generated to find the "elbow" is below.

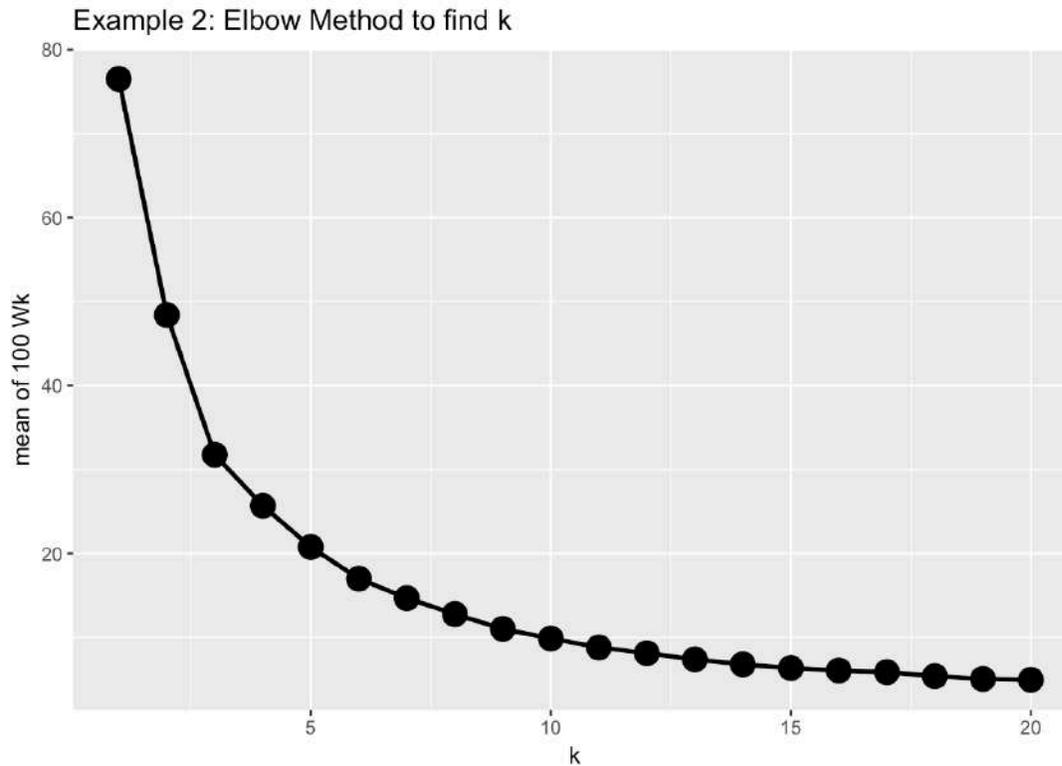

*Figure 14*

We find from the plot that the Wk is still significantly decreasing after k = 3. Therefore we should pick 3 as our k. Then we use Gap Statistic to find the K. Again, we will use the function provided by package "cluster". The result it provided is also 3. Again the code and the Gap-statics calculated is in the appendix and Rpubs.

We learned from the result that the best k is 3. We are going to repeat k-means clustering 100 times with different initial points and plot the result with the smallest within-group sum of squares while k is set as 3. The code and plot are shown below.

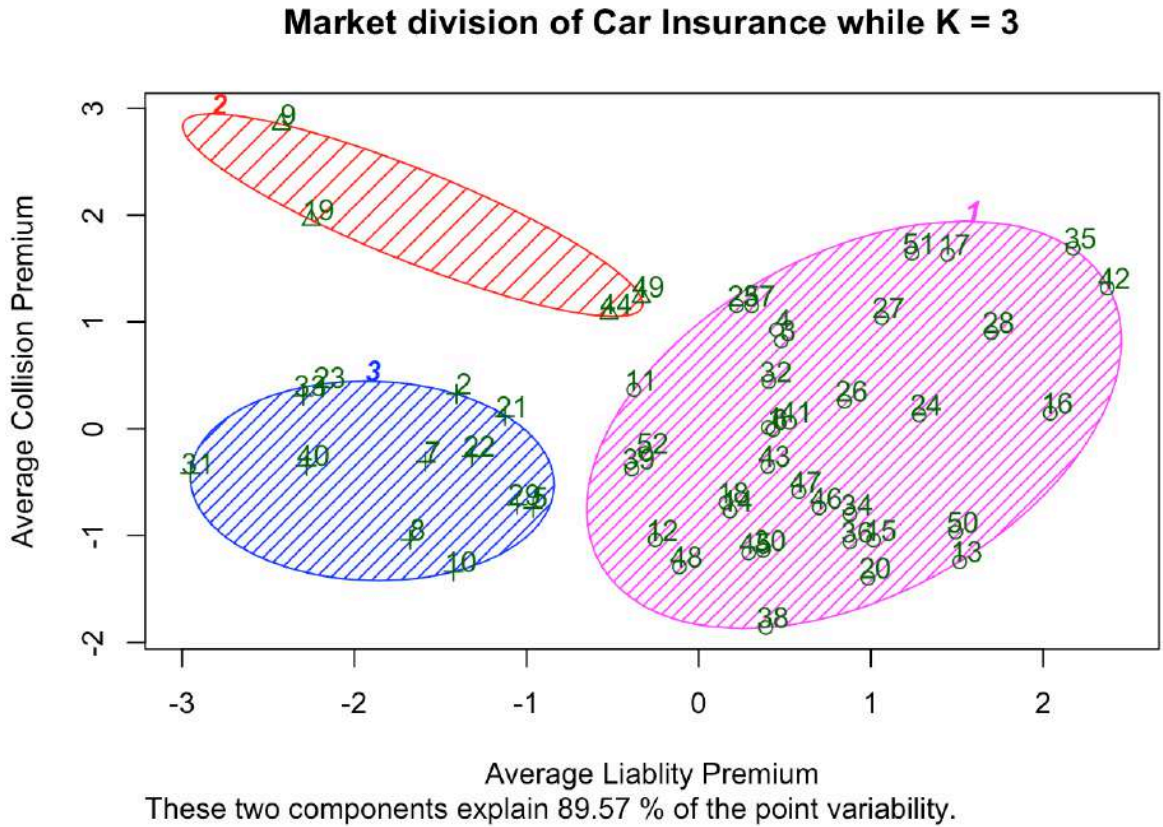

*Figure 15*

Other relevant aspects of the result are shown below.

```
## label of clusters
kmCI3$cluster
```

```
##  [1] 1 3 1 1 3 1 3 3 2 3 1 1 1 1 1 1 1 1 2 1 3 3 3 1 1 1 1 1 3 1 3 1 3 1 1
## [36] 1 1 1 1 3 1 1 1 2 1 1 1 1 2 1 1 1
```

```
## cordinates of centers
kmCI3$centers
```

```
##           lap         cap        cpap
## 1 -0.5193289 -0.4909840  0.02171412
## 2  0.7130070  1.5820433  1.40981372
## 3  1.3203177  0.9456041 -0.53508027
```

```
## total squared distance from data points to their centers
## it is equal to sum of squares because the center is the mean
## in each dimensions
kmCI3$withinss
```

```
## [1] 58.344648  6.037484 15.133394
```

```
## total within-cluster sum of distances
kmCI3$tot.withinss
```

```
## [1] 79.51553
```

When we compare the results, we find that clusters do not have a lot in common. However, since the first tow dimension explains 89.57% percent fo the point variance, we can look at the plot direct to get some sense of the actual distance between the points. The result with k = 3 has more significant gaps between each cluster. However with k = 4, we have smaller distance between data points within groups. As a result, I believe k = 3 and k = 4 are both relatively reasonable. However, since k = 3 is suggested by the methods which fit our context better, we should set k equals to 3. Again, in a legal context, we can decide between the result with legal reasonings based on specific situations.

Here, the insurance providers are repetitive service providers. They have more reason to bind unnecessary terms when there is less competition. In the context of premerger investigation, we know that all of the cases we get in HSR premerger investigation are reaching a threshold that indicates the danger of less competitive market. Therefore, we should use a more strict standard. In the context of premerger investigation, a more strict standard means a smaller relevant market. Therefore, we should pick a bigger k.

One may wonder if we are still not sure about which result to take, then how could we recommend our model to the practice. It is important to remember we are not sure if the result of a Hypothetical Monopolist Test is right either. In fact, we can even say we are sure it is wrong, based on how many different SSNIPs we should try to actually perform the Hypothetical Monopolist Test as intended. Not even mentioning all of the unrealistic assumptions and the other problems we have mentioned in Chapter 3, compared to the Hypothetical Monopolist Test, our approach at least has many ways to verify and re-examine the result, not leaving the Agencies with no choice. Our method can also ensure that we come up with the most scientific result based on the need of different market facts.

## 3.3 Conclusions based on the examples

We can learn from the two examples that:

1.  K-means clustering can find the pattern underneath the data and group the data points accordingly, in a more accurate way than the Hypothetical Monopolist Test with CLA.

2.  From the application of the data to k-means clustering we verified the fact that we do not need to make assumptions to perform this model. Everything is based on our purpose according to the facts of the market.

3.  The elbow method and the gap statistic are more reliable methods to find the best k, because they are consistent to our purpose as well.

4.  Hierarchical clustering might work for market segmentation, but it does not work well in the context of premerger investigation.

5.  When the k-means clustering concludes multiple results that we cannot choose via a technical way, we can still use legal and factual reasoning based on our context to pick a result.

6.  Since the Merger Guidelines permits the usage of models other than hypothetical monopolist test, we may at least try to use k-means clustering as a supplement in the legal proceedings to define the relevant product market along with the Hypothetical Monopolist Test.

# CHAPTER 5: CONCLUSION

In the first part of Chapter 1, we introduced the background of the premerger investigation. Coordinated effects and unilateral effects are both considered when analyzing the possible effects of a merger. We then illustrated the framework of the premerger investigation and the major aspects of unilateral effects and the coordinated effects.

The methods used to assess the unilateral effects have been developing in the past 20 years. In regarding the unilateral effects, we summarized the upward pricing pressure test (UPP) and the merger simulation models. In contrast, the model used to assess the coordinated effects have not evolved for decades. The 1982 Merger Guidelines stipulated the Hypothetical Monopolist Test (SSNIP test) as the test for the range of products belonging to the relevant product market. The SSNIP test was not well implemented until the Critical Loss Analysis (CLA) was established in 1989 to solve the problem of how to determine when the SSNIP (price) is profitable.

The critical loss analysis was adopted by the 2010 Merger Guidelines and officially becomes a part of the premerger analysis when necessary data is available. However, the current way to perform the SSNIP test and the CLA by the Agencies has several problems, which also caused the criticism on CLA for many years. The essential error in the application of the models is that the contribution margin (CM) used in the CLA model is not a match with the price increase in percentage (Y), due to the fact that if the hypothetical monopolist does exist, its profit margin should be significantly wider than the merging parties's. An abnormally small CM induces an abnormally big critical loss(CL), which will result in a narrower market. Another major problem is the definition of SSNIP. It is unreasonable to define the SSNIP as a price increase of all the products in the candidate market, and in this case, the real CM of the hypothetical monopolist is hard to estimate as well.

In Chapter 2, we clarified some problematic criticisms against the CLA and suggested better ways to define the SSNIP and conduct the models based on a progressive analysis of the SSNIP test and the CLA. We further introduced the frequently used demand models and analyzed the strength and limit of them. Nevertheless, there are still some unresolvable problems in the SSNIP test and CLA. We addressed those problems at the end of Chapter 2 and proposed another model in Chapter 3.

In Chapter 3, we first discussed the course and progress of the SSNIP test and the CLA's development, and found the essential purpose of the SSNIP test -- quantifying the substitutability of the products. The SSNIP cleverly used the change in demand to reflect the substitutive relations, but this way has unavoidable problems as well. We then introduced a new model based on the substitutive relations directly and ensured the level of substitutability within each group to be as small as possible. We also discussed the data collection for our model and used simulated and real data to perform the models. The results are reasonable under our context and flexible based on the specific needs of the product type.

Based on Chapter 2's elaboration of the practical problems of the SSNIP test and the CLA and Chapter 3's expounding on the limited function of a one-dimensional substitutability level on market definition, we conclude that the SSNIP test is not easy to apply or as accurate as it seems to be. The conclusion it intends to reach is that the hypothetical monopolist can profit from the SSNIP, but the implementation of the model cannot properly get to this conclusion. On the other hand, the conclusion itself is not meaningful since the merging parties may not profit from the SSNIP even if the analysis shows that the hypothetical monopolist can profit from the SSNIP.

Our method can make sure every product in the same group is a close substitute of each other. It compares all the aspects of product features at the same time. It clusters the products in a way that the substitutability inside each group is a local optimum result in every single run of the model. We also summarized the methods available to help us get a universal optimum result, which is certain to divide the market in a way that the substitutability within each group is the maximum. Our method to define the relevant market does not rely on any unrealistic assumptions. After the definition of the relevant market, we can use the Herfindahl-Hirschman Index (HHI) or other methods to measure the effect of a merger.

The significance of our method is that it gives an alternative approach to define a relevant product market which is directly based on substitutability between the products. Our method does not directly predict if the merger will cause a price increase because this is the purpose of the models assessing the unilateral effect. In contrast, our method helps to show a big picture of the market structure and to analyze the merging companies' market power on both demand and supply. On the other hand, since the Guidelines does not preclude the application of models other than the ones suggested by it, our method can at least participate as a supplement to help the Agencies and the court to get a better understanding of the substitutive relations between the products in the candidate relevant product market.

# APPENDIX: CODE USED IN

# EXAMPLES

## K-means example 1: Wholesales durable goods

industry group: Motor Vehicle and Motor Vehicle Parts and Supplies Merchant Wholesalers: NAICS 4231

### Genetic K-means clustering with two-steps approach
### Simulation Example of Merchant Build the simulated dataset

```
set.seed(1)
```

The first variable we simulate is the number of categories of products available in each wholesaler. We put 3 patterns in the data. In each pattern, we put 10 simulated values. The code is:

```
ctgr <- c(sample(1:50, 10, replace = TRUE),
          sample(100:200, 15, replace = TRUE),
          sample(250:300, 5, replace = TRUE))
ctgrscore <- (ctgr - mean(ctgr))/sd(ctgr)
```

The second variable we simulate is average number of products in each category available in each wholesaler. We put 2 patterns in the data. In each pattern, we put 15 simulated values. The code is:

```
nCtgr <- c(sample(1:10, 15, replace = TRUE),
           sample(20:30, 15, replace = TRUE))
nCtgrscore <- (nCtgr - mean(nCtgr))/sd(nCtgr)
```

The third variable we simulate is the number of brands usually used by customers. We put 2 patterns in the data. In the first pattern, we put 20 simulated values. In the second pattern we put 10 simuated values. The code is:

```
nBrnd <- c(sample(10:30, 20, replace = TRUE),
           sample(1:10, 10, replace = TRUE))
nBrndscore <- (nBrnd - mean(nBrnd))/sd(nBrnd)
```

The forth variable we simulate is the number of models used by customers. We put 2 patterns in the data. In the first pattern, we put 20 simulated values. In the second pattern, we put 10 simuated values. The code is:

```
nMdl <- c(sample(50:100, 20, replace = TRUE),
```

```
          sample(1:50, 10, replace = TRUE))
nMdlscore <- (nMdl - mean(nMdl))/sd(nMdl)
```

The fifth variable we simulate is the price deviation from average of a product which has an around-average number of sales. We put only 1 pattern in the data. The code is:

```
pPrdct <- sample(1:300, size = 30, replace = TRUE)
pPrdctscore <- (pPrdct - mean(pPrdct))/sd(pPrdct)
```

The sixth variable we simulate is price deviation from average of a category which has an around-average number of sales. We put only 1 pattern in the data. The code is:

```
pCtgr <- sample(1:100, size = 30, replace = TRUE)
pCtgrscore <- (pCtgr - mean(pCtgr))/sd(pCtgr)
```

The seventh variable we simulate is the percentage of non-parts sold. We put 2 patterns in the data. In the first pattern, we put 20 simulated values. In the second pattern, we put 10 simuated values. The code is:

```
pNp <- c(sample(30:70, size = 20, replace = TRUE),
          sample(1:100, size = 10, replace = TRUE))
pNpscore <- (pNp - mean(pNp))/sd(pNp)
```

The eighth variable we simulate is percentage of uncommon transportation products & parts. We put 3 patterns in the data. In each pattern, we put 10 simulated values. The code is:

```
larget <- sample(c(rep(0, times = 5),
                  sample(10:30, size = 5, replace =
TRUE)), size = 10)
mediant <- sample(c(rep(0, times = 6),
                  rep(100, times = 2), 20, 10), size =
10)
smallt <- sample(c(rep(0, times = 6), rep(100, times = 4)),
size = 10)
pUt <- c(larget, mediant, smallt)
pUtscore <- (pUt - mean(pUt))/sd(pUt)
```

The ninth variable we simulate is percentage of used products. We put 3 patterns in the data. In each pattern, we put 10 simulated values. The code is:

```
largeu <- sample(c(rep(0, times = 8), 3, 5), size = 10)
medianu <- sample(c(rep(0, times = 7), 3, 5, 8), size = 10)
smallu <- sample(c(rep(0, times = 6),
                  sample(1:20, size = 4, replace = TRUE)),
size = 10)
pUp <- c(largeu, medianu, smallu)
```

```
pUpscore <- (pUp - mean(pUp))/sd(pUp)
```

Now we build our simulated data into a data frame, so it can be later used in clustering analysis.

```
simMV <- cbind(ctgrscore, nCtgrscore, nBrndscore,
nMdlscore,
           pPrdctscore, pCtgrscore, pNpscore, pUtscore,
pUpscore)
set.seed(Sys.time())
```

## Step 1: hierachical clustering to find a proper k.

```
library(cluster)
hcMV <- hclust(dist(simMV))
plot(hcMV, main = "Hierarchical Clustering to find K (MV)",
     xlab = "Numbered Wholesalers")
abline(h = 4.6, col = "blue")
abline(h = 5.1, col = "red")
abline(h = 5.5, col = "green")
```

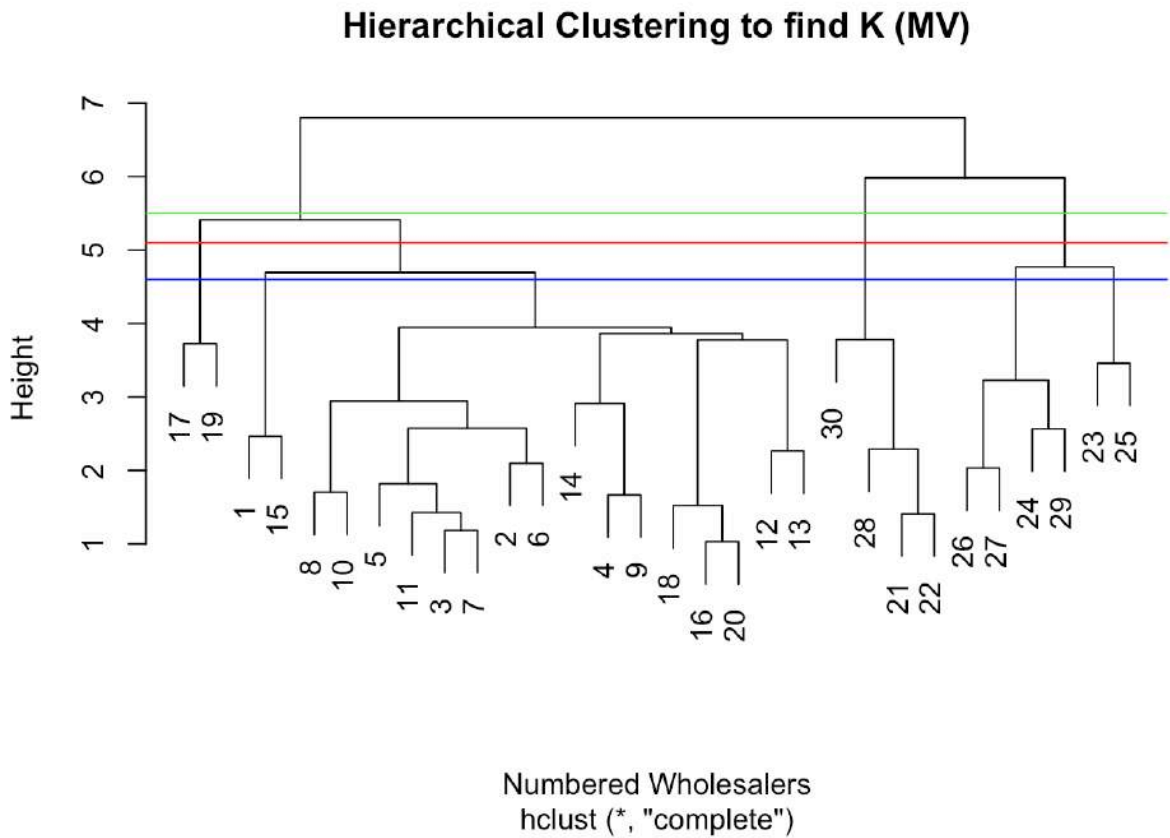

**Hierarchical Clustering to find K (MV)**

Numbered Wholesalers
hclust (*, "complete")

## Step 2: Genetic k-means clustering

Conduct k-means with k equals to 3 from 100 different starting points for 100 times

```r
## k-means with k = 3
## repeat k-means clustering for 100 times with different
start points

totWithinSq <- vector()
for (i in 1:100){
        set.seed(i)
        kmMV3 <- kmeans(simMV, 3)
        totWithinSq <- c(kmMV3$tot.withinss, totWithinSq)
}
## find the result with the smallest within-group sum of
squares
sed <- c(1:100)[totWithinSq == min(totWithinSq)][1]
## redo the one with the smallest within-group sum of
squares
set.seed(sed)
kmMV3 <- kmeans(simMV, 3)
```

Plot the result with the smallest within-group sum of squares

```r
## plot the result with the smallest within-group sum of
squares
clusplot(simMV, kmMV3$cluster, color=TRUE, shade=TRUE,
labels=2,
        lines=0, main = "Market division of Motor Vehicle
while K = 3",
        xlab = "Number of product categories available",
        ylab = "Number of products in each category")
```

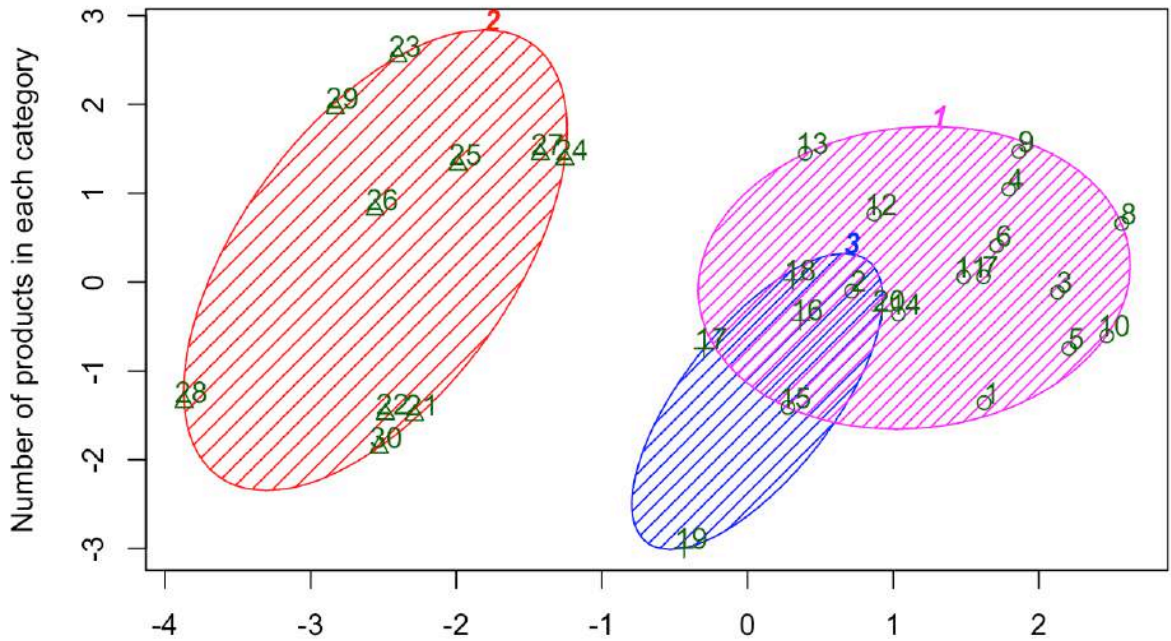

**Market division of Motor Vehicle while K = 3**

Number of product categories available
These two components explain 56.08 % of the point variability.

## Other relevant aspects of the result

```
## label of clusters
kmMV3$cluster
## [1] 1 1 1 1 1 1 1 1 1 1 1 1 1 1 1 1 3 3 3 3 3 2 2 2 2 2 2
2 2 2 2
## cordinates of centers
kmMV3$centers
##     ctgrscore nCtgrscore nBrndscore   nMdlscore
pPrdctscore pCtgrscore
## 1 -0.7217317 -0.9487876  0.4389097  0.6538996
-0.10778829 -0.2323299
## 2  0.8701004  0.8935854 -1.1928434 -1.2187335
-0.07775567  0.6471109
## 3  0.4249945  1.0591919  1.0689576  0.4757683
0.47887620 -0.5972322
##      pNpscore    pUtscore    pUpscore
## 1  0.3242067 -0.3780789 -0.27590332
```

```
## 2 -0.6220715   0.3780789   0.38459250
## 3  0.2715231   0.3780789   0.05852495
## total squared distance from data points to their centers
## it is equal to sum of squares because the center is the
mean
## in each dimensions
kmMV3$withinss
## [1] 56.57548 73.02529 21.88084
## total within-cluster sum of distances
kmMV3$tot.withinss
## [1] 151.4816
```

Conduct k-means with k equals to 4 from 100 different starting points for 100 times

```
## k-means with k = 4
## repeat k-means clustering for 100 times with different
start points

totWithinSq <- vector()
for (i in 1:100){
        set.seed(i)
        kmMV4 <- kmeans(simMV, 4)
        totWithinSq <- c(kmMV4$tot.withinss, totWithinSq)
}
## find the result with the smallest within-group sum of
squares
sed <- c(1:100)[totWithinSq == min(totWithinSq)][1]
## redo the one with the smallest within-group sum of
squares
set.seed(sed)
kmMV4 <- kmeans(simMV, 4)
```

Plot the result with the smallest within-group sum of squares

```
## plot the result with the smallest within-group sum of
squares
clusplot(simMV, kmMV4$cluster, color=TRUE, shade=TRUE,
labels=2,
```

```
        lines=0, main = "Market division of Motor Vehicle
while K = 4",
        xlab = "Number of product categories available",
        ylab = "Number of products in each category")
```

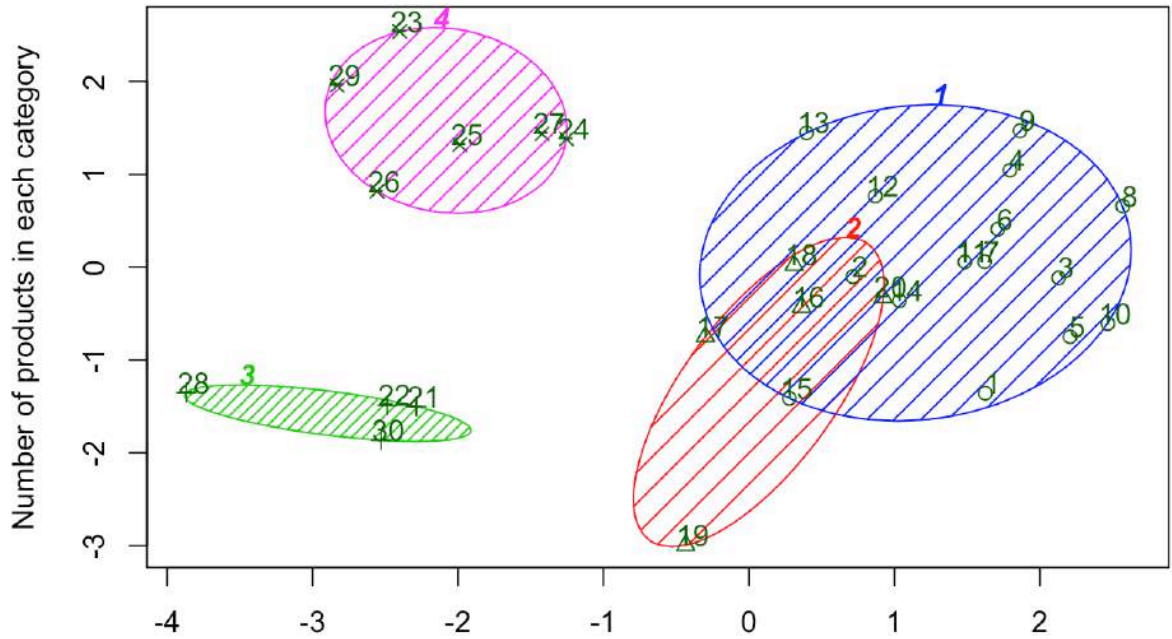

**Market division of Motor Vehicle while K = 4**

Number of product categories available
These two components explain 56.08 % of the point variability.

## Other relevant aspects of the result

```
## label of clusters
kmMV4$cluster
##  [1] 1 1 1 1 1 1 1 1 1 1 1 1 1 1 1 2 2 2 2 2 3 3 4 4 4 4
4 3 4 3
## cordinates of centers
kmMV4$centers
##     ctgrscore nCtgrscore nBrndscore   nMdlscore
pPrdctscore pCtgrscore
## 1 -0.7217317 -0.9487876  0.4389097  0.6538996
-0.1077883 -0.2323299
## 2  0.4249945  1.0591919  1.0689576  0.4757683
0.4788762 -0.5972322
```

```
## 3   0.9389722   1.0747176  -1.2406279  -0.9498455
0.8065608   1.0074191
## 4   0.8241858   0.7728306  -1.1609870  -1.3979922
-0.6673000   0.4069054
##        pNpscore     pUtscore     pUpscore
## 1   0.3242067  -0.3780789  -0.27590332
## 2   0.2715231   0.3780789   0.05852495
## 3  -0.8165955   1.9039576  -0.49328169
## 4  -0.4923888  -0.6391737   0.96984197
## total squared distance from data points to their centers
## it is equal to sum of squares because the center is the
mean
## in each dimensions
kmMV4$withinss
## [1] 56.57548 21.88084 12.33399 32.95276
## total within-cluster sum of distances
kmMV4$tot.withinss
## [1] 123.7431
```

Conduct k-means with k equals to 5 from 100 different starting points for 100 times

```
## k-means with k = 5
## repeat k-means clustering for 100 times with different
start points

totWithinSq <- vector()
for (i in 1:100){
        set.seed(i)
        kmMV5 <- kmeans(simMV, 5)
        totWithinSq <- c(kmMV5$tot.withinss, totWithinSq)
}
## find the result with the smallest within-group sum of
squares
sed <- c(1:100)[totWithinSq == min(totWithinSq)][1]
## redo the one with the smallest within-group sum of
squares
set.seed(sed)
kmMV5 <- kmeans(simMV, 5)
```

## Plot the result with the smallest within-group sum of squares

```
## plot the result with the smallest within-group sum of
squares
clusplot(simMV, kmMV5$cluster, color=TRUE, shade=TRUE,
labels=2,
        lines=0, main = "Market division of Motor Vehicle
while K = 5",
        xlab = "Number of product categories available",
        ylab = "Number of products in each category")
```

**Market division of Motor Vehicle while K = 5**

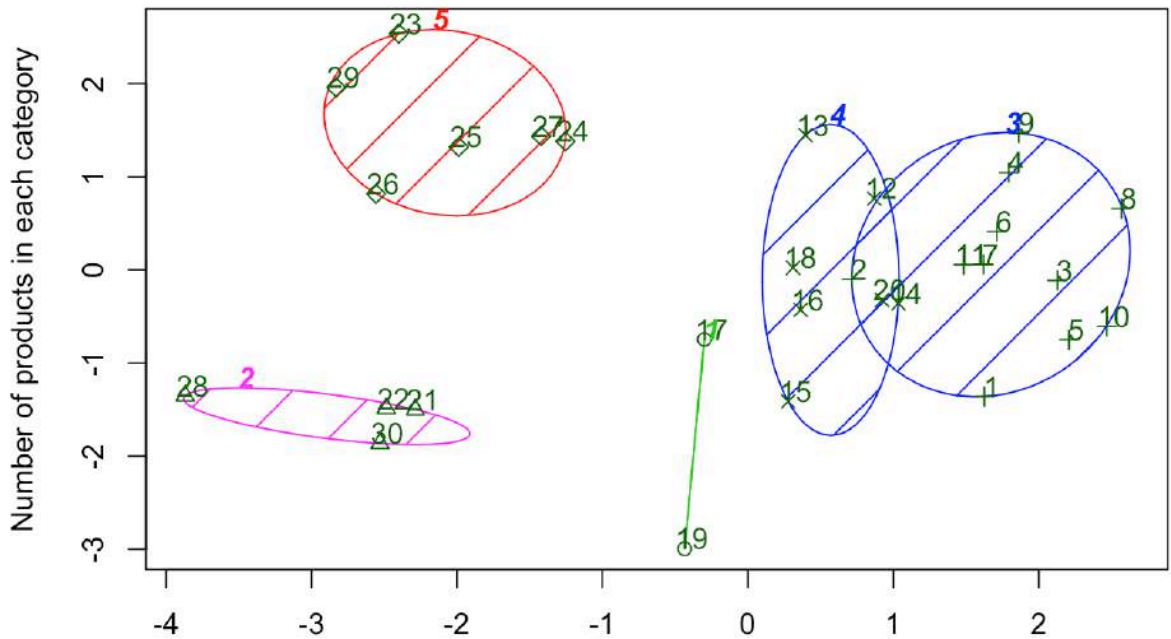

Number of product categories available
These two components explain 56.08 % of the point variability.

## Other relevant aspects of the result

```
## label of clusters
kmMV5$cluster
## [1] 3 3 3 3 3 3 3 3 3 3 3 4 4 4 4 1 4 1 4 2 2 5 5 5 5
5 2 5 2
## cordinates of centers
kmMV5$centers
```

```
##      ctgrscore    nCtgrscore nBrndscore   nMdlscore
pPrdctscore pCtgrscore
## 1   0.2815589  0.893585379   1.1751454   0.5806177
0.98860783 -0.5184763
## 2   0.9389722  1.074717551  -1.2406279  -0.9498455
0.80656082  1.0074191
## 3  -1.0669018 -0.969488385   0.4318305   0.6590242
-0.03186524 -0.6366101
## 4   0.3531141 -0.008378903   0.6897153   0.5395477
-0.12130590  0.2240793
## 5   0.8241858  0.772830598  -1.1609870  -1.3979922
-0.66730000  0.4069054
##       pNpscore     pUtscore    pUpscore
## 1 -0.5126518  1.9039576   0.8862349
## 2 -0.8165955  1.9039576  -0.4932817
## 3  0.4406263 -0.3524934  -0.3108663
## 4  0.3427327 -0.5301823  -0.3141237
## 5 -0.4923888 -0.6391737   0.9698420
## total squared distance from data points to their centers
## it is equal to sum of squares because the center is the mean
## in each dimensions
kmMV5$withinss
## [1]   6.943531 12.333986 28.328260 31.108963 32.952755
## total within-cluster sum of distances
kmMV5$tot.withinss
## [1] 111.6675
```

# K-means example 2: Example of Insurance Carriers and Related Activities

Industry Group: Car Insurance

## Genetic K-means clustering with two-steps approach

Read in data. Adjust it into the required form. Normalize the data

```
cidata <- read.csv(file = "/Users/YanYang/code/
Car_Insurance_Example/2014 Auto Insurance Average Premium
Data.csv")
lap <- sapply(1:nrow(cidata), function(x)
mean(as.numeric(cidata[x, 3:7])))
cap <- sapply(1:nrow(cidata), function(x)
mean(as.numeric(cidata[x, 8:12])))
cpap <- sapply(1:nrow(cidata), function(x)
mean(as.numeric(cidata[x, 13:17])))
simCI <- scale(cbind(lap, cap, cpap))
```

**Step 1: hierachical clustering to find a proper k.**

```
hcCI <- hclust(dist(simCI))
plot(hcCI, main = "Hierarchical Clustering to find K (CI)",
     xlab = "State Numbers")
abline(h = 3.5, col = "red")
abline(h = 2.5, col = "blue")
```

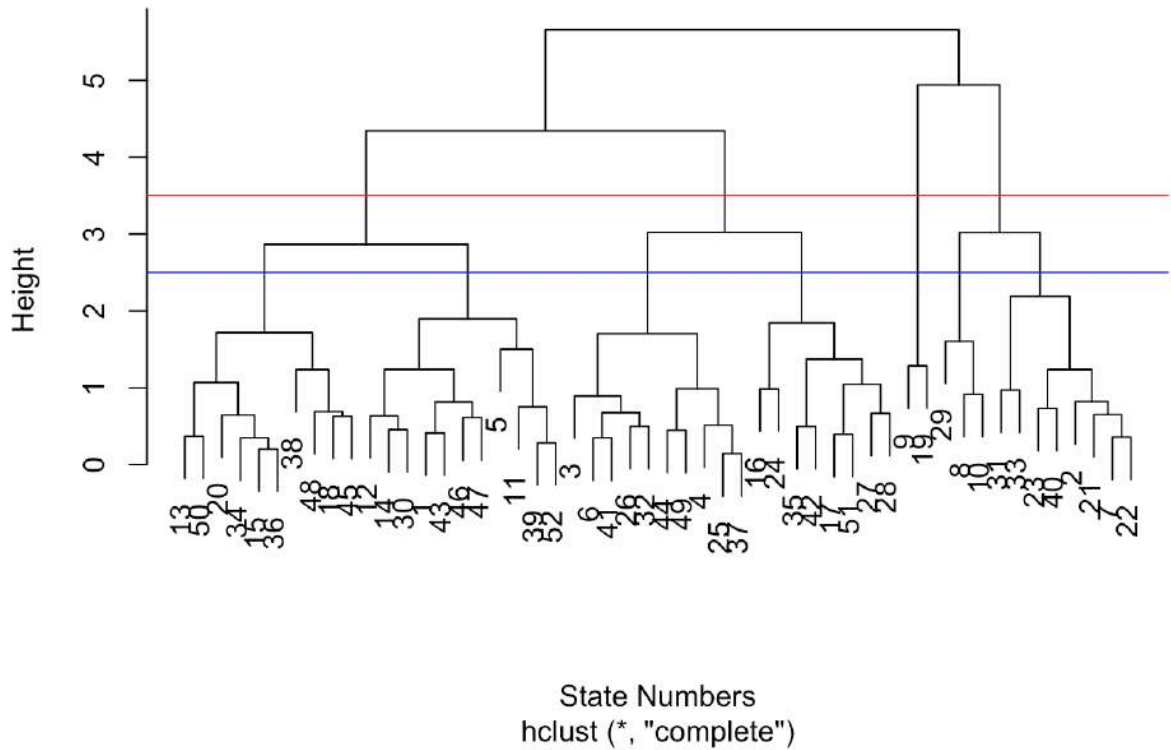

**Hierarchical Clustering to find K (CI)**

State Numbers
hclust (*, "complete")

## Step 2: Genetic k-means clustering

Conduct k-means with k equals to 4 from 100 different starting points for 100 times

```
## k-means clustering when k = 4
## repeat k-means clustering for 100 times with different
start points
totWithinSq <- vector()
for (i in 1:100){
        set.seed(i)
        kmCI4 <- kmeans(simCI, 4)
        totWithinSq <- c(kmCI4$tot.withinss, totWithinSq)
}
## find the result with the smallest within-group sum of
squares
sed <- c(1:100)[totWithinSq == min(totWithinSq)][1]
```

```
## redo the one with the smallest within-group sum of
squares
set.seed(sed)
kmCI4 <- kmeans(simCI, 4)
```

## Plot the result with the smallest within-group sum of squares

```
## plot the result with the smallest within-group sum of
squares
clusplot(simCI, kmCI4$cluster, color=TRUE, shade=TRUE,
labels=2,
         lines=0, main = "Market division of Car Insurance
while K = 4",
         xlab="Average Liablity Premium", ylab="Average
Collision Premium")
```

### Market division of Car Insurance while K = 4

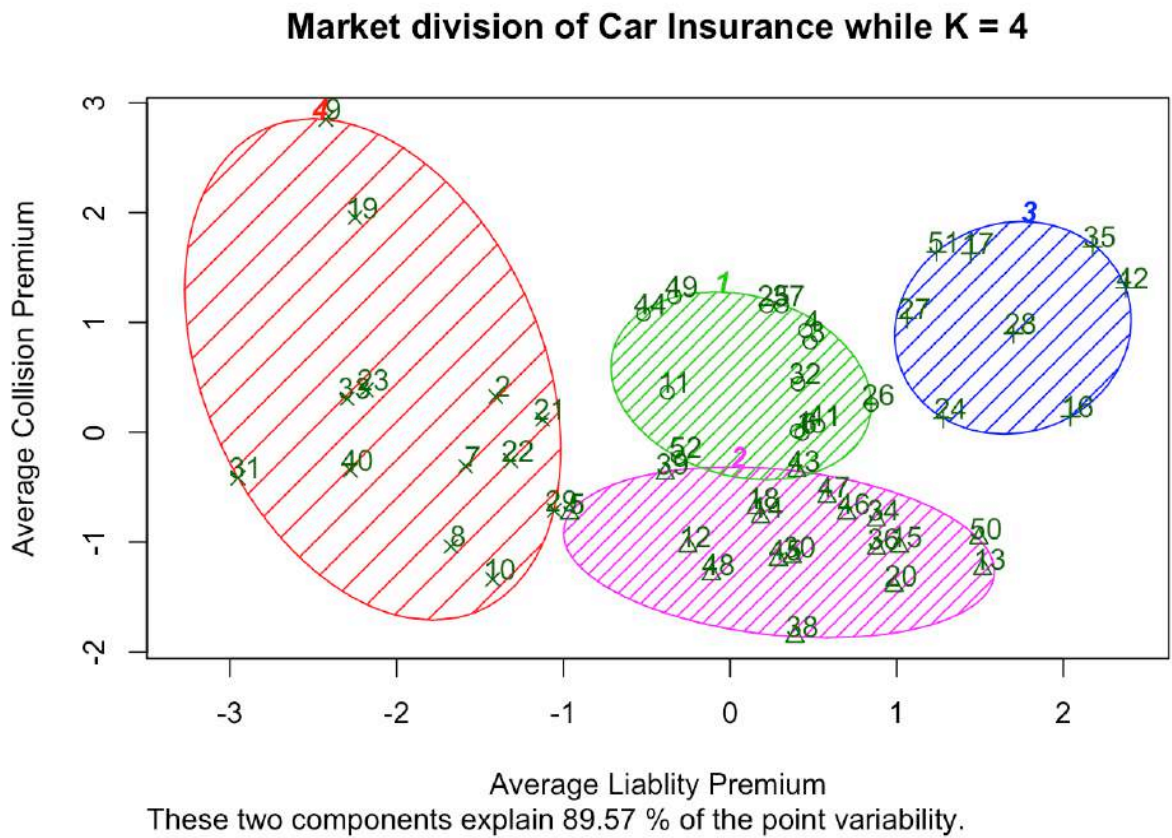

These two components explain 89.57 % of the point variability.

## Other relevant aspects of the result

```
## label of clusters
kmCI4$cluster
```

```
## [1] 1 4 1 1 2 1 4 4 4 4 1 2 2 2 3 3 2 4 2 4 4 4 3 1 1
3 3 4 2 4 1 4 2 3
## [36] 2 1 2 2 4 1 3 2 1 2 2 2 2 1 2 3 1
## cordinates of centers
kmCI4$centers
##          lap        cap        cpap
## 1 -0.2162251  0.0671289  0.5416765
## 2 -0.4250923 -0.4167178 -0.9077268
## 3 -1.0233950 -1.0358827  1.3526659
## 4  1.4345960  1.1473312 -0.1172338
## total squared distance from data points to their centers
## it is equal to sum of squares because the center is the
mean
## in each dimensions
kmCI4$withinss
## [1]  7.330818 13.474020  4.984122 25.873355
## total within-cluster sum of distances
kmCI4$tot.withinss
## [1] 51.66232
```

Conduct k-means with k equals to 7 from 100 different starting points for 100 times

```
## k-means clustering when k = 7
## repeat k-means clustering for 100 times with different
start points
totWithinSq <- vector()
for (i in 1:100){
        set.seed(i)
        kmCI7 <- kmeans(simCI, 7)
        totWithinSq <- c(kmCI7$tot.withinss, totWithinSq)
}
## find the result with the smallest within-group sum of
squares
sed <- c(1:100)[totWithinSq == min(totWithinSq)][1]
## redo the one with the smallest within-group sum of
squares
set.seed(sed)
kmCI7 <- kmeans(simCI, 7)
```

## Plot the result with the smallest within-group sum of squares

```
## plot the result with the smallest within-group sum of
squares
clusplot(simCI, kmCI7$cluster, color=TRUE, shade=TRUE,
labels=2,
        lines=0, main = "Market division of Car Insurance
while K = 7",
        xlab="Average Liablity Premium", ylab="Average
Collision Premium")
```

**Market division of Car Insurance while K = 7**

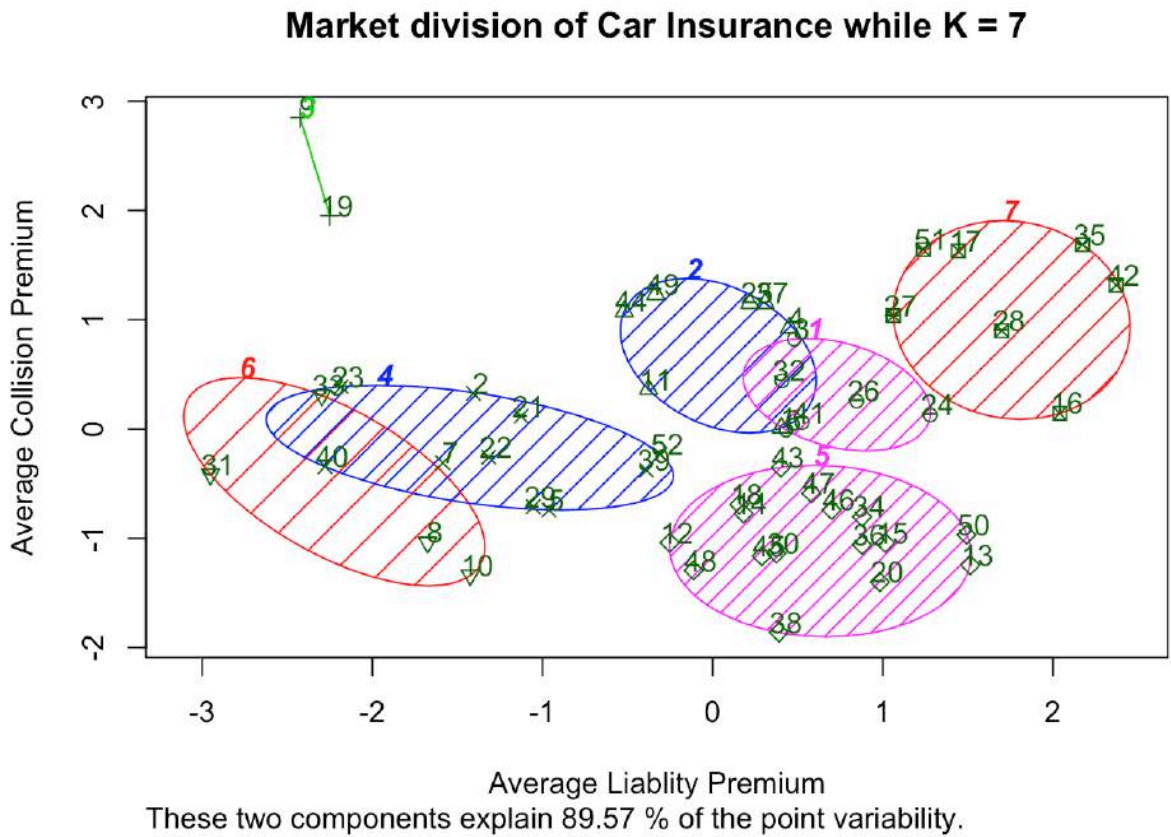

These two components explain 89.57 % of the point variability.

## Other relevant aspects of the result

```
## label of clusters
kmCI7$cluster
##  [1] 2 4 1 2 4 1 4 6 3 6 2 5 5 5 5 7 7 5 3 5 4 4 4 1 2 1
7 7 4 5 6 1 6 5 7
## [36] 5 2 5 4 4 1 7 5 2 5 5 5 5 2 5 7 4
## cordinates of centers
kmCI7$centers
```

```
##             lap         cap        cpap
## 1 -0.2043081 -0.6370889  0.4854048
## 2 -0.3011256  0.4464633  0.7098097
## 3  1.3140978  2.4377592  1.8389374
## 4  0.7083975  0.9803677 -0.4775267
## 5 -0.4727496 -0.5727570 -0.9136164
## 6  2.2576671  0.5235082 -0.6085935
## 7 -1.1207312 -0.9874039  1.4669332
## total squared distance from data points to their centers
## it is equal to sum of squares because the center is the
mean
## in each dimensions
kmCI7$withinss
## [1] 1.5108276 2.8056016 0.8263802 7.0045401 8.8901085
3.8771411 3.5907569
## total within-cluster sum of distances
kmCI7$tot.withinss
## [1] 28.50536
```

# Use Elbow Method and Gap Statistic to find K

## Elbow Method Function

**For easier use later,we write a function to calculate Wk in elbow method with input, and the largest meaningful k**

The input "data" should be a dataframe. maxk is the largest k.

```r
elbow <- function(data, maxk){
        vc <- vector()
        for(j in 1:maxk){
                km <- kmeans(data, j)
                datak <- cbind(data, km$cluster)
                v <- vector()
                for(i in 1:j){
                        count <- sum(datak[ ,ncol(datak)]
== i)
                        if(count == 1){
                                sdgi == 0
                        }else{
                                dgi <-
dist(datak[datak[ ,ncol(datak)] == i,
                                                c(1:3)],

method = "euclidean",

                                                diag
= TRUE,

upper = TRUE)^2
                                sdgi <-
sum(as.numeric(dgi))/(2*count)
                        }
                        v <- c(v, sdgi)
                }
                vc <- c(vc, sum(v))
        }
        vc
}
```

# Example 1: whole seller dataset
## Use elbow method to find k

Calculate Wk for from 100 different start points when k changes from 1 to 20

Since the random start points will affect Wk we get, we use 100 sets of random numbers provided by R to avoid the effect of randomness.

```r
vv <- vector()
for(i in 1:100){
        set.seed(i)
        wk <- elbow(simMV, 20)
        vv <- rbind(vv, wk)
}
meanWkMV <- colMeans(vv)
```

Observe the figure below to find the k where Wk decrease fastest.

```r
cc <- as.data.frame(cbind(1:20, meanWkMV))
library(ggplot2)
g = ggplot(cc, aes(x = V1, y = meanWkMV))
g = g + geom_point(size = 5)
g = g + geom_line(size = 1)
g = g + labs(x = "k", y = "mean of 100 Wk",
             title = "Example 1: Elbow Method to find k")
g = g + scale_color_gradient2(low = "purple", high = "red")
g
```

Example 1: Elbow Method to find k

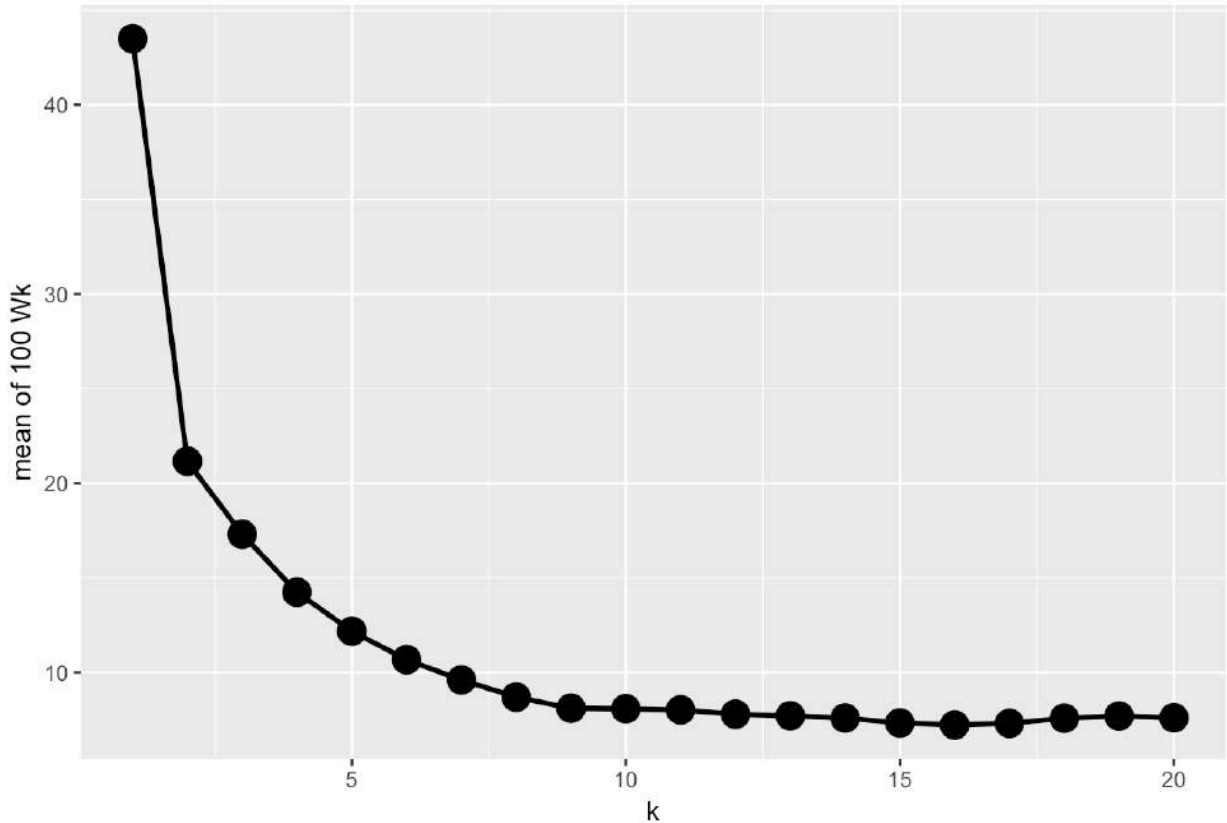

From the plot we know k should be 2.

## Calculate Gap Statistic to determind K

Since R already has a package to calculate Gap Statistic, we will use the function from package cluster directly.

```
library(cluster)
gap_clust <- clusGap(simMV, kmeans, K.max = 20, B = 30)
gap_clust
## Clustering Gap statistic ["clusGap"] from call:
## clusGap(x = simMV, FUNcluster = kmeans, K.max = 20, B =
30)
## B=30 simulated reference sets, k = 1..20;
spaceH0="scaledPCA"
##  --> Number of clusters (method 'firstSEmax',
SE.factor=1): 2
##            logW     E.logW       gap      SE.sim
##  [1,] 3.382692 3.519929 0.1372375 0.03233278
##  [2,] 3.172895 3.360477 0.1875819 0.04082617
```

```
##  [3,] 3.089048 3.248258 0.1592098 0.03538789
##  [4,] 2.950934 3.154275 0.2033417 0.03680279
##  [5,] 2.878570 3.069781 0.1912115 0.03568238
##  [6,] 2.803485 2.986658 0.1831729 0.03518157
##  [7,] 2.674556 2.910726 0.2361702 0.04006084
##  [8,] 2.593718 2.834799 0.2410807 0.03378752
##  [9,] 2.492509 2.760599 0.2680900 0.03932611
## [10,] 2.442075 2.688316 0.2462415 0.04048860
## [11,] 2.351627 2.611193 0.2595656 0.03915989
## [12,] 2.284554 2.533258 0.2487041 0.04585996
## [13,] 2.191623 2.456227 0.2646039 0.04196856
## [14,] 2.188607 2.374776 0.1861685 0.03743426
## [15,] 2.034108 2.295494 0.2613855 0.05008610
## [16,] 1.912181 2.205446 0.2932655 0.03806879
## [17,] 1.887987 2.106953 0.2189663 0.04487602
## [18,] 1.743643 2.007366 0.2637225 0.04176051
## [19,] 1.580676 1.901965 0.3212900 0.04092718
## [20,] 1.554539 1.798900 0.2443611 0.05964913
```
From the result we know Gap statistic suggests the best k is 2

## Both method suggests k to be 2

k-means with k = 2

repeat k-means clustering for 100 times with different start points

```
totWithinSq <- vector()
for (i in 1:100){
        set.seed(i)
        kmMV2 <- kmeans(simMV, 2)
        totWithinSq <- c(kmMV5$tot.withinss, totWithinSq)
}
## find the result with the smallest within-group sum of
squares
sed <- c(1:100)[totWithinSq == min(totWithinSq)][1]
```

```
## redo the one with the smallest within-group sum of
squares
set.seed(sed)
kmMV2 <- kmeans(simMV, 2)
```

Plot the result with the smallest within-group sum of squares

```
clusplot(simMV, kmMV2$cluster, color=TRUE, shade=TRUE,
labels=2,
         lines=0, main = "Market division of Motor Vehicle
while K = 2",
         xlab = "Number of product categories available",
         ylab = "Number of products in each category")
```

**Market division of Motor Vehicle while K = 2**

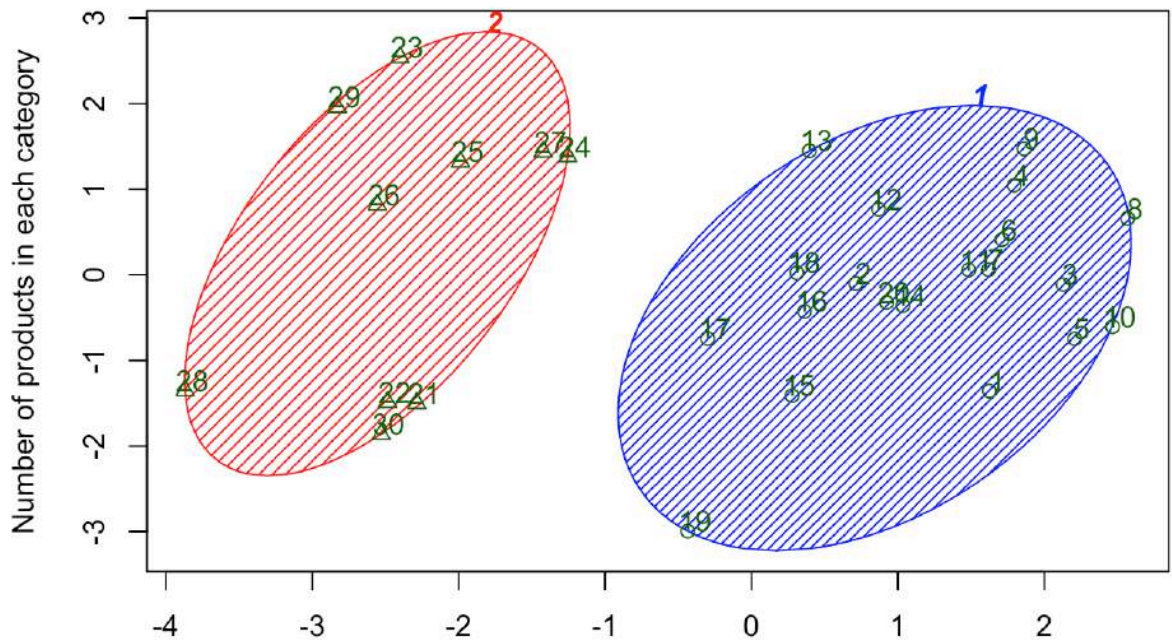

Number of product categories available
These two components explain 56.08 % of the point variability.

### Other relevant aspects of the result

```
## label of clusters
kmMV2$cluster
##  [1] 1 1 1 1 1 1 1 1 1 1 1 1 1 1 1 1 1 1 1 1 2 2 2 2 2 2
2 2 2 2
## cordinates of centers
```

```
kmMV2$centers
##       ctgrscore nCtgrscore nBrndscore   nMdlscore
pPrdctscore pCtgrscore
## 1 -0.4350502 -0.4467927  0.5964217  0.6093667
0.03887784 -0.3235554
## 2  0.8701004  0.8935854 -1.1928434 -1.2187335
-0.07775567  0.6471109
##       pNpscore   pUtscore   pUpscore
## 1  0.3110358 -0.1890394 -0.1922963
## 2 -0.6220715  0.3780789  0.3845925
## total squared distance from data points to their centers
## it is equal to sum of squares because the center is the
mean
## in each dimensions
kmMV2$withinss
## [1] 104.47897  73.02529
## total within-cluster sum of distances
kmMV2$tot.withinss
## [1] 177.5043
```

# Example 2: car insurance dataset

Calculate Wk for from 100 different start points when k changes from 1 to 20

```
vv <- vector()
for(i in 1:100){
        set.seed(i)
        wk <- elbow(simCI, 20)
        vv <- rbind(vv, wk)
}
meanWk <- colMeans(vv)
```

Observe the figure below to find the k where Wk decrease fastest

```
cc <- as.data.frame(cbind(1:20, meanWk))
library(ggplot2)
g = ggplot(cc, aes(x = V1, y = meanWk))
g = g + geom_point(size = 5)
g = g + geom_line(size = 1)
g = g + labs(x = "k", y = "mean of 100 Wk",
            title = "Example 2: Elbow Method to find k")
g = g + scale_color_gradient2(low = "purple", high = "red")
g
```

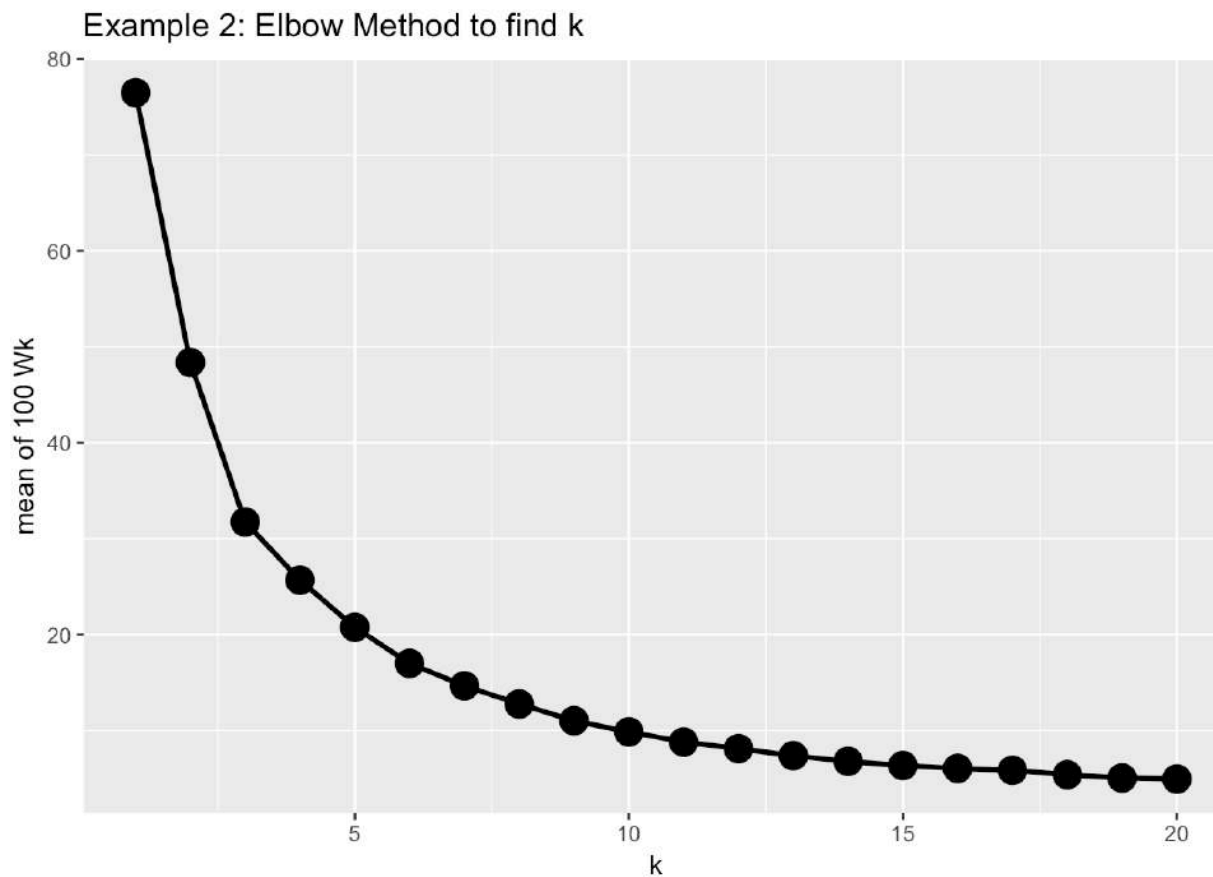

Example 2: Elbow Method to find k

From the plot we know k should be 3

---

## Calculate Gap Statistic to determind K

```
library(cluster)
gap_clust <- clusGap(simCI, kmeans, K.max = 20, B = 52)
gap_clust
## Clustering Gap statistic ["clusGap"] from call:
## clusGap(x = simCI, FUNcluster = kmeans, K.max = 20, B =
52)
## B=52 simulated reference sets, k = 1..20;
spaceH0="scaledPCA"
##  --> Number of clusters (method 'firstSEmax',
SE.factor=1): 3
##           logW    E.logW      gap      SE.sim
##  [1,] 3.339359 3.615619 0.2762602 0.03735108
##  [2,] 3.086954 3.363500 0.2765454 0.04025360
##  [3,] 2.860584 3.194636 0.3340520 0.05045086
##  [4,] 2.764690 3.054210 0.2895200 0.04720784
```

```
##  [5,] 2.691983 2.949296 0.2573135 0.05029899
##  [6,] 2.544050 2.855352 0.3113014 0.04983119
##  [7,] 2.526056 2.768268 0.2422118 0.04872225
##  [8,] 2.366667 2.688637 0.3219698 0.04943543
##  [9,] 2.284573 2.618430 0.3338567 0.05184171
## [10,] 2.230376 2.540910 0.3105346 0.05384160
## [11,] 2.142184 2.489855 0.3476711 0.05281161
## [12,] 2.090391 2.413626 0.3232352 0.05258721
## [13,] 2.045172 2.357836 0.3126637 0.06204036
## [14,] 1.962233 2.292731 0.3304978 0.05859114
## [15,] 1.927258 2.228948 0.3016906 0.06271220
## [16,] 1.877180 2.163823 0.2866432 0.05876341
## [17,] 1.796337 2.109715 0.3133781 0.06329356
## [18,] 1.763847 2.050724 0.2868765 0.06414906
## [19,] 1.681757 1.999434 0.3176773 0.06064812
## [20,] 1.692030 1.938623 0.2465932 0.07002069
```

From the result we know Gap statistic suggests the best k is 3

**Both method suggests k to be 3**

## k-means clustering when k = 3

Repeat k-means clustering for 100 times with different start points

```
totWithinSq <- vector()
for (i in 1:100){
        set.seed(i)
        kmCI3 <- kmeans(simCI, 3)
        totWithinSq <- c(kmCI3$tot.withinss, totWithinSq)
}
## find the result with the smallest within-group sum of
squares
sed <- c(1:100)[totWithinSq == min(totWithinSq)][1]
## redo the one with the smallest within-group sum of
squares
set.seed(sed)
kmCI3 <- kmeans(simCI, 3)
```

Plot the result with the smallest within-group sum of squares

```
clusplot(simCI, kmCI3$cluster, color=TRUE, shade=TRUE,
labels=2,
        lines=0, main = "Market division of Car Insurance
while K = 3",
        xlab="Average Liablity Premium", ylab="Average
Collision Premium")
```

**Market division of Car Insurance while K = 3**

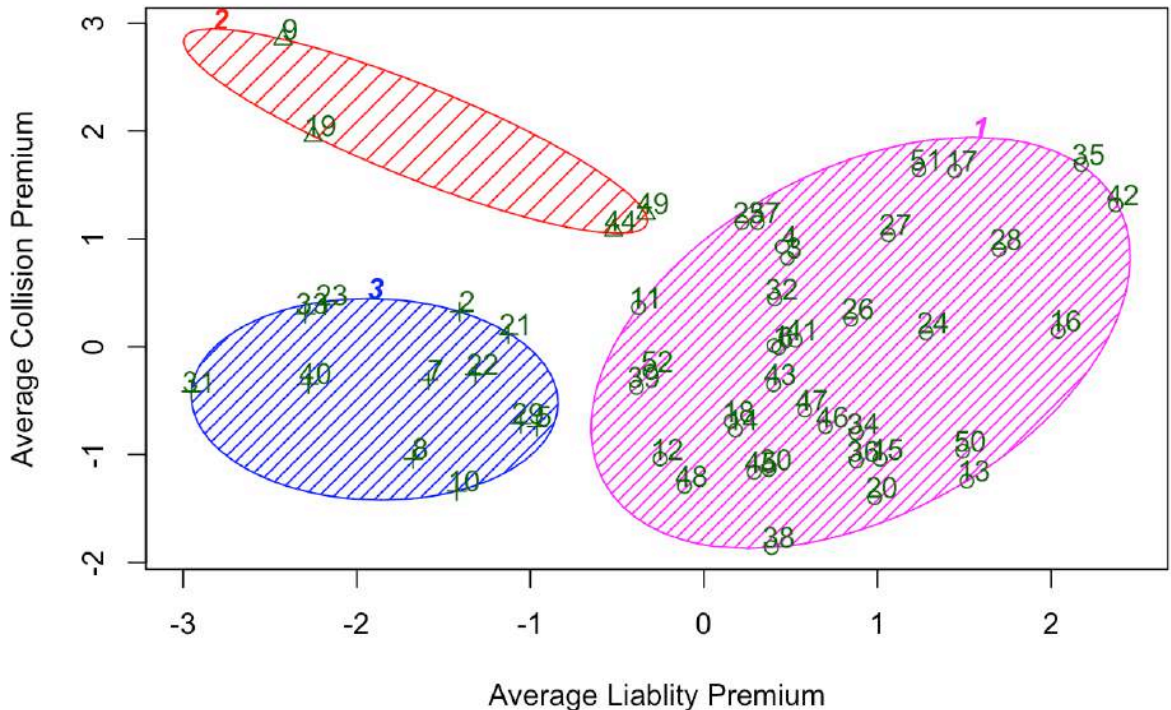

Average Liablity Premium

These two components explain 89.57 % of the point variability.

###Other relevant aspects of the result

```
## label of clusters
kmCI3$cluster
##  [1] 1 3 1 1 3 1 3 3 2 3 1 1 1 1 1 1 1 1 2 1 3 3 3 1 1 1
1 1 3 1 3 1 3 1 1
## [36] 1 1 1 1 3 1 1 1 2 1 1 1 1 2 1 1 1
## cordinates of centers
kmCI3$centers
##           lap        cap        cpap
## 1 -0.5193289 -0.4909840  0.02171412
## 2  0.7130070  1.5820433  1.40981372
```

```
## 3  1.3203177  0.9456041 -0.53508027
## total squared distance from data points to their centers
## it is equal to sum of squares because the center is the mean
## in each dimensions
kmCI3$withinss
## [1] 58.344648  6.037484 15.133394
## total within-cluster sum of distances
kmCI3$tot.withinss
## [1] 79.51553
```

# ACKNOWLEDGEMENTS

At the end of this dissertation, I would like to say thank you to my dissertation committee -- Professor Drobak, Professor De Geest and Professor Cramer for all the suggestions and help in accomplishing this dissertation. I would also say thank you to the staff members in Washington University School of Law who have been very supportive of me for all these years in Washington University School of Law.

To my husband, Jonathan Mailoa, who have helped me proofread my dissertation: I still have so much to learn in English, just as I have so much to learn to be a wife. Thank you for always being so patient and providing me a real home in the U.S.

The most gratitude goes to my beloved parents and my grandma: No words can express how grateful I am to you. I am never smart enough to succeed in the first try for any of the important tasks in my life. Thank you for always having my back and giving me the support to try again until I make it. Thank you for having faith in me no matter how many times I fail. My cute, selfless and robust mom is always the best mom in my eyes. My father's quiet but attentive support has always been my armor. My grandma's kindness and magnanimousness have taught me what kind of person I want to be. I would also say thank you to my beloved brother: Thank you for being a role model for me for all those years, even though I am so much less smart than you that your suggestions are usually not that useful to me (sigh). Thank you for taking care of mom and dad when I am not around. You are the most awesome brother in my eyes.

I also want to say thank you to my best friends, Xi Chen, Sisi Zhou, and Yun Ding:

Thank you for spending so much time listening to me when I was down, and staying on my side when I was confused and distressed. A special thanks to Sisi: you saved me from the edge of being depressed. Your company in my darkest days means a lot to me.

Finally, to all of my friends and the extended family members: it is a bliss to have you all in my life. Thank you all for being a part of my life!